\documentclass{aastex631}

\usepackage{subfigure}
\usepackage{booktabs}
\usepackage{multirow}
\usepackage{amsmath}
\usepackage{array}
\usepackage{soul}

\newcommand{\TabFirstGrid}{}
\newcommand{\TabSecondGrid}{}
\newcommand{\TabThirdGrid}{}

\usepackage{gensymb}
\usepackage{hyperref}

\shorttitle{Estimation of Photometric redshifts}
\shortauthors{Lee et al.}

\begin{document}

\title{Estimation of Photometric Redshifts. I. Machine Learning Inference for 
Pan-STARRS1 Galaxies Using Neural Networks}

\author[0000-0003-2510-4132]{Joongoo Lee}
\affiliation{Korea Astronomy and Space Science Institute, 776,
Daedeokdae-ro, Yuseong-gu,                                                      
Daejeon 34055, Republic of Korea}
\affiliation{Department of Physics and Astronomy, 
Seoul National University, 1, Gwanak-ro, Gwanak-gu,
Seoul 08826, Republic of Korea}
\email{goolee5286@gmail.com}

\author[0000-0002-9934-3139]{Min-Su Shin}
\correspondingauthor{Min-Su Shin}
\affiliation{Korea Astronomy and Space Science Institute, 776,
Daedeokdae-ro, Yuseong-gu,                                                      
Daejeon 34055, Republic of Korea}
\email{msshin@kasi.re.kr}

\begin{abstract}

We present a new machine learning model for estimating photometric redshifts 
with improved accuracy for galaxies in Pan-STARRS1 data release 1. 
Depending on the estimation range of redshifts, this model based on neural 
networks can handle the difficulty for inferring photometric redshifts. 
Moreover, to reduce bias induced by the new model's ability to deal with estimation 
difficulty, it exploits the power of ensemble learning. We extensively examine the 
mapping between input features 
and target redshift spaces to which the model is validly applicable  
to discover the strength and weaknesses of trained model. Because our trained model 
is well calibrated, our model produces 
reliable confidence information about objects with non-catastrophic estimation. While our model 
is highly accurate for most test examples residing in the input space, 
where training samples are densely populated, its accuracy quickly diminishes 
for sparse samples and unobserved objects (i.e., unseen samples) 
in training. 
We report that out-of-distribution (OOD) samples for our model contain 
both physically OOD objects (i.e., stars and quasars) and 
galaxies with observed properties not represented by training 
  data. The code for our model is available at \href{https://github.com/GooLee0123/MBRNN}{https://github.com/GooLee0123/MBRNN} for 
other uses of the model and retraining the model with different data.
\end{abstract}

\section{Introduction} \label{sec:intro}

For various astronomical studies, the photometric redshifts of galaxies are critical. 
Representative research areas include cosmological model testing 
\citep{10.1111/j.1365-2966.2005.09526.x,10.1093/mnras/sty1624} 
and dark energy survey \citep{10.1111/j.1365-2966.2008.13095.x,10.1093/mnras/stu1836}.
In terms of accuracy, although the spectroscopic estimation of redshifts is 
the most appropriate method,   
acquiring spectroscopic redshifts is significantly more expensive than estimating 
photometric redshifts \citep{2019NatAs...3..212S}. 
In terms of cost, at a tolerable expense of accuracy, photometric redshifts can be 
a suitable substitute for spectroscopic redshifts.

Modern photometric redshift estimation approaches are split into two large 
branches: the spectral energy distribution (SED) fitting based on SED models 
(including spectral templates) and 
machine learning inference \citep{2017MNRAS.465.1959C,2019NatAs...3..212S}. 
These two methods are mutually complementary with different pros and cons.
The template-based SED fitting may provide photometric redshifts in a wide redshift 
and photometric range; moreover, using Bayesian inference improves 
the effectiveness of the method \citep{2000AA...363..476B}. However, 
this approach heavily relies on the prior knowledge of SEDs and the understanding 
of related physics determining SEDs. This dependency 
may result in biased results \citep{2011ApSS.331....1W,2015ApJ...801...20T}. However, 
the machine learning method can quickly retrieve 
accurate photometric redshifts without dependence on prior knowledge 
\citep{2017MNRAS.465.1959C}. Nonetheless, most of the machine learning models 
suffer from performance degradation for few or unseen data during their training 
since these methods are induction models for the provided data 
\citep{liang2017enhancing,DBLP:conf/iclr/HendrycksMD19}.

Samples drawn from out-of-distribution (OOD) (i.e., few or 
unseen data in training) are well-known distress 
in a reliable application of neural networks (NNs). 
\citet{DBLP:conf/iclr/HendrycksG17} demonstrate that an ML model's 
accuracy degrades for OOD samples for several training datasets: e.g.,
MNIST \citep{lecun-mnisthandwrittendigit-2010}, CIFAR-10, and CIFAR-100 \citep{cifar}, 
and they suggest a baseline model for OOD detection in an NN. 
\citet{10.5555/3327757.3327819} propose 
an advanced method for detecting OOD 
examples using Mahalanobis distance, thus assuming the trained network parameters 
can be fitted by a class-conditional Gaussian distribution. 
The unsupervised approach introduced by \citet{Yu_2019_ICCV} 
uses unlabeled samples as training data to equip NNs with 
the functionality of scoring and detecting OOD examples. 
A parameter-free OOD score is proposed 
by \citet{Serra2020Input} to handle the OOD issue 
in generative models, thus posing that 
the problem is attributed to the excessive effect of input complexity. 
These recent studies emphasize that, for achieving a more robust NN model, 
the OOD instance is a practical and important limitation.

The appropriate warning on OOD samples and handling their impact on models 
should be offered in machine learning inference of photometric redshifts. 
The machine learning method for photometric redshift estimation has been 
explored in many past studies 
\citep{10.1046/j.1365-8711.2003.06271.x,ball2008robust,
2011PASP..123..615S,Brescia_2013,10.1093/mnras/stx3055,
bilicki2018photometric,2019EPJWC.20609006C}. 
In these past studies, although the superior performance of machine learning 
methods has been demonstrated, the input/target feature spaces regarding OOD examples that 
the models are unable to describe for accurate and reliable prediction 
of photometric redshifts have not been quantitatively investigated. 

In this series of studies, we propose a machine learning method to improve 
the accuracy and reliability of photometric redshift inference, thus 
exploiting the well-known flexibility of NN models. 
The NNs are renowned for their capacity of mapping nonlinear 
relationships between input and target (i.e., redshift) 
as a universal approximator and handling 
a massive amount of data. 
NNs have been one of the most popularly used machine learning 
algorithms throughout a wide range of fields and tasks including 
natural language processing \citep{NIPS2017_3f5ee243,gpt-3}, 
image classification \citep{NIPS2012_c399862d,7298594}, 
autonomous vehicles \citep{levinson2011towards,pmlr-v78-dosovitskiy17a}, 
and protein structure prediction \citep{alphafold,TORRISI20201301}.

This study, the first study in the series, focuses on improving accuracy in inferring 
photometric redshifts. In our NN models, we adopt anchor loss 
\citep{9009013}, which considers 
the difficulty of inferring photometric redshifts with respect to 
the estimated redshift, i.e., target. The primary cause of the inference difficulty is 
imbalanced training samples for redshifts and complex patterns 
in mapping from input features to the target redshift space. 
Because a new loss can cause systematic 
bias effects, we use an ensemble learning approach, 
which combines multiple base models into a unified model, to reduce 
the bias of models and improve accuracy \citep{Zhou2009}.

Photometric data used are collected from 
the Panoramic Survey Telescope and Rapid Response System (Pan-STARRS) public data 
release \citep{2016arXiv161205560C}. We intend to use our trained model 
to infer the photometric redshifts 
of objects that correspond to extragalactic transient and variable sources. 
For transient sources such as supernovae and variable 
sources such as quasars, Pan-STARRS photometric catalogs can be useful to infer the photometric 
redshifts of hosts. Because photometric information used by our model is 
not much different from that acquired in other surveys, such as the 
SkyMapper Southern Sky Survey \citep{2007PASA...24....1K} and 
Legacy Survey of Space and Time 
\citep{2002SPIE.4836...10T,2019ApJ...873..111I}, 
using the Pan-STARRS data for the model has potential benefits 
for applications with other data. Moreover, there are not many 
previous studies on photometric redshifts with Pan-STARRS data 
\citep{2021MNRAS.500.1633B}.

This study is organized as follows. 
Section \ref{sec:data} provides the overview of training data and 
pre-processing of input features for machine learning applications. 
In Section \ref{sec:method}, machine learning approaches and performance 
evaluation metrics are elaborated in detail. Section \ref{sec:result} 
focuses on the performance analysis of our machine learning model and 
its comparison with baseline models. In Section \ref{sec:model_validation}, 
we explore the mapping between the input feature space and the target redshift 
space for which our model is validly applicable using comparison data. 
Finally, we discuss our results and provide the conclusion in Section \ref{sec:discon}. 

\section{Data} \label{sec:data}

\begin{table*}
\centering
\caption{Spectroscopic galaxy redshift samples. \label{tab:spec_z_samples}}
  \hspace*{-2.5cm}\begin{tabular}{c|c p{6cm} c}
\tableline\tableline
Dataset name & Number of objects & Selection conditions & Reference \\
\tableline\tableline
  SDSS DR15 & 1294042 & (CLASS $==$ GALAXY) and (ZWARNING $==$ 0 or 16) and (Z\_ERR $>=$ 0.0) & \citet{2019ApJS..240...23A} \\
  LAMOST DR5 & 116186 & (CLASS $==$ GALAXY) and (Z $>$ -9000) & \citet{2012RAA....12.1197C} \\
  6dFGS & 45036 & (QUALITY\_CODE $==$ 4) and (REDSHIFT $<=$ 1.0) & \citet{2009MNRAS.399..683J} \\
  PRIMUS & 11012 & (CLASS $==$ GALAXY) and (ZQUALITY $==$ 4) & \citet{2013ApJ...767..118C} \\
  2dFGRS & 7000 & (Q\_Z $>=$ 4) and (O\_Z\_EM $<$ 1) and (Z $<$ 1) & \citet{2001MNRAS.328.1039C} \\
  OzDES & 2159 & (TYPES $!=$ RadioGalaxy or AGN or QSO or Tertiary) and (FLAG $!=$ 3 and 6) and (Z $>$ 0.0001) & \citet{2017MNRAS.472..273C} \\
  VIPERS & 1680 & (4 $<=$ ZFLG $<$ 5) or (24 $<=$ ZFLG $<$ 25) & \citet{2018AA...609A..84S} \\
  COSMOS-Z-COSMOS & 985 & ((4 $<=$ CC $<$ 5) or (24 $<=$ CC $<$ 25)) and (REDSHIFT $>=$ 0.0002) & \citet{2007ApJS..172...70L,2009ApJS..184..218L}\\
  VVDS & 829 & ZFLAGS $==$ 4 or 24 & \citet{2013AA...559A..14L} \\
    DEEP2 & 540 & (ZBEST $>$ 0.001) and (ZERR $>$ 0.0) and (ZQUALITY $==$ 4) and (CLASS $==$ GALAXY) & \citet{2013ApJS..208....5N} \\
    COSMOS-DEIMOS & 517 & (REMARKS $!=$ STAR) and (QF $<$ 10) and (Q $>=$ 1.6) & \citet{2018ApJ...858...77H} \\
    COMOS-Magellan & 183 & (CLASS $==$ nl or a or nla) and (Z\_CONF $==$ 4) & \citet{2009ApJ...696.1195T} \\
    C3R2-Keck & 88 & (REDSHIFT $>=$ 0.001) and (REDSHIFT\_QUALITY $==$ 4) & \citet{2017ApJ...841..111M,2019ApJ...877...81M} \\
    MUSE-Wide & 3 & No filtering conditions. & \citet{2019AA...624A.141U} \\
    UVUDF & 2 & Spectroscopic samples. & \citet{2015AJ....150...31R} \\
\tableline\tableline
\end{tabular}
\end{table*}

Training samples comprise 1,480,262 galaxy objects 
with known spectroscopic redshifts. 
We compile the samples from multiple spectroscopic redshift catalogs with 
the condition of reliable redshift estimation. Table \ref{tab:spec_z_samples} 
summarizes the spectroscopic redshift samples that satisfy 
the required selection conditions and have acceptable photometric 
data as subsequently described. Most selection conditions adopted here 
help us use only samples with highly reliable redshifts.

\begin{figure}
\centering
\includegraphics{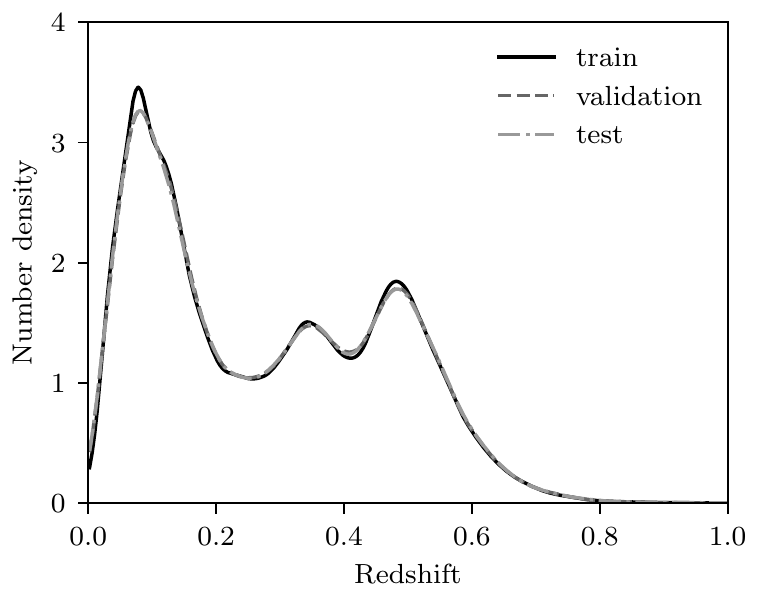}
\caption{
Redshift distributions of training, validation, and test sets. We use an adaptive kernel density estimation for density estimation of redshifts.
\label{fig:distribution_dataset}
}
\end{figure}

We use the photometric data of Pan-STARRS project as input 
photometric data for the machine learning model. The Pan-STARRS1 (PS1) 
is the first 1.8m telescope of 
the Pan-STARRS project \citep{2010SPIE.7733E..0EK}, and 
the $3\pi$ Steradian Survey is one of the primary surveys covering 
75\% of the sky \citep{2016arXiv161205560C}.
The PS1 survey provides photometry in five {\it grizy} bands with 
limiting magnitudes of 23.3, 23.2, 23.1, 22.3, and 21.4, 
respectively \citep{2016arXiv161205560C}. 

We retrieve the photometric data of spectroscopic samples from  
the public data release 1 \citep[DR1,][]{2020ApJS..251....7F} using the 
Vizier table\footnote{\href{http://cdsarc.unistra.fr/viz-bin/cat/II/349}{http://cdsarc.unistra.fr/viz-bin/cat/II/349}}, 
which has objects included in the ObjectThin and StackObjectThin tables 
with nDetections $>$ 2 option. For given positions of spectroscopic 
samples, we search the nearest object in the photometric table using 
the search radius of 0.5 arcseconds with the condition of 
{\it ObjectQualityFlags} $==$ {\it QF\_OBJ\_GOOD} (i.e., 
good-quality measurement in the PS1) \citep{2020ApJS..251....7F}. 
A single photometric object can be matched to multiple spectroscopic samples. 
We use the median redshift if more than two spectroscopic samples correspond 
to a single PS1 DR1 object. We use the average of two redshifts if two 
spectroscopic samples are matched to a single photometric object and 
their redshift difference is less than or equal to 0.005. Objects with 
a difference of greater than 0.005 are excluded from training samples.

We purposely restrict our training input data to color-related features 
that allow the easy interpretation of the model results 
rather than exploring all possible combinations of input features 
\citep{2018A&A...616A..97D}. 
The input features comprise four colors $(g - r)$, $(r - i)$, $(i - z)$, and $(z - y)$ 
in PSF measurement, their uncertainties derived in the quadrature rule, 
and the same quantities in Kron measurement. 
The training data include 
objects only if all four colors are valid in the data release.

Magnitudes of samples are excluded as input; rather, we include the 
color and color difference uncertainties that implicitly depend on 
the magnitude and SED of sources. As expected, fainter objects have 
larger uncertainties. Although uncertainties implicitly contain information 
correlated with the source magnitudes and SEDs, 
how NN models treat these inputs may be different.

Furthermore, the $E(B-V)$ value is one of the input features. It is common 
to apply Galactic dust extinction correction to photometric data 
as a pre-processing step before photometric redshifts are estimated by 
machine learning methods. In our case, we let the machine learning models 
consider the effect of the Galactic dust extinction in the training 
stage \citep[][]{2021MNRAS.500.1633B}. We decide to use 
the $E(B-V)$ values based on the dust emission model of 
the Plank cosmic microwave background observation 
\citep{2014A&A...571A..11P}, which provides a wide coverage of the sky.

The input features are transformed to improve the performance of the 
machine learning model. In training machine learning models, 
data pre-processing is required when input features have different 
ranges \citep{589532}. 
Diversely ranged features have a different effect on the loss function 
adopted in a machine learning model, thus making results biased. 
From the geometrical standpoint, the data points of differently ranged 
features form a multi-dimensional asymmetric volume in the input 
space. Data pre-processing handles this geometrical 
asymmetry and makes data more symmetric. Moreover, it smooths out pointy 
regions that might exist on the surface of the volume. 
Our data pre-processing restricts each feature to a comparable range. 
We test two data pre-processing methods: min-max normalization and 
standardization. Both methods process data in a feature-wise manner. 
The former method uses the minimum and maximum values of each 
feature to restrict feature ranges. For standardization, 
each feature is re-scaled to have zero mean and unit variance. 

After the pre-processing step, we randomly select 80\% of our samples 
as the training set and each 10\% of samples as the test and validation sets, respectively. 
Because we randomly split the data and have many samples, we assume that 
the samples allocated to each set are drawn from the same distribution. 
Figure \ref{fig:distribution_dataset} shows the redshift distribution of the 
training, test, and validation sets. To avoid any confusion, we sometimes refer 
to redshifts as targets in training the machine learning models.

\section{Method \label{sec:method}}

In this section, we describe the baseline models based on regression, 
our machine learning approach, ensemble methods for better machine 
learning performance, and metrics to quantitatively assess 
the photometric redshift quality.

\subsection{Baseline Models Based on Regression \label{subsec:Baseline}}

For both classification and regression, K-nearest neighbors (KNN) is a relatively 
simple non-parametric model that can be used  
\citep{doi:10.108000031305.1992.10475879}. Because of its simplicity, the model 
has been used to estimate the photometric redshifts of 
galaxies and quasars or used as a baseline to evaluate the performances 
of novel approaches \citep{2013AJ....146...22Z,2018AA...611A..97P}. The model 
estimates targets (or labels) of given samples based on the averaged 
target values of $k$ nearest samples in a training set.

Random forest (RF) is an ensemble method using multiple 
decision trees \citep{breiman2001random}. The model has been extensively used 
for astronomical classification, regression, and other tasks 
\citep{2015DatSJ..14...11Z}. RF is essentially optimized 
to use feature-based inputs because it recursively splits 
high-dimensional features by generating multiple root nodes 
and their succeeding child nodes. RF, in addition to KNN, is frequently used 
as a baseline because of its usage of ensemble learning resulting in statistical 
robustness and its split-rule-learning characteristics, 
which can return the importance of input features.

A NN is a representative machine learning model 
inspired by the functioning of the human brain \citep{Hopfield2554}. 
Generally, NN architecture comprises one input layer, 
multiple hidden layers, and one output layer. Moreover, these layers are 
composed of artificial neurons or perceptrons \citep{Rosenblatt58theperceptron:}, 
which are the basic units forming the model and mimicking biological neurons. 
Artificial neurons are interconnected to ones in adjacent layers 
with randomly initialized connection weights. Vectors provided to neurons 
in the input layer are transformed by neurons in the hidden layers 
using connection weights with nonlinear activation functions 
and are then transmitted to the output layer. Then, neurons 
in the output layer finally produce scalar or vector outputs. 
Although the definition of the term is not stringent, 
the network is usually referred to as a vanilla NN (VNN) 
when the model generates a scalar output. We emphasize that the regression NN, 
hereafter, is referred to as VNN because, in our study, 
it produces a scalar output --- a photometric redshift.

We use the mean squared error (MSE) as a loss function to train the parameters of 
RF and VNN for redshift regression. KNN does not require 
a loss function. Moreover, we also have tested the adaptive robust loss 
function proposed by \cite{Barron_2019_CVPR}, thus reflecting the anchor loss 
to be explained later. However, we found that a simpler MSE outperforms 
the loss in this study.

\subsection{Multiple-Bin Regression with NN \label{subsec:mbrnn}}

The MSE for regression problems with the heterogeneous target 
is not an optimal option \citep{L2Reg}. MSE is the most commonly used 
loss function to train machine learning models for regression 
because minimizing the MSE is generally identical for maximizing log-likelihood 
from a probabilistic standpoint. In most real-life cases, however, 
the target is heterogeneous rather than homogeneous. If it is the case, 
using the MSE may result in an undesirable performance of the regression model 
because the loss function fosters the model to minimize the error throughout 
all modes while not considering the multi-modal behavior of the target.

Because our target, i.e., redshift, can have a multi-modal distribution 
particularly because of the degeneracy of redshifts in terms of input features, 
we consider multiple-bin regression with a NN (hereafter, MBRNN) 
to bypass the limitation of the MSE for a heterogeneously behaving target. 
The MBRNN model has been previously explored in several studies 
for viewpoint estimation \citep{7410665,BMVC2016_91}, 
bounding-box estimation \citep{Mousavian_2017_CVPR}, pose estimation 
\citep{3DRCNN_CVPR18}, and redshift estimation in astronomy 
\citep{2018AA...611A..97P,2019AA...621A..26P}. Compared with regression using the MSE, 
these studies demonstrated 
the performance enhancement of the approach. Furthermore, the property of 
the probabilistic model enables deeper scrutiny on the causes of 
poor inference performance 
for specific samples and examination of model calibration, 
which are discussed in Section \ref{subsec:photoz}.

We first discretize spectroscopic redshifts and divide them 
into $n$ independent bins. We test two types of redshift bins: 
uniform and non-uniform\footnote{Some previous studies have considered 
overlapping bins; however we did not find any advantages of using this strategy 
for our purpose.}. For uniform bins, we equally discretize the redshifts 
of the training data with a constant bin width. However, we make 
each bin have an almost uniform number of samples in the non-uniform binning. 
Because most objects reside in low redshift ranges, the non-uniform 
bin width becomes wider as the redshift increases even though it is not monotonic.

The MBRNN model using the softmax function estimates probabilities $p_{i}$ 
that the photometric redshifts of objects lie in $i^{th}$ redshift bin. 
That is, we modify the regression problem into a classification problem 
with multiple redshift bins. For point estimation, we can compute the photometric 
redshift $z_{phot}$ either by selecting a redshift bin center 
with peak probability (i.e., mode) as $z_{mode}$ 
or by averaging with the output 
probabilities and central values of the bins as $z_{avg}$ as follows:
\begin{equation}
\begin{split}
  z_{mode} & = c_{j} ~ {\rm for} ~ j = \underset{k}{\operatorname{argmax}}(p_{k}), \\
    {\rm or} \\
    z_{avg} & = \sum p_{i} \, c_{i}, \\
\end{split}
\end{equation}
where $c_{i}$ is the central value of ${i^{th}}$ redshift bin. 
We compare the prediction accuracy of the two different 
point-estimation methods in Section \ref{sec:result}. 

We use the anchor loss \citep{9009013} as a classification 
loss function for training the model\footnote{Although \cite{9009013} 
reported that they obtained the highest performance using sigmoid output, 
we stick to softmax output for MBRNN since we found that the softmax 
performed better in our case.}. The loss is designed to measure 
the disparity of two given probability distributions 
considering prediction difficulties, which can be attributed to various reasons such 
as the scarcity of data or the similarities between samples drawn 
from different distributions. This function evaluates the prediction 
difficulty using the difference between network-estimated probabilities 
for the true and other classes. In an easy prediction case, 
the network-estimated probability of the true class is higher than 
those of the other classes, whereas it is lower in a difficult case. 
The difficulty is used to weigh the loss of a sample. 
For the given two discrete probability distributions, 
$g$ and $p$, the anchor loss $\ell(g, p)$ is defined as follows:
\begin{equation}
    \ell(g, p) = - \underset{k}{\sum} \, g_{k} \, log(p_{k}) + (1 - g_{k})(1 + p_{k} - p_{*})^{\gamma}log(1 - p_{k}),
\end{equation}
where $g_{k}$ and $p_{k}$ represent the label (i.e., $g_{k} = 1$ 
for the correct redshift bin and $g_{k} = 0$ for other bins) 
and network-estimated probabilities for class (i.e., redshift bin) $k$, respectively; 
$\gamma$ represents an exponent governing the weights of prediction difficulties; 
$p_{*}$ represents the anchor probability which 
is set to the network-estimated true class probability (i.e., the classification 
probability for the correct redshift bin). 
Therefore, $p_{*}$ determines the prediction difficulties of samples. 
In the cases of easy prediction, 
the trained model assigns high probability on the true class (i.e., the 
correct redshift bin), leading to higher $p_{*}$ than $p_{k}$ 
(see the appendix of \citet{9009013} and its equation A-3). 
Note that the anchor loss approaches a binary cross-entropy loss, 
which is one of the most commonly used classification losses, as $\gamma$ goes to $0$.

The baseline VNN and our proposed model MBRNN share 
the same NN structures except for output layers 
because these networks generate differently shaped outputs. 
The networks are composed of fully connected layers only: 
an input layer with 17 neurons, eight hidden layers sequentially 
with 128, 256, 512, 1024, 512, 256, 128, and 32 neurons, 
and the output layer with the number of neurons 
corresponding to the output size. Each layer is followed 
by a batch normalization layer \citep{pmlr-v37-ioffe15} 
and softplus activation function \citep{7280459}.

\subsection{Ensemble of multiple-bin method}

Our adoption of ensemble methods combines the 
outputs of multiple MBRNN models and generates an integrated model. 
Ensemble methods have been extensively used in the field of machine learning 
to overcome the limitations of models trained for a limited set of tasks.
This strategy often results in better or more generalized model performance 
by weighting the models depending on their performance or 
properly assigning each model to the task where the model performs best.

To identify a suitable ensemble approach for our purpose, we evaluate four different 
methods: plain model averaging ensemble (E1), weighted model ensemble (E2), 
bin ensemble (E3), and bin-wise selective ensemble (E4) \citep{5631385}. 
E1 is simply averaging results from each model. For E1, 
the performance of the integrated model might be downgraded 
when a poorly performing model is included because this method does 
not consider the specialty of each model. E2 uses 
weighted averaging of predictions from each model. 
Assigning higher and lower weights to high- and low-performance models, 
respectively, this method is frequently expected to yield higher performance 
than a single model or E1. Optimal weights can be found by various methods 
such as Bayesian optimization methods \citep{10.5555/2999325.2999464} and 
gradient descent methods \citep{pub.1017229575}. The gradient descent method 
with the validation set is used in our study. Moreover, we consider another weighted 
ensemble method E3, which allocates individual weight to each model and 
redshift bin combination. Because it provides different weights to each bin in addition 
to models, this approach has additional parameters to be tuned and more flexibility 
than E2. Finally, we evaluate the selective ensemble method E4, which selects  
a single model for each redshift bin where the model shows the highest point 
estimation accuracy using the validation set. In E4, we find 
redshift bins to which point-estimated redshifts of each test object belong 
using the vote of the single models. Then, the model allocated to the bin 
is used to estimate probability distributions for objects.

\subsection{Metric \label{subsec:metric}}

We use multiple metrics to assess performance in estimating photometric redshifts. 
We refer to \cite{2017MNRAS.465.1959C} for the detailed description of metrics. 
The following is a brief explanation of metrics:

\begin{itemize}
    \item {\it Bias}: the absolute mean of redshift differences defined by 
      $\frac{1}{N}  |\sum_{i}^{N} \Delta z_{i}|$ 
      where $N$ represents the number of samples in the dataset, 
      $i$ represents the sample index, and $\Delta z_{i}=(z_{i, spec}-z_{i, phot})/(1+z_{i, spec})$, 
    \item {\it MAD}: the mean of absolute differences defined by $\frac{1}{N} \sum_{i}^{N} |\Delta z_{i}|$,
    \item {\it $\sigma$}: the standard deviation of the difference $z_{i, spec}-z_{i, phot}$,
    \item {\it $\sigma_{68}$}: the 68th percentile of the absolute difference, i.e., $|z_{i, spec}-z_{i, phot}|$,
    \item {\it NMAD}: the normalized median absolute deviation of the differences, 
      which is $1.4826 \times Median(\Delta z)$,
    \item {\it $R_{cat}$}: catastrophic (hereafter, \textit{cat}) error, which 
      corresponds to $|\Delta z| > 0.15$, fraction.
\end{itemize}

Using these metrics, we evaluate the quality of photometric redshifts 
and perform a grid search to identify the empirically optimal configuration of 
the MBRNN model. For the numerical metrics, the lower the values, 
the better the model performance.

\section{Result \label{sec:result}}

\subsection{Single Model Performance Test \label{subsec:photoz}}

\begin{figure}
\centering
\includegraphics[]{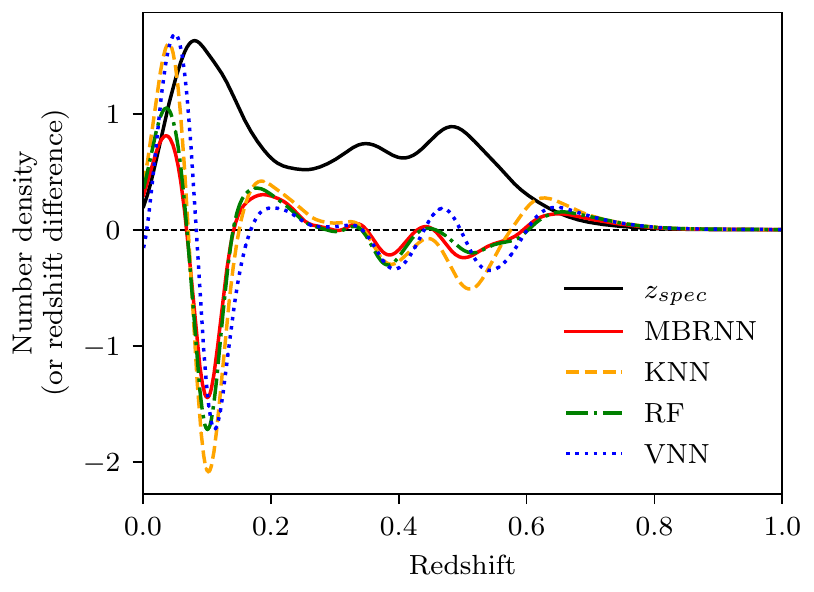}
\caption{Redshift difference distributions of the MBRNN and baseline 
models. The distribution of spectroscopic redshifts is also drawn here 
for comparison. The spectroscopic redshift distribution is scaled down 
by half for visual clarity. Positive and negative differences indicate 
under- and over-estimation of the distribution, respectively.
\label{fig:distributionResid}}
\end{figure}
\begin{figure*}
\centering

\includegraphics[]{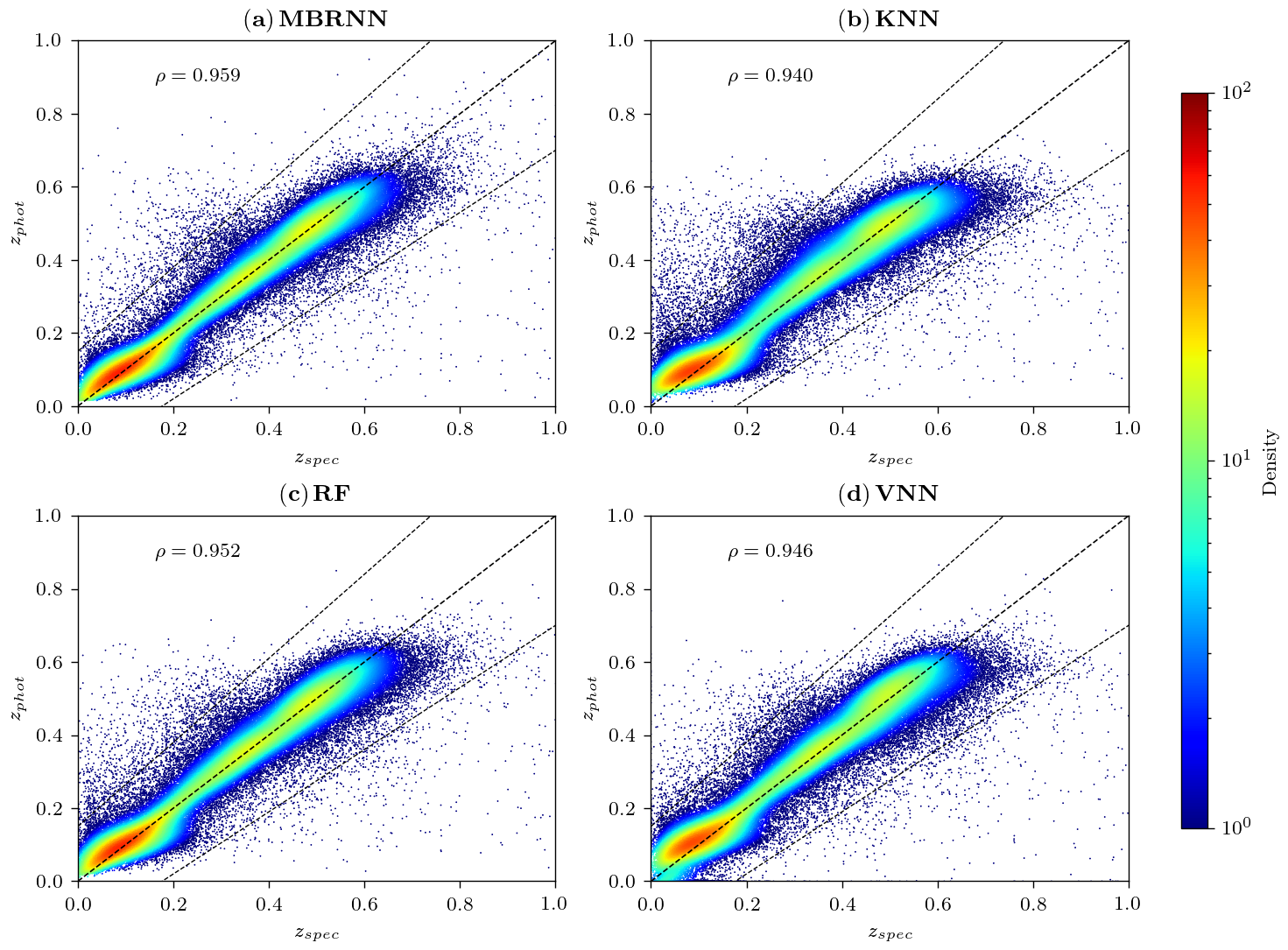}
\caption{
Distributions of spectroscopic and photometric redshifts obtained by  
the MBRNN and baseline models. The markers are color-coded with 
number density, and the color is normalized in the log scale 
for better contrast. 
Samples outside the dashed lines on both sides 
correspond to {\it cat} objects, and the central dashed line 
has a slope of 1. The Pearson correlation coefficients $\rho$ 
between spectroscopic and photometric redshifts are also presented 
in each panel.
\label{fig:ScatterComp}}
\end{figure*}
\begin{figure*}
\centering
\includegraphics[width=1.0\textwidth]{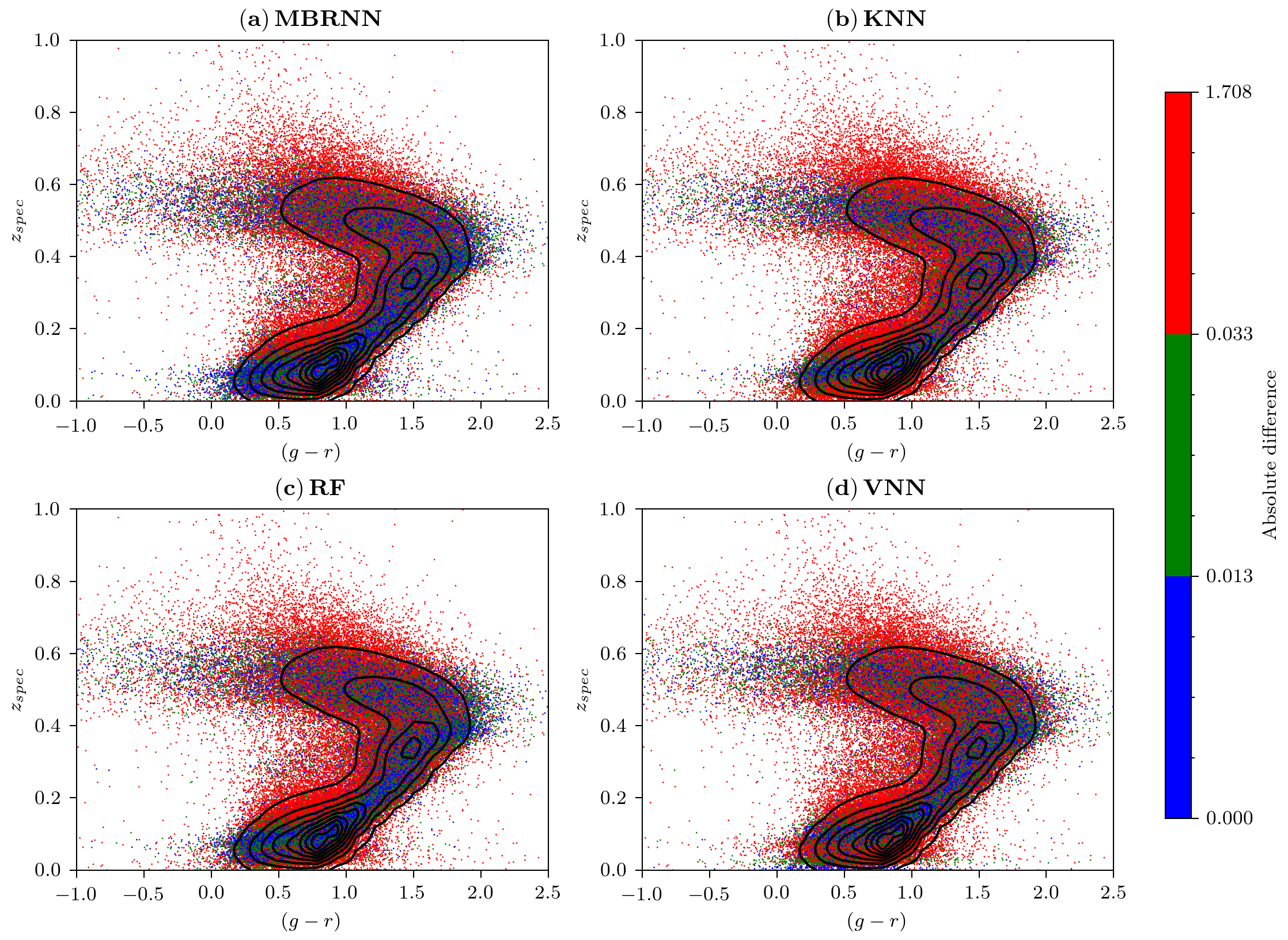}
\caption{
Distribution of the absolute difference between spectroscopic and 
photometric redshifts in the space of $(g - r)$ in Kron measurement 
and spectroscopic redshifts. 
The contour lines show the densities of the samples in this space. 
We assign samples to low-, middle-, and high-difference groups, 
and the samples in each group are represented in different colors. 
Providing the smallest area of the high-difference region in the MBRNN model
compared with the other baseline models, the MBRNN model reproduces 
spectroscopic redshifts with the highest accuracy 
in the outskirt area of the contours.
\label{fig:ColorRedshiftScatter}}
\end{figure*}
\begin{figure*}
\centering
\includegraphics[width=1.0\textwidth]{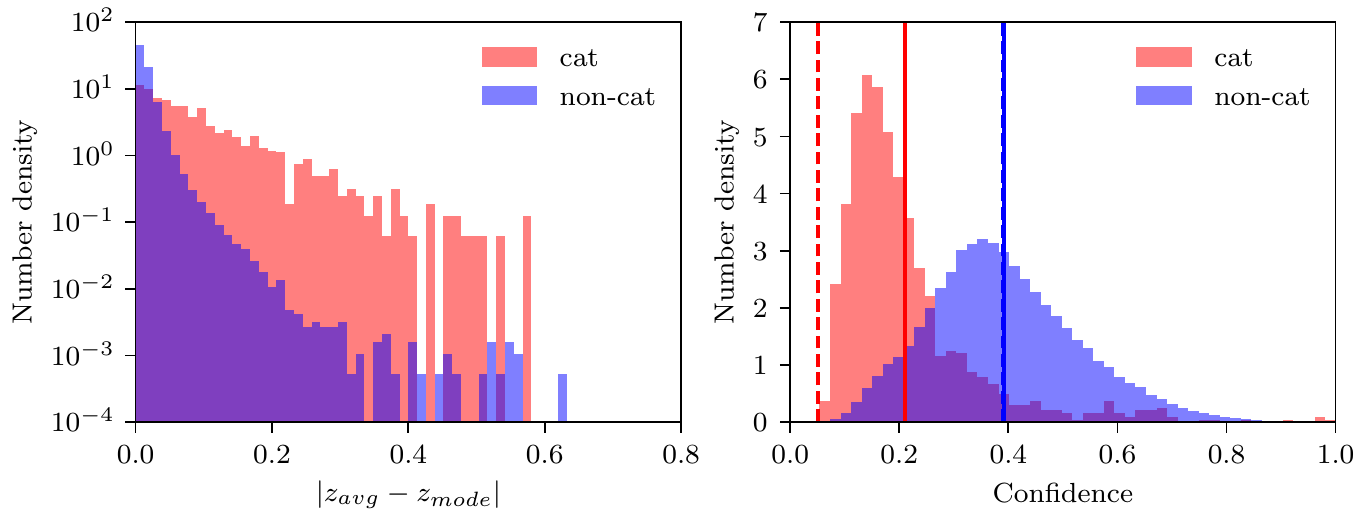}
\caption{{\it Left}: Distribution of the difference between the average 
  and mode photometric redshifts for the {\it cat} and 
  {\it non-cat} cases. {\it Right}: Distribution of the confidence 
  estimation for photometric redshifts. 
  The vertical solid and dashed lines mark the mean confidence and accuracy, 
  respectively, for each case.
\label{fig:distributionAvgMode}}
\end{figure*}

\begin{table*}
\centering
\caption{Metrics comparison between the average and mode point estimation of 
the MBRNN and baseline models. The best performing cases in each search 
are presented in boldface. \label{tab:comp}}
\hspace*{-2.5cm}\begin{tabular}{c|c c c c c c c}
\tableline\tableline
Model & Bias & MAD & $\sigma$ & $\sigma_{68}$ & {\it NMAD} & $R_{cat}$ \\
\tableline\tableline
MBRNN (mode) & 0.0036 & 0.0266 & 0.0430 & 0.0274 & {\bf 0.0256} & 0.0111 \\
MBRNN (avg) & {\bf 0.0017} & {\bf 0.0254} & {\bf 0.0394} & {\bf 0.0272} & 0.0256 & {\bf 0.0084} \\
KNN & 0.0055 & 0.0324 & 0.0487 & 0.0352 & 0.0340 & 0.0158 \\
RF & 0.0027 & 0.0277 & 0.0430 & 0.0294 & 0.0277 & 0.0116 \\
VNN & 0.0046 & 0.0307 & 0.0450 & 0.0338 & 0.0327 & 0.0113 \\
\tableline\tableline
\end{tabular}
\end{table*}

We quantitatively assess and visually inspect 
photometric redshifts obtained by the MBRNN model using test set samples. 
The empirically optimal model configuration for inference is selected 
in the grid search (see appendix \ref{AppendSec:grid-search}). 
The default model setup adopts the 
anchor loss with $\gamma$ of 0 and 64 uniform redshift bins. 
Henceforth, unless otherwise stated, we shall restrict ourselves 
to the MBRNN model with this setup. 

Because the baseline models perform regression, we require a way to 
derive the point estimation of redshifts with 
the MBRNN model for a fair comparison with the baseline models. 
We first juxtapose the metrics of mode and average photometric redshifts, 
as shown in Table~\ref{tab:comp}, to determine which estimation is more suitable 
for point estimation. Average redshifts evince lower metrics than mode redshifts. 
Because this result indicates that the average estimation accompanies higher point estimation 
accuracy than the mode estimation, we use average redshifts in the rest 
of the analysis related to point estimation, except cases 
specifying the usage of the mode redshift. 
Using the average estimation of redshifts, we now can equitably compare the 
MBRNN model with the baseline models.

The MBRNN model shows a higher prediction accuracy for overall redshift ranges 
than baseline models. Figure \ref{fig:distributionResid} shows 
the distribution of differences between spectroscopic and photometric redshifts 
in the MBRNN and baseline models. We obtain the distributions of the differences by 
subtracting the distribution of photometric redshifts from that of spectroscopic 
redshifts. The MBRNN model's relatively close-to-zero differences indicate that it 
best captures the multi-modal behavior of targets, i.e., redshifts. 
The KNN, RF, and VNN models show poorer descriptions for the heterogeneous property 
of targets, particularly for redshift ranges with distribution peaks. 
These results are naively expected from the lowest metrics of the MBRNN model 
compared with the baseline models as presented in Table \ref{tab:comp}, even though  
the metrics, which are marginalized into one scalar throughout multiple dimensions, 
cannot represent locally different behaviors of the distribution.

The distributions of spectroscopic and photometric redshifts 
shown in Figure \ref{fig:ScatterComp} summarize 
well the differences reported in the metrics. The density peaks of the MBRNN and RF 
models are aligned with the slope-one line, whereas those of the KNN and VNN models 
are misaligned. This difference reflects lower deflection-related 
metrics (i.e., bias and MAD) of the MBRNN and RF models. 
As expected from the dispersion-related metrics (i.e., $\sigma$ and $\sigma_{68}$), 
the redshift distribution of the MBRNN model shows the smallest dispersion around 
the slope-one line. This characteristic is conspicuous in redshift regions 
with the distribution peaks (i.e., $z_{spec} \sim$ 0.125, 0.35, and 0.5), 
as presented in Figure \ref{fig:distribution_dataset}, where scatters 
spread more extensively than other redshift regions. 
The small bias and dispersion values of the MBRNN model contribute to 
the highest Pearson correlation coefficient.

The lack of objects in the input space is one of the 
primary causes that induce prediction difficulty. 
Figure \ref{fig:ColorRedshiftScatter} 
shows the distribution for the absolute difference of redshifts 
between spectroscopic and photometric 
redshifts in the space of the input $(g - r)$ in Kron 
measurement and target 
redshifts. Although this visualization only shows the projected 
distribution in the input $(g - r)$ space with degeneracy of the other 
input features in the higher dimensional space, the $(g - r)$ color has 
a well known correlation with redshifts, and its interpretation is usually 
straightforward in examining models \citep[e.g.][]{2019ApJS..245...26K}. 
When grouping the samples into low-, middle-, and high-difference groups 
corresponding to about 33\% of the samples per group in the MBRNN model,
a large number of samples with significant redshift discrepancy are found 
independently in models. 
Intriguingly, the high-difference groups are similarly distributed 
in every model, whereas the area of the regions corresponding to this group 
is the smallest in the MBRNN model. These regions are the outskirts of the density 
contour lines in which the objects sparsely reside. 
As presented in Appendix \ref{AppendSec:Cat}, the distribution of the {\it cat} 
samples in the MBRNN model is similar to that of baseline models. 
These model-independent patterns also can be found in 
Figure \ref{fig:distributionResid}. The results 
indicate that the samples populating over the specific regions in the input space 
are likely to bring high prediction difficulty regardless of the model. 

The high-difference samples possibly have multi-modal 
model probability distributions. 
Benefiting from the MBRNN model's property as a probabilistic classification 
model, we examine the distribution of the difference between the average 
and mode photometric redshifts for the {\it cat} and {\it non-cat} samples. 
Figure \ref{fig:distributionAvgMode} shows that the {\it cat} error samples 
have higher differences than {\it non-cat} ones. 
The higher differences between the average and mode estimation 
of the {\it cat} samples indicate that 
the model-estimated probabilities of {\it cat} samples are possibly multi-modal.

Furthermore, the confidence distribution of the MBRNN model 
shown in Figure \ref{fig:distributionAvgMode}  
endorses our interpretation of the {\it cat} samples. 
Confidence expresses a model's level of certainty about the classification result 
for a given sample. The measure of confidence is defined as the maximum value 
in the probability output of a model \citep{DBLP:conf/iclr/HendrycksG17}. 
The low confidence of {\it cat} samples indicates 
a high possibility that these samples have multiple probability peaks.

It is worthwhile to refer to the practical guide 
for flagging objects that require caution in analysis. 
In the confidence histogram presented in 
Figure \ref{fig:distributionAvgMode}, 
the difference 
between mean confidence and accuracy of {\it cat} samples 
is higher than that of {\it non-cat} samples. 
We may refer that the model is well-calibrated for {\it non-cat} samples 
because mean confidence and accuracy are comparable 
\citep{pmlr-v70-guo17a}. 
However, the model is overconfident about the {\it cat} samples 
because the mean confidence of the model is significantly higher than 
mean accuracy. In this case, caution is required 
because overconfident estimation leads to high prediction error 
and may lead to an incorrect interpretation. 
We recommend being cautious 
about objects outside the dense regions in the input space 
because the model is overconfident about {\it cat} samples, 
and most of the {\it cat} samples reside in the low-density region of 
input space (see Figure \ref{fig:ColorRedshiftScatter}).

\subsection{Ensemble Model Performance Test \label{subsec:ensemble}}

\begin{figure*}
\centering
\includegraphics[width=1.0\textwidth]{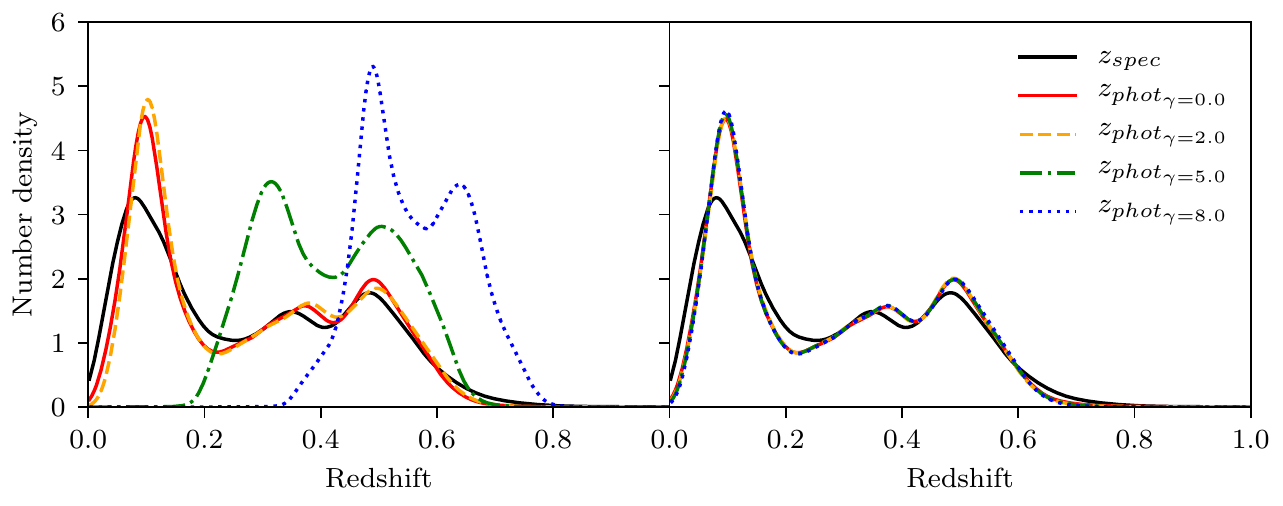}
\caption{Distribution of photometric redshifts in the MBRNN model 
  with 64 uniform bins ({\it left}) and 512 non-uniform bins ({\it right}) 
  with the anchor loss $\gamma$ of 0, 2, 5, and 8. 
  The distribution of spectroscopic redshifts is also drawn for comparison. 
\label{fig:distributionGamma}}
\end{figure*}
\begin{table*}
\begin{center}
\caption{The metrics of the ensemble learning. We also summarize the results of the single model for comparison. \label{tab:ensemble}}
\hspace*{-2.5cm}\begin{tabular}{c|c|c c c c c c}
\tableline\tableline
Bin type & Case & Bias & MAD & $\sigma$ & $\sigma_{68}$ & {\it NMAD} & $R_{cat}$ \\
\tableline\tableline
\multirow{5}{*}{64 uniform bins} & Single Model & 0.0017 & 0.0254 & 0.0394 & 0.0272 & 0.0256 & 0.0084 \\
& E1 & 0.0023 & 0.0255 & 0.0393 & 0.0274 & 0.0258 & 0.0084 \\
& E2 & 0.0019 & 0.0254 & 0.0392 & 0.0273 & 0.0256 & 0.0083 \\
& E3 & {\bf 0.0010} & {\bf 0.0253} & {\bf 0.0389} & {\bf 0.0272} & {\bf 0.0255} & {\bf 0.0082} \\
& E4 & 0.0020 & 0.0255 & 0.0394 & 0.0273 & 0.0257 & 0.0086 \\
\cline{1-8}
\multirow{5}{*}{512 non-uniform bins} & Single Model & 0.0023 & 0.0256 & 0.0404 & 0.0272 & 0.0255 & 0.0090 \\
& E1 & 0.0021 & 0.0255 & 0.0400 & 0.0272 & 0.0255 & 0.0088 \\
& E2 & 0.0021 & 0.0255 & 0.0400 & {\bf 0.0272} & 0.0255 & 0.0088 \\
& E3 & {\bf 0.0021} & {\bf 0.0255} & {\bf 0.0400} & 0.0272 & {\bf 0.0255} & {\bf 0.0087} \\
& E4 & 0.0022 & 0.0256 & 0.0405 & 0.0272 & 0.0255 & 0.0090 \\
\tableline
\end{tabular}
\end{center}
\end{table*}

Performance elevation of the integrated model through ensemble learning 
emerges from diverse specialties of individual models. 
During grid-search, we find that the anchor loss assigns different specialties 
to models based on $\gamma$. As the value of $\gamma$ increases, 
the distributions of photometric redshifts shift toward the higher redshift 
region in the uniform bin case (see Figure \ref{fig:distributionGamma}). 
Because the loss with larger $\gamma$ allocates more difficult objects for prediction 
with bigger weights, this model bias is attributed to 
the higher redshift region which corresponds to the most difficult part for prediction 
from the model's perspective \footnote{It is anticipated because our dataset 
rarely has high-redshift samples, and hence the equal-width bins 
in the high-redshift region have a comparably small number of objects. 
Besides, we have already confirmed that data sparsity contributes to 
the prediction difficulties in Section \ref{subsec:photoz}.}.

We examine four ensemble methods explained in Section \ref{sec:method}. 
Because the models with high anchor loss $\gamma$ focus excessively 
on the high redshift region, we perform ensemble learning with 
models trained with moderate values of $\gamma$: 0, 0.2, 0.5, and 1. 
Furthermore, for comparison, we test ensemble methods with 512 non-uniform bins trained with 
the same set of $\gamma$ values. A set of models used for each ensemble 
combination has the same architecture and number of redshift bins. 
The ensemble experiments with different numbers of bins and sets of $\gamma$ 
values can be found in Appendix \ref{AppendSec:DiffEnsemble}.

We report that E3 with 64 uniform bins has the best performance metrics. 
Moreover, the E3 ensemble model is well-calibrated. 
Appendix \ref{AppendSec:ModelCalibration} provides the calibration study 
of the ensemble model. 
In other words, E3 successfully considers the varied specialties of the individual 
models with different values of $\gamma$. Table \ref{tab:ensemble} compares 
the results of the four different ensemble methods using 64 uniform bins and 
512 non-uniform bins.

For the 512 non-uniform bin cases, all methods outperform the single model; 
however, the performance disparity between  
the ensemble methods is not as pronounced as it is in the uniform bin case. 
We recognize that it is because improvement 
is attributed to the stochastic differences 
between models, and not diverse specialties. 
The non-uniform bins are designed 
to make each bin have an almost uniform number of samples. Since the 
prediction difficulty mostly stems from the scarcity of data, 
the non-uniform bin cases have no particularly challenging parts 
to predict. Therefore, it results in monotonous specialties of individual 
models being insensitive to the variation of $\gamma$. Figure \ref{fig:distributionGamma} 
shows that the redshift distributions 
in the MBRNN model with the non-uniform bins do not vary significantly  
as the $\gamma$ increases. Consequently, it results in a small improvement 
in all ensemble methods because of the stochastic difference among the single models.

\begin{figure*}
\centering
\includegraphics[width=0.98\textwidth]{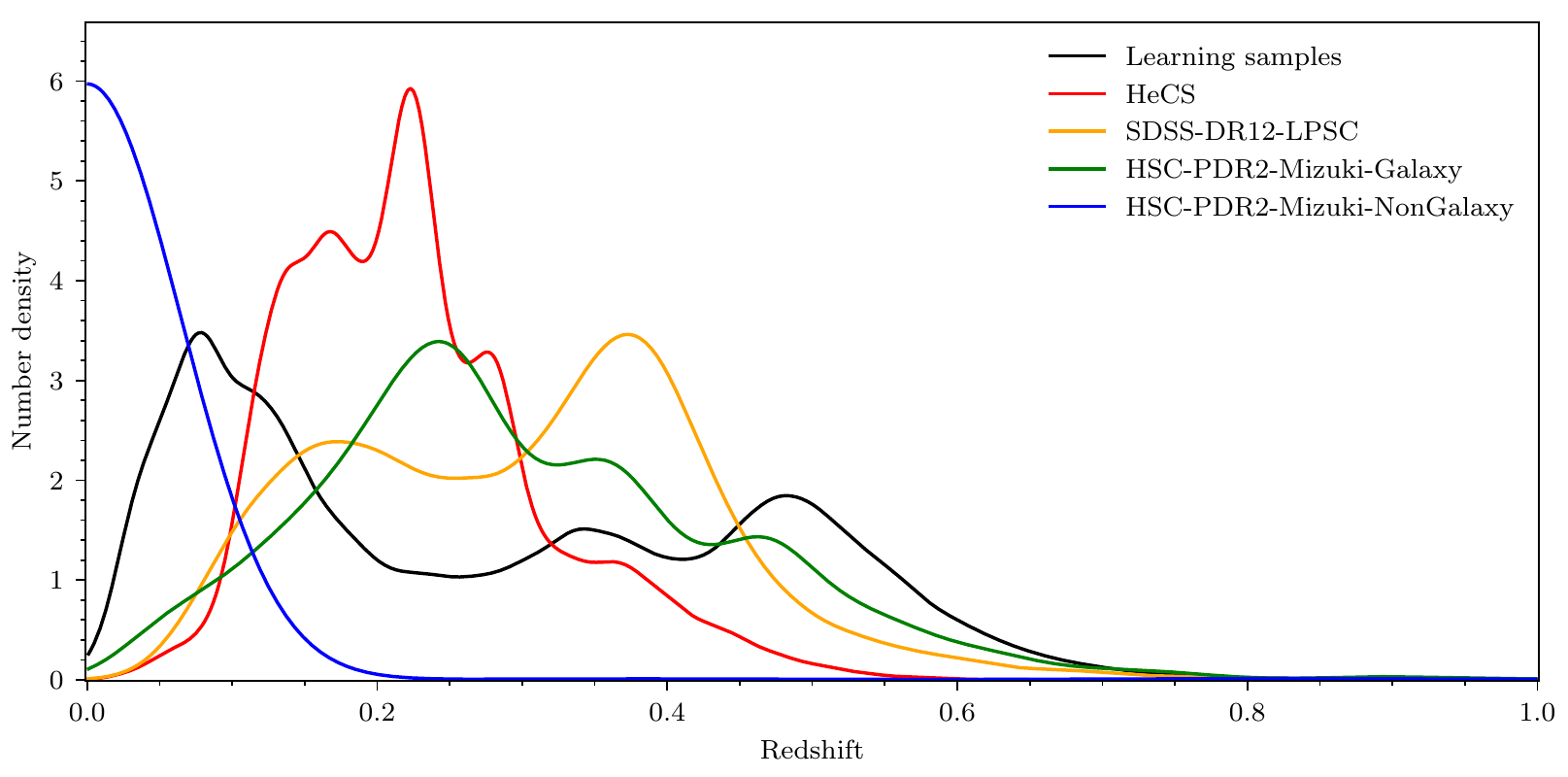}
\caption{
  Redshift distribution of the comparison samples for model validation as 
  summarized in Table \ref{tab:photoz_comp}. The plot displays redshifts 
  up to approximately 1 even though the maximum redshift is 
  4.15 in the HSC-PDR2-Mizuki-NonGalaxy.
  \label{fig:photoz_comp_redshift}
}
\end{figure*}
\begin{table*}
\centering
\caption{Comparison data with either spectroscopic or photometric redshifts. \label{tab:photoz_comp}}
\begin{tabular}{c|c c c}
\tableline\tableline
Name & Number of objects & Maximum redshift & Reference\\
\tableline\tableline
HeCS & 20,544 & 0.72 & \citet{2013ApJ...767...15R}\\
  SDSS-DR12-LPSC & 38,818 & 1.00 & \href{https://www.sdss.org/dr12/algorithms/photo-z/}{https://www.sdss.org/dr12/algorithms/photo-z/}\\
HSC-PDR2-Mizuki-Galaxy & 6,996 & 1.47 &\multirow{2}{*}{\citet{2020arXiv200301511N}}\\
HSC-PDR2-Mizuki-NonGalaxy & 3,267 & 4.15 &\\
\tableline\tableline
\end{tabular}
\end{table*}
\begin{figure*}
  \plottwo{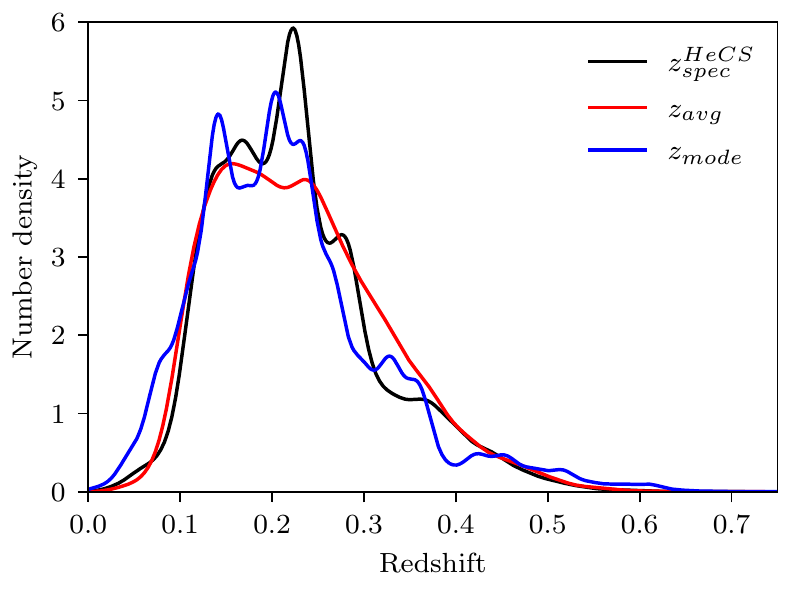}{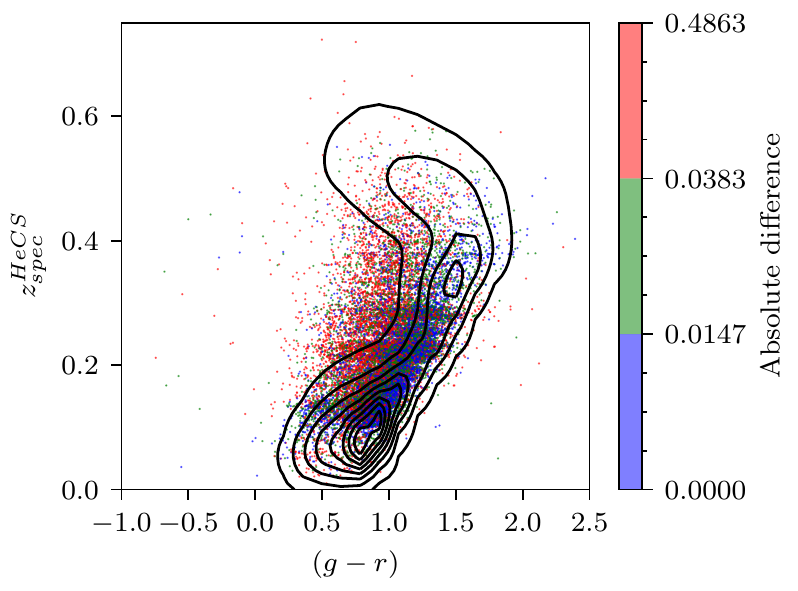}
  \caption{{\it Left}: Distribution of the comparison HeCS spectroscopic redshifts and 
  derived photometric redshifts $z_{avg}$ as the average value, and $z_{mode}$ 
  as the mode value. {\it Right}: The distribution of the absolute difference 
  between the HeCS spectroscopic redshifts and the derived $z_{avg}$ in the space of 
  $(g - r)$ and HeCS spectroscopic redshifts. 
  Following Figure \ref{fig:ColorRedshiftScatter}, the samples are color-coded, 
  and the distribution of the training samples is 
  drawn as contour lines for comparison. The distribution of the absolute difference 
  between $z_{spec}$ and $z_{avg}$ with respect to $(g-r)$ in Kron measurement  
  evidently shows that the incorrect estimation of $z_{avg}$ is 
  due to the lack of sufficient 
  training samples to occupy the input space. 
  \label{fig:HeCS_redshift_and_input_space}}
\end{figure*}

\section{Model Validation \label{sec:model_validation}}

We compare our photometric redshift results with other existing 
spectroscopic or photometric redshifts from various sources to investigate 
when the trained models fail or are not trustworthy. 
In particular, the data collected from the OOD 
with respect to the training data distribution should have highly 
uncertain photometric redshifts because the training data have never 
allowed the model to acquire the relevant information 
about the OOD data \citep[e.g.,][]{NIPS2019_9611}. 
Table \ref{tab:photoz_comp} summarizes the comparison data, 
and Figure \ref{fig:photoz_comp_redshift} shows the redshift distribution of 
the comparison samples. 
We match the PS1 DR1 data to comparison data with a search radius 
of 0.5\arcsec, and we apply the same selection and filtering rules 
to the PS1 DR1 data as we do to the training samples.

\subsection{Validation with Spectroscopic Redshift Samples \label{subsec:HeCS}}

We estimate the photometric redshifts of the matched PS1 DR1 objects 
for the dataset HeCS. The HeCS dataset includes objects with the spectroscopic 
redshifts of galaxies found in galaxy cluster areas over 
z = 0.1 -- 0.3 \citep{2013ApJ...767...15R}. Therefore, this dataset 
contains both cluster members and line-of-sight field galaxies. 
We use objects with a redshift quality flag of Q, thus indicating a secure redshift. 
Moreover, we exclude objects with redshifts $<$ 0.009 because they are not galaxies 
generally.

The comparison with spectroscopic samples allows us to unbiasedly evaluate  
the performance of our trained machine learning model although 
the comparison sample size is not as large as 
that of training samples. As shown in Figure 
\ref{fig:photoz_comp_redshift}, the true redshift range is well matched 
to the redshift range over which we intend to use the trained model. 
However, the distribution of true redshifts differs from 
that of training samples. Moreover, this comparison helps us assess 
how well the trained model can be used in identifying potential galaxy groups and 
clusters with the PS1 DR1 data \citep[see][for discussion]{2019A&A...627A..23E}.

As shown in Figure \ref{fig:HeCS_redshift_and_input_space}, 
the photometric redshifts are well estimated for the redshift range 
of the HeCS dataset. However, we find that certain objects with 
the biased estimation of photometric redshifts over 0.1 $< ~ z ~ <$ 0.3. As 
the distribution of the absolute difference between the spectroscopic 
and photometric redshifts in the space of $(g - r)$ and 
the spectroscopic redshifts highlights, 
the biased estimation of the photometric redshifts is attributed to 
the lack of blue training samples for given redshifts 
over 0.1 $< ~ z ~ <$ 0.3. Although the distribution 
in the input space such as $(g - r)$ does not completely 
show the distribution mismatch with respect to the 
model because of the model's nonlinearity, 
the mismatched training and test HeCS distributions 
lead to the model's biased results. 
This result indicates that our model may require to be updated with 
sufficient samples of blue galaxies to estimate photometric 
redshifts in galaxy groups and clusters over 0.1 $< ~ z ~ <$ 0.3.

\subsection{Validation with Photometric Redshift Samples \label{subsec:sdss}}

\begin{figure*}
  \plottwo{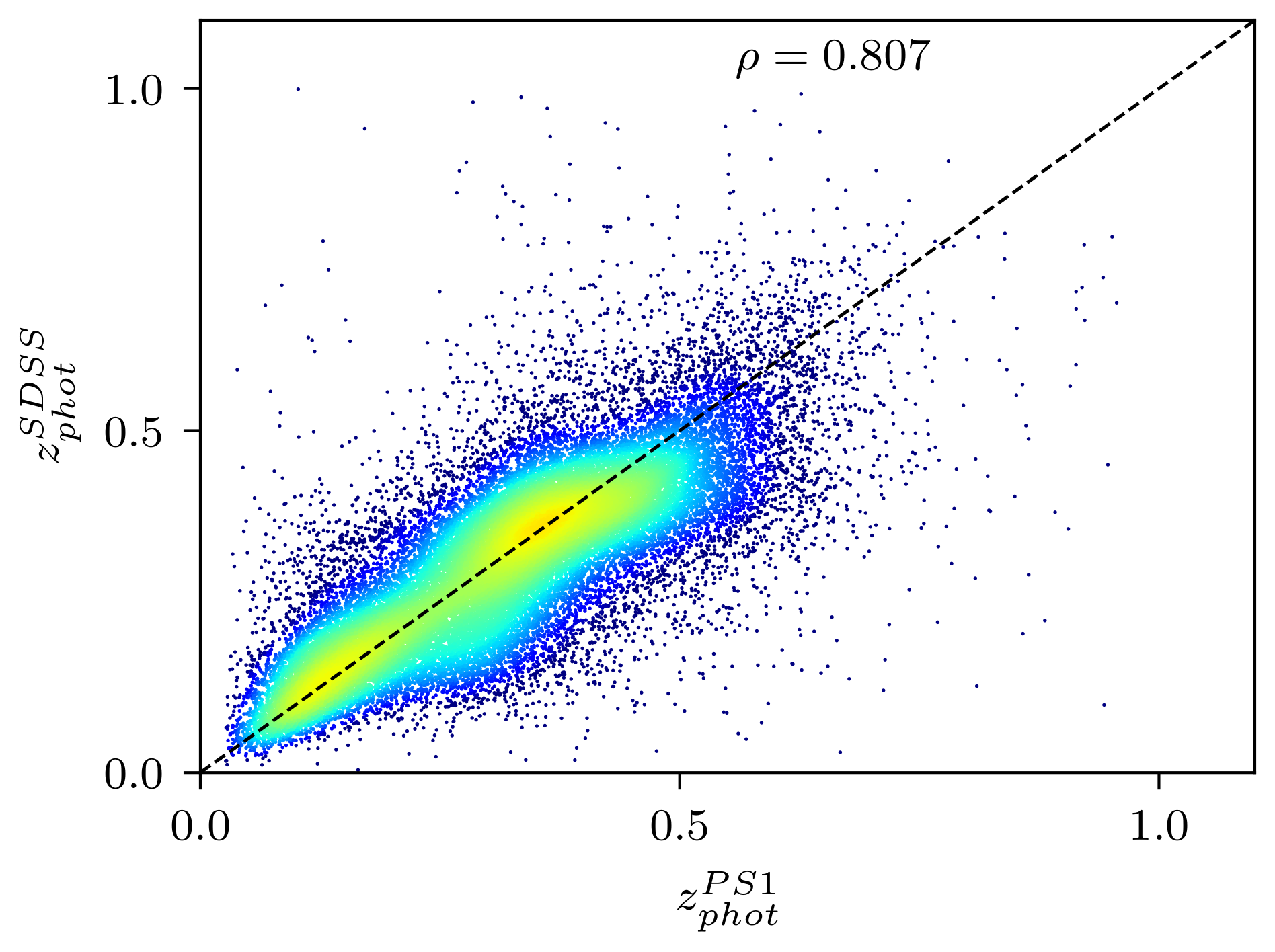}{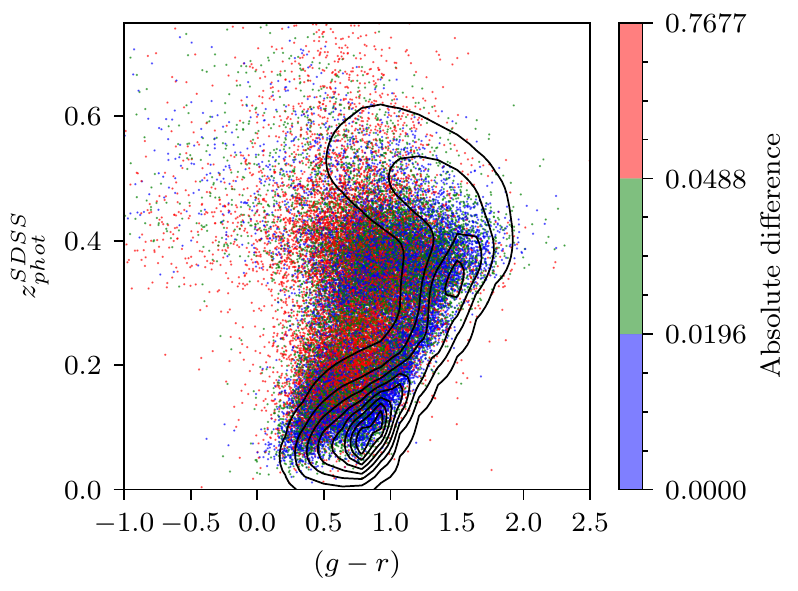}
  \caption{{\it Left}: Distribution of the comparison SDSS photometric redshifts and 
  derived photometric redshifts $z_{avg}$ for the PS1 photometric data. 
  {\it Right}: The distribution of the absolute difference 
  between SDSS photometric redshifts and the derived $z_{avg}$ in the space of 
  $(g - r)$ (Kron measurement) and SDSS photometric redshifts. The scatters are color-coded, as 
  in Figure~\ref{fig:HeCS_redshift_and_input_space}, and the distribution of PS1 training 
  samples are drawn as contour lines for comparison. The distribution of the absolute 
  difference between SDSS photometric redshifts and the derived $z_{avg}$ with 
  respect to $(g - r)$ shows that the significant fraction of samples with a large 
  difference of photometric redshifts have $(g - r)$ out of the main training samples. 
  However, we still find some consistent estimation of photometric redshifts between 
  the two different methods/data for $(g - r) \sim 0.7$ and $z_{phot}^{SDSS} \sim 0.3$.
  \label{fig:SDSS_redshift_and_input_space}}
\end{figure*}

Our trained model is compared with other photometric redshift estimation 
methods. Because there are more galaxy objects with known photometric redshifts 
than those with spectroscopic redshifts, we intend to use this comparison to 
assess the validity of the trained model over a wide range of redshifts and 
input space although photometric redshifts are not as precise 
and accurate as spectroscopic redshifts. 

As summarized in Table \ref{tab:photoz_comp}, we use 
the SDSS Data Release 12 (DR12) \citep{2015ApJS..219...12A}, which 
contains photometric redshifts derived from a nearest-neighbor fit 
with a kd-tree structure of training samples \citep{2007AN....328..852C}. 
We extract SDSS photometric redshifts for objects found in the area 
centered at RA 26.25\degree, DEC -7.5\degree with a radius of 2.5\degree, and 
we extract the PS1 DR1 objects corresponding to the SDSS objects that were identified. 
After discarding objects with SDSS spectroscopic redshifts, the sample 
size becomes 44,890 for the same filtering rule that we use for the 
training data of the PS1 DR1 data. 
We examine point-source scores of the 44,890 objects in the PS1 DR1 catalog 
\citep{2018PASP..130l8001T}, leaving out only extended sources 
(i.e., galaxies) in the PS1 image data with the condition of 
the point-source score $<$ 0.1. The filtered data, including 
38,818 objects commonly found in the SDSS and PS1, should comprise only 
galaxy-like objects for which our trained model should be able to estimate 
photometric redshifts.

Generally, two different estimations of photometric redshifts are consistent 
although the methods and training data are completely different. Figure 
\ref{fig:SDSS_redshift_and_input_space} shows the comparison between the two 
photometric redshift inferences where $z_{phot}^{PS1}$ corresponds to $z_{avg}$ 
in our estimation and $z_{phot}^{SDSS}$ represents 
estimation by robust fit to nearest neighbors in the SDSS reference set. 
Pearson's correlation coefficient $\rho \sim 0.81$ confirms 
that the two methods are consistent for most samples. A dominant fraction of
objects with discrepant redshifts seems to have a range of colors not properly 
represented by training samples; however, we report certain fraction of 
objects to have consistent redshifts although their colors do not follow 
the color of the training samples 
(see Figure \ref{fig:SDSS_redshift_and_input_space}). 

We examine possible causes of certain objects showing a large discrepancy in the estimated 
redshifts. For example, \object{SDSS J014232.69-090324.5} corresponding 
to PS1 DR1 objid $=$ 97130256361222090 seems blended in the PS1 DR1 and 
the SDSS images. The photometric redshifts of this object are 0.203 and 0.858 
as $z_{phot}^{SDSS}$ and $z_{phot}^{PS1}$, respectively. 
The \object{SDSS J014228.47-072343.0} (i.e., PS1 DR1 objid $=$ 99120256186386095) 
shows possible effects of blending in their images with 
the photometric redshifts $z_{phot}^{SDSS} = 0.018$ and 
$z_{phot}^{PS1} = 0.391$. The blending of sources is a well-known problem 
for estimating photometric redshifts \citep{2019MNRAS.483.2487J,
2021ApJS..253...31L}. We also find certain objects have conflicting 
colors between the SDSS DR12 and PS1 DR1 catalogs. For example, 
\object{SDSS J014126.01-065056.2} matched to the PS1 DR1 objid $=$ 
99780253584271692 has $(g - r) = 1.1 \pm 0.01$ (Kron color) and 
$0.7 \pm 0.32$ (aperture color) in the SDSS DR12 and PS1 DR1 catalogs, 
respectively. The photometric redshifts of this object are 
$z_{phot}^{SDSS} = 0.941$ and $z_{phot}^{PS1} = 0.172$.

\subsection{Model Outcomes for Non-galaxy Objects \label{subsec:hsc_nongalaxy}}

\begin{figure*}
  \plottwo{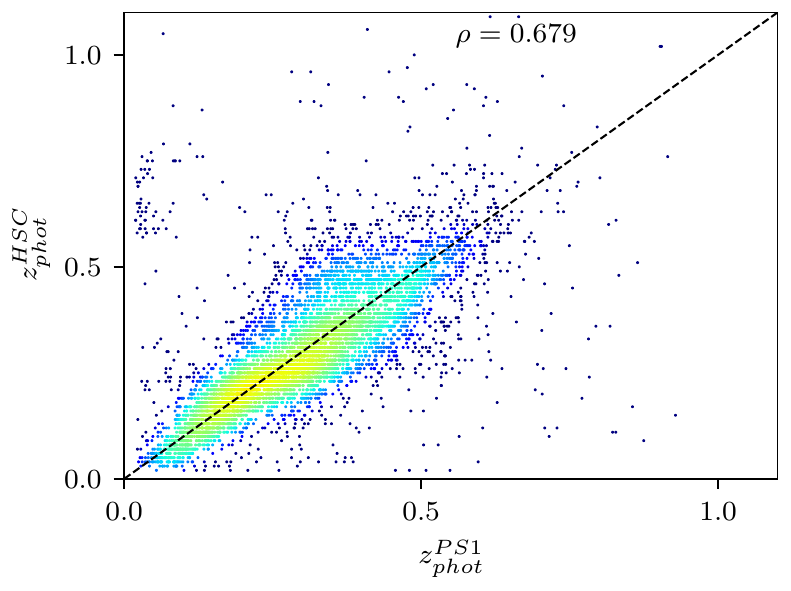}{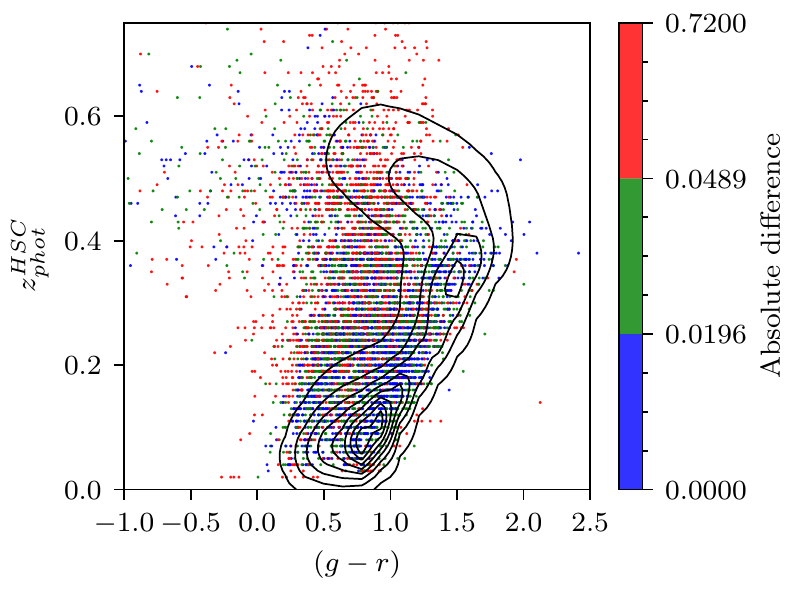}
  \caption{{\it Left}: Distribution of the comparison HSC galaxy photometric redshifts and 
  derived photometric redshifts $z_{avg}$ for the PS1 photometric data. {\it Right}: The 
  distribution of the absolute difference between the HSC photometric redshifts and the derived 
  $z_{avg}$ in the space of $(g - r)$ (Kron measurement) and HSC photometric redshifts. The color-coding of scatters 
  and contour lines are as those of Figure~\ref{fig:SDSS_redshift_and_input_space}. The pattern of 
  the discrete $z_{phot}^{HSC}$ distribution appears clearly in both plots. The correlation $\rho$ 
  between the two photometric redshifts is lower than that in for the comparison with the SDSS 
  photometric redshift presented in Figure \ref{fig:SDSS_redshift_and_input_space} partly due to the 
  discrete distribution of $z_{phot}^{HSC}$. 
  \label{fig:HSC_galaxy}}
\end{figure*}

The validity of trained machine learning models depends on the assumption 
that the training and test data follow the same distribution 
from the learning model's view. Therefore, when the trained model 
infers photometric redshifts for objects obtained from the OOD 
data, the estimation should be highly \textit{uncertain} and/or 
\textit{inaccurate}. We already present the case showing this effect 
in Section \ref{subsec:HeCS} for the slightly different distribution of input features 
for galaxies between training samples and HeCS test data.

The application of the trained model on the 
datasets HSC-PDR2-Mizuki-Galaxy and HSC-PDR2-Mizuki-NonGalaxy enables us 
to evaluate the results for the physically OOD objects, 
i.e., non-galaxy objects. Photometric redshifts in these datasets 
are results acquired in running a template fitting-code MIZUKI 
\citep{2015ApJ...801...20T,2018PASJ...70S...9T,2020arXiv200301511N}. 
The estimation products have probabilities of being stars, quasars, and 
galaxies. We select objects with photometric redshifts around 
RA 31.25\degree, DEC -2.5\degree with a radius of 2.5\degree. Following 
the filtering criteria used for the training data, we extract 6,996 objects as 
galaxy objects and 3,267 objects as non-galaxy objects with the PS1 DR1 data. 
The number of stars is 3,171 among 3,267 non-galaxy objects. 
Therefore, $\sim$100 objects are classified as quasars in the HSC test data.

The estimated photometric redshifts of galaxy objects in the HSC comparison data 
are similarly estimated to what our trained model estimates as shown in 
Figure \ref{fig:HSC_galaxy}. In general, our 
machine-learning estimation of photometric redshifts seems consistent with 
those derived by the template fitting-code. However, we report  
certain systematic difference patterns such as objects with 
$z_{phot}^{HSC} \sim 0.6$ for $z_{phot}^{PS1} \sim 0.1$. The discrete distribution 
of $z_{phot}^{HSC}$ appears to be a systematic pattern embedded in the HSC Mizuki 
inference of photometric redshifts. As shown in Figures 
\ref{fig:HeCS_redshift_and_input_space} and 
\ref{fig:SDSS_redshift_and_input_space}, we report that most objects with 
a large difference between the two photometric redshifts 
can be considered OOD samples based on the input data 
properties (see Figure \ref{fig:HSC_galaxy}).

\begin{figure*}
  \plottwo{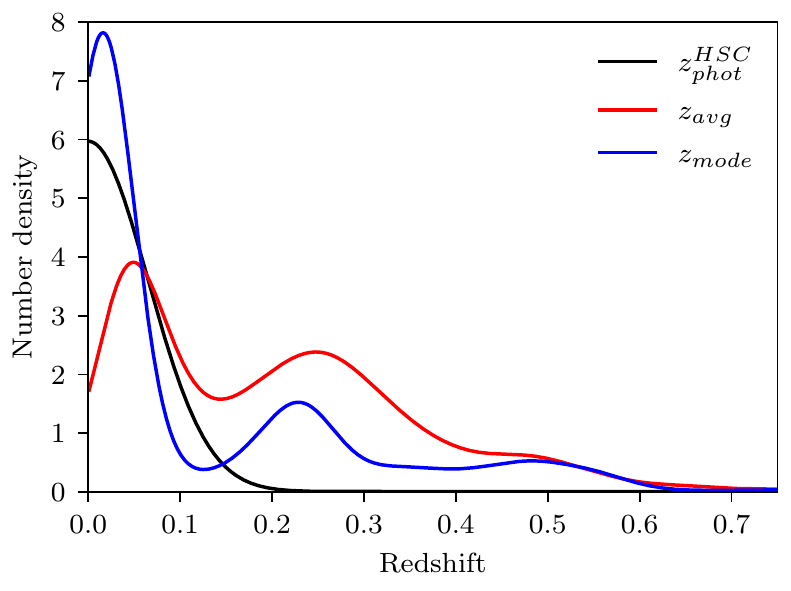}{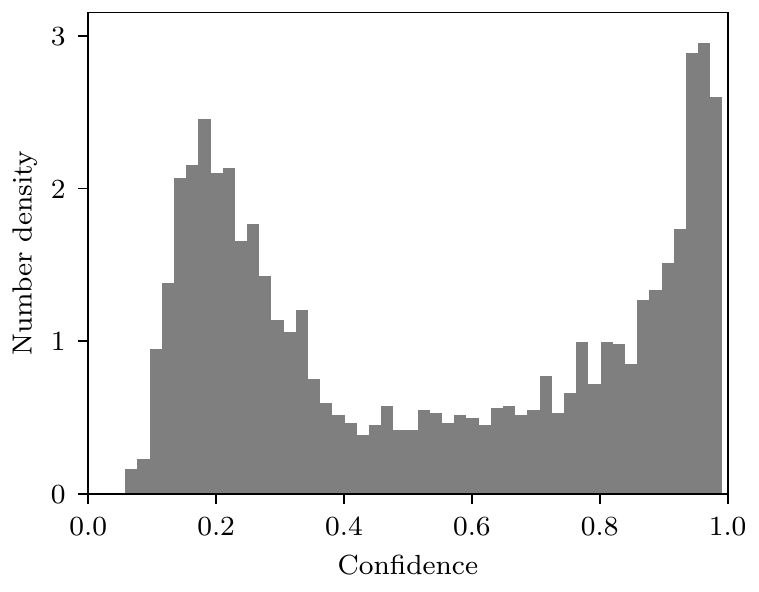}
  \caption{Distribution of the comparison HSC non-galaxy spectroscopic redshifts, 
  derived photometric redshifts $z_{avg}$ as the average value, and $z_{mode}$ 
  as the mode value ({\it left}). Confidence distribution of the non-galaxy samples 
  ({\it right}) shows that our trained model is overconfident on some of the physically 
  OOD samples.
  \label{fig:HSC_nongalaxy}}
\end{figure*}

We examine how our trained model estimates 
photometric redshifts of non-galaxy objects. If the model is 
well-trained to inductively infer the galactic photometric redshifts, 
photometric redshifts of the physically OOD samples (i.e., 
stars and galaxies) should follow the overall redshift estimation of galaxies 
when the OOD samples have similar input values as training galaxies. The useful 
machine learning model should result in a highly uncertain estimation of 
photometric redshifts for data samples from the 
OOD with respect to the input space and the trained model even when the 
OOD samples are galaxies. 

Figure \ref{fig:HSC_nongalaxy} shows the distribution of photometric redshifts 
for the non-galaxy HSC objects. 
Because the majority of non-galaxy HSC objects 
are stars, their redshift distribution has a peak at z $=$ 0. Comparing the 
distribution to that of the training samples 
(see Figure \ref{fig:photoz_comp_redshift}), 
the derived photometric redshifts of the physically OOD samples share a similar distribution 
to that of the training samples at z $\sim$ 0.1 and 0.5. However, the 
derived photometric redshifts of the physically OOD samples 
show a concentrated distribution around z $=$ 0.25 where the input 
values and photometric redshifts of these physically OOD samples are 
distributed in a manner similar to the galaxy OOD samples as shown in Figure 
\ref{fig:HSC_galaxy} (i.e., the galaxy OOD objects at $(g - r) \sim 0.7$ and 
$z_{phot}^{HSC} \sim 0.25$).

As shown in Figure \ref{fig:HSC_nongalaxy}, the trained model produces 
overconfident results on certain physically OOD samples. 
We anticipate that the well-trained model will output nearly identical probability 
distributions for OOD samples as random guesses, i.e., uniform distribution. 
In such a case, the confidence distribution should be nearly uni-modal with 
a peak in the low confidence range and then monotonically diminish as the confidence 
increases. However, the confidence distribution of physically OOD 
samples is multi-modal and has the tallest peak with high confidence. 
Therefore, the high confidence of the trained model's outputs does not guarantee that 
the tested sample has the same distribution as the training samples, particularly for 
the physically OOD samples. Unless the physically OOD samples such 
as stars and quasars are separately classified \citep[e.g.,][]{2018A&A...619A..14F}, 
the application of our model to these OOD samples can produce incorrect inference 
results with overconfidence.

\section{Discussion and Conclusion \label{sec:discon}}

The machine learning model presented in this study shows that 1) 
the machine-learning estimation of photometric redshifts 
can incorporate the uncertainty of color measurements and $E(B-V)$ as input 
data\footnote{In Appendix \ref{AppndSec:EBVcorr}, we explore the influence 
of the Galactic dust extinction correction on the model performance and find 
that the correction as a pre-processing of the data causes undesirable accuracy 
declination and bias according to $E(B-V)$.}, 2) the anchor loss adopted in the 
framework of NNs can be useful to handle difficult training 
data for target prediction generally linked to the sample imbalance problem, 
thereby improving the trained model for the difficult problem domains, and 3) 
the ensemble learning approach with various anchor loss weighting parameters 
is an effective way of maintaining the balance between the increased bias 
caused by the anchor loss and the improved overall accuracy.

We expect other researchers to use our trained model in various ways with the 
publicly available Python code on 
GitHub\footnote{\href{https://github.com/GooLee0123/MBRNN}{https://github.com/GooLee0123/MBRNN}} 
under an MIT License \citep{lee_and_shin_2021_5529452}, 
which is based on PyTorch \citep{NEURIPS2019_9015}. 
The code can be used for both training new data such as 
the PS1 data release 2 with photometric colors in different types like 
aperture magnitudes and inference 
of photometric redshifts with the PS1 DR1 data. 
As planned, the inference made by our model can be combined with 
the conventional fitting-based 
inference. Users of our model can consider 
using the time-consuming 
model-fitting codes to estimate the photometric redshifts only 
for objects with low confidence values or a large difference 
between $z_{avg}$ and $z_{mode}$ 
in our machine learning inference (see Figure \ref{fig:distributionAvgMode}). 
The model can be adopted and retrained with new training samples, which 
focus on specific science goals such as studies of low-z red galaxies 
\citep[e.g.,][]{2011MNRAS.417.1891A}. In particular, retraining the model 
with more fine redshift bins than those used in our study at low redshift might be useful 
for science goals related to low-redshift galaxies with better performance 
(i.e., higher accuracy and lower variance). 

Our application of ensemble learning and comparison to the other machine 
learning method results encourages others to combine their machine learning model 
results with ours in different ensemble learning forms (for example, 
dynamical ensemble, instead of the static ensemble adopted in this study). 
As shown in Section \ref{subsec:sdss}, the consensus between our model and other 
machine learning models indicates that combining multiple model results, 
which are generated by either different data or methods, can 
be a useful approach to improve the reliability of machine learning inference 
\citep[e.g.][]{2018AJ....156..201S}.

The performance of ensembling models with the anchor loss 
with respect to the given redshift bins hints that it might be powerful 
to combine models trained with the observed spectroscopic training samples, 
which are mainly low-redshift bright objects, and models trained with 
the {\it simulated} high-redshift or faint training samples. 
Using the simulated mock 
catalog data in the machine learning models is not a new concept, and 
the application of the NNs with the mock data has already been 
presented in the past \citep[e.g.][]{10.1046/j.1365-8711.2003.06271.x}. 
Because the observed training samples have the redshift-dependent 
distribution biased toward low-redshift space 
(see Figure \ref{fig:distribution_dataset}), the lack 
of high-redshift learning samples can be handled by incorporating 
high-redshift mock/simulated data if the simulated data is 
well-calibrated to the observed/expected properties (particularly, 
photometric properties) of galaxies 
at high redshifts with a correct distribution of the properties 
\citep[][]{1997AJ....113....1S,2017AJ....154..277K}. 
Our method can naturally combine the multiple models 
with redshift bins where a fraction of models trained with mock data 
perform well for a limited range of redshift bins.

Our examination of the trained model's limitations leads us to issues that 
should be addressed to improve machine learning inference of 
photometric redshifts. First, additional spectroscopic samples are required 
to improve the trained model's accuracy. 
The accuracy of machine learning inference 
might be degraded for specific parts of mapping between the input and 
redshift space because of the lack of sufficient training samples, which are generally 
considered close to OOD samples \citep{2017MNRAS.468.4323B}. 
The mismatch between the training sample and test 
data distributions results in biased estimation of photometric redshifts 
\citep{2018MNRAS.477.4330R}. 
Even when the test data contain only galaxies, the distribution of 
input features in the test data can become the OOD case, 
as we examined in Section \ref{subsec:HeCS}. Therefore, new machine learning 
models require to include additional spectroscopic samples as training 
data covering large input and output (i.e., redshift) spaces.

The physically OOD objects become problematic as well as reported in 
Section \ref{subsec:hsc_nongalaxy}. Some fraction of physically  
OOD objects can be seen as valid in-distribution galaxies with respect to 
input space and models even though quasars and stars are physically different 
from galaxies. Therefore, the algorithm detecting the OOD samples in terms of 
the input-output mapping cannot handle the physically OOD samples 
with valid input values. The most straightforward solution 
to address this issue might be using a separate machine learning model 
to classify test objects \citep[e.g.,][]{2015MNRAS.453..507K,
2020A&A...639A..84C,2021MNRAS.503.4136G} and forward only galaxy objects 
to the trained model for the inference of photometric redshifts.

Future studies, including our next study, will require to tackle the OOD problem. 
In particular, the improved model should be able to produce a quantitative 
evaluation of how much the test objects deviate from the in-distribution samples. 
The quantitative OOD score of test samples will allow us to use 
computationally expensive procedures such as spectral modeling 
for only objects with high OOD scores.

\begin{acknowledgments}

We thank David Parkinson for his careful reading and thoughtful comments. We are 
also grateful to Kyungmin Kim and Hyung Mok Lee for insightful discussions. 
We also thank the anonymous referee for helpful comments. 
This research has made use of the VizieR catalog access tool, CDS,
Strasbourg, France (DOI : 10.26093/cds/vizier). The original description 
of the VizieR service was published in 2000, A\&AS 143, 23. 
The Pan-STARRS1 Surveys (PS1) and the PS1 public science archive have been made 
possible through contributions by the Institute for Astronomy, the University 
of Hawaii, the Pan-STARRS Project Office, the Max-Planck Society and its 
participating institutes, the Max Planck Institute for Astronomy, Heidelberg 
and the Max Planck Institute for Extraterrestrial Physics, Garching, The Johns 
Hopkins University, Durham University, the University of Edinburgh, the Queen's 
University Belfast, the Harvard-Smithsonian Center for Astrophysics, the Las 
Cumbres Observatory Global Telescope Network Incorporated, the National Central 
University of Taiwan, the Space Telescope Science Institute, the National 
Aeronautics and Space Administration under Grant No. NNX08AR22G issued through 
the Planetary Science Division of the NASA Science Mission Directorate, 
the National Science Foundation Grant No. AST-1238877, the University of Maryland, 
Eotvos Lorand University (ELTE), the Los Alamos National Laboratory, and 
the Gordon and Betty Moore Foundation.

\end{acknowledgments}

\appendix

\setcounter{table}{0}
\renewcommand{\thetable}{\Alph{section}.\arabic{table}}
\setcounter{figure}{0}
\renewcommand{\thefigure}{\Alph{section}.\arabic{figure}}

\section{Search for the Optimal Configuration of the MBRNN Model\label{AppendSec:grid-search}}
\renewcommand{\TabFirstGrid}{
\hspace*{-2.5cm}\begin{tabular}{c|c|c|c c c c c c}
\tableline\tableline
  Data scaling & Color uncertainties & $E(B-V)$ & Bias & MAD & $\sigma$ & $\sigma_{68}$ & {\it NMAD} & $R_{cat}$ \\
\tableline\tableline
\multirow{4}{*}{Min-max} & \multirow{2}{*}{Yes} & Yes & {\bf 0.0019} & {\bf 0.0256} & {\bf 0.0398} & {\bf 0.0273} & {\bf 0.0256} & {\bf 0.0088} \\
\cline{3-3}
&  & No & 0.0021 & 0.0267 & 0.0412 & 0.0287 & 0.0270 & 0.0096 \\
\cline{2-3}
& \multirow{2}{*}{No} & Yes & 0.0028 & 0.0316 & 0.0496 & 0.0329 & 0.0310 & 0.0188 \\
\cline{3-3}
& & No & 0.0030 & 0.0336 & 0.0533 & 0.0346 & 0.0327 & 0.0226 \\
\cline{1-3}
\multirow{4}{*}{Standardization} & \multirow{2}{*}{Yes} & Yes & 0.0022 & 0.0256 & 0.0399 & 0.0273 & 0.0257 & 0.0086 \\
\cline{3-3}
& & No & 0.0023 & 0.0268 & 0.0415 & 0.0288 & 0.0270 & 0.0095 \\
\cline{2-3}
& \multirow{2}{*}{No} & Yes & 0.0030 & 0.0317 & 0.0498 & 0.0329 & 0.0310 & 0.0191 \\
\cline{3-3}
& & No & 0.0031 & 0.0336 & 0.0535 & 0.0346 & 0.0328 & 0.0229 \\
\tableline
\end{tabular}
}

\renewcommand{\TabSecondGrid}{
\hspace*{-2.5cm}\begin{tabular}{c|c|c c c c c c}
\tableline\tableline
\multicolumn{2}{c|}{Loss configuration} & Bias & MAD & $\sigma$ & $\sigma_{68}$ & {\it NMAD} & $R_{cat}$ \\
\tableline\tableline
\multicolumn{2}{c|}{Binary Cross Entropy ($\gamma=0$)} & {\bf 0.0019} & {\bf 0.0256} & 0.0398 & {\bf 0.0273} & {\bf 0.0256} & 0.0088 \\
\cline{1-2}
\multirow{5}{*}{Anchor Loss} & $\gamma=0.2$ & 0.0025 & 0.0257 & 0.0398 & 0.0275 & 0.0259 & 0.0087 \\
\cline{2-2}
& $\gamma=0.5$ & 0.0027 & 0.0256 & {\bf 0.0396} & 0.0274 & 0.0258 & {\bf 0.0086} \\
\cline{2-2}
& $\gamma=1$ & 0.0027 & 0.0256 & 0.0396 & 0.0275 & 0.0260 & 0.0086 \\
\cline{2-2}
& $\gamma=2$ & 0.0039 & 0.0260 & 0.0399 & 0.0280 & 0.0266 & 0.0087 \\
\cline{2-2}
& $\gamma=5$ & 0.0327 & 0.0483 & 0.0611 & 0.0536 & 0.0507 & 0.0451 \\
\tableline
\end{tabular}
}

\renewcommand{\TabThirdGrid}{
\hspace*{-2.5cm}\begin{tabular}{c|c|c c c c c c}
\tableline\tableline
\multicolumn{2}{c|}{Bin type} & Bias & MAD & $\sigma$ & $\sigma_{68}$ & {\it NMAD} & $R_{cat}$\\
\tableline\tableline
\multirow{5}{*}{Uniform Bins} & 32 & 0.0020 & 0.0257 & 0.0399 & 0.0274 & 0.026 & 0.0086 \\
\cline{2-2}
& 64 & {\bf 0.0017} & {\bf 0.0254} & {\bf 0.0394} & {\bf 0.0272} & 0.0256 & {\bf 0.0084} \\
\cline{2-2}
& 128 & 0.0019 & 0.0256 & 0.0398 & 0.0273 & 0.0256 & 0.0088 \\
\cline{2-2}
& 256 & 0.0024 & 0.0254 & 0.0396 & 0.0272 & {\bf 0.0255} & 0.0084 \\
\cline{2-2}
& 512 & 0.0024 & 0.0255 & 0.0396 & 0.0273 & 0.0256 & 0.0086 \\
\cline{1-2}
\multirow{5}{*}{Non-uniform Bins} & 32 & 0.0144 & 0.0337 & 0.0577 & 0.0310 & 0.0281 & 0.0342  \\
\cline{2-2}
& 64 & 0.0075 & 0.0285 & 0.0470 & 0.0288 & 0.0266 & 0.0168 \\
\cline{2-2}
& 128 & 0.0045 & 0.0266 & 0.0429 & 0.0278 & 0.0259 & 0.0117 \\
\cline{2-2}
& 256 & 0.0028 & 0.0259 & 0.0415 & 0.0274 & 0.0255 & 0.0096 \\
\cline{2-2}
& 512 & 0.0023 & 0.0256 & 0.0404 & 0.0272 & 0.0255 & 0.0090 \\
\tableline
\end{tabular}
}

\begin{table*}
\centering
\caption{Metrics of the grid search for the best model configuration of the MBRNN model. As mentioned in Section~\ref{sec:result}, the lower metrics indicate a higher accuracy of the model.}
  \subfigure[][Metrics for different data scaling methods, and the existence of color uncertainties and $E(B-V)$ as input features.]{\TabFirstGrid}\qquad
\subfigure[][Metrics for different anchor loss $\gamma$ values.]{\TabSecondGrid}
\qquad
\subfigure[][Metrics for different redshift bin configurations.]{\TabThirdGrid}
\label{tab:grid_search}
\end{table*}

We perform a grid search by varying model configurations with the validation data to find 
the empirically optimal configuration of the MBRNN model. The grid search includes 
three different sets of configuration variations. 
First, when setting 
the number of redshift bins to 128 and using the anchor loss with $\gamma ~=~ 0$, 
we examine the changes in the model's accuracy in terms of 
input data scaling methods (i.e., standardization vs. min-max normalization), 
and the existence of color uncertainties and $E(B-V)$ as input features.

The model using the min-max normalization, color uncertainties, and $E(B-V)$ 
shows the best point-estimation accuracy as summarized in 
Table \ref{tab:grid_search} (a). Including the color uncertainties as input features 
has the largest impact on point-estimation accuracy among 
three factors. The min-max normalization and inclusion of 
$E(B-V)$ in the input have small positive effects on 
point-estimation accuracy.

Second, we perform the grid-search for the anchor loss parameter $\gamma$ 
fixing the number of redshift bins to 128. Because 
the anchor loss requires a prior setup of $\gamma$ before training, 
we examine a set of $\gamma$ values of 0 (i.e., binary cross entropy loss), 
0.2, 0.5, 1, 2, and 5. The results of this second grid search are 
presented in Table \ref{tab:grid_search} (b). 
The model with $\gamma$ of 0 outperforms the others in terms of 
the overall accuracy as naively expected.

Using the maximum performance configuration, we finally examine 
the various strategies of redshift binning and the number 
of bins. The search includes a comparison of results 
between uniform and non-uniform binning methods as well as the results for 
the 32, 64, 128, 256, and 512 redshift bins.
In the non-uniform redshift binning, we set the bin edges such that each 
bin contains nearly the same number of samples, as explained in Section \ref{subsec:mbrnn}.
As illustrated in Table \ref{tab:grid_search} (c), the MBRNN model 
with 64 uniform bins outperforms the other configurations for most metrics, 
although the difference is not significant. Moreover, we discover 
that the uniform binning case outperforms the non-uniform one.

\section{Catastrophic Samples \label{AppendSec:Cat}}
\begin{figure*}
\centering
\includegraphics[width=1.0\textwidth]{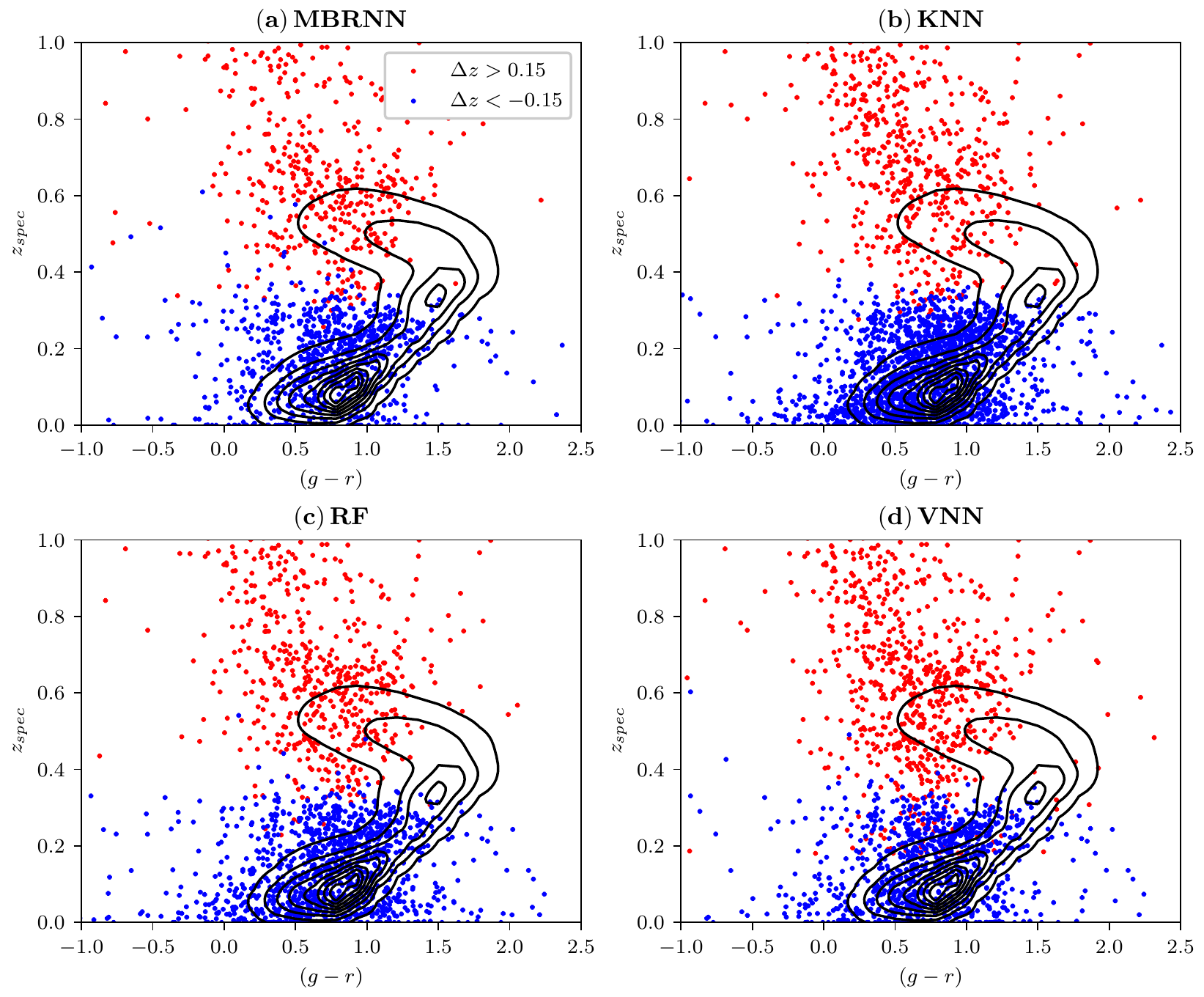}
\caption{Distributions of the {\it cat} samples in the space of $(g-r)$ 
in Kron measurement and spectroscopic 
redshifts. The {\it cat} samples are shown with different colors 
for $\Delta z > 0.15$ (red) and $\Delta z < -0.15$ (blue). The lower fraction of 
  the {\it cat} samples in the MBRNN than in other models is mainly 
  due to the reduction of these samples at low redshifts. \label{fig:cat_dstrb}}
\end{figure*}

We examine the distributions of the {\it cat} samples in the input space 
according to models and find that these samples populate in the similar regions 
of the input space. Figure \ref{fig:cat_dstrb} presents the distributions of under 
and overestimated {\it cat} samples in the MBRNN and baseline models. As shown in the figure, 
the point-estimated photometric redshifts of the {\it cat} samples in MBRNN 
with true spectroscopic redshifts larger and lower than $z_{spec} \sim 0.4$ 
tend to be under- and over-estimated, respectively. 
This pattern appears in the other baseline models too, and 
the under- and over-estimated photometric redshifts of the {\it cat} samples 
are similarly distributed in the input space. However, the area taken by 
the {\it cat} samples in the MBRNN model is smallest among the models, being 
consistent with the fact that the $R_{cat}$ is smallest in the MBRNN model 
(see Table \ref{tab:comp}).

We also find that MBRNN works better than the baseline models 
particularly in the low redshift range. Specifically, we compare 
how many {\it cat} samples found in the RF model, 
which shows the best performance among the baseline models, become 
{\it non-cat} samples in the MBRNN model. The RF model has 766 {\it cat} samples 
in its test, and the entire 543 and 223 samples with over- and under-estimated 
photometric redshifts in the RF model are the {\it non-cat} samples in the MBRNN. 
Especially, samples with $z_{spec} \sim 0$ mainly turn into the {\it non-cat} 
objects in the MBRNN model (see Figure \ref{fig:cat_dstrb}).

The {\it cat} samples in the models include various cases. When 
inspecting the {\it cat} samples in the MBRNN model, we find that some objects 
such as \object{SDSS J014904.18+243502.4}, 
\object{SDSS J104307.62+084059.2}, and \object{SDSS J101223.88+161313.4} 
might not have reliable photometric input features due to neighbor 
objects. Star-forming galaxies with emission lines 
are also found as the {\it cat} samples. For example, 
\object{SDSS J085139.46+455518.4} and \object{SDSS J091022.97+164534.9} 
have emission-lines representing star formation at redshifts of 
0.28 and 0.30, respectively. Objects like \object{SDSS J150912.92+344418.1} 
at $z_{spec} ~ = ~ 0.06028$ have a large apparent size, and their photometry 
might not be reliable.

\section{Results With Different Ensemble Learning Configurations \label{AppendSec:DiffEnsemble}}
\begin{table*}
\centering
\caption{Metrics for the ensemble cases with different redshift bins. \label{tab:EnsembleDiffBins}}
\hspace*{-2.5cm}\begin{tabular}{c|c c c c c c}
\tableline\tableline
Number of bins & Bias & MAD & $\sigma$ & $\sigma_{68}$ & {\it NMAD} & $R_{cat}$ \\
\tableline\tableline
\cline{1-7}
32 & 0.0015 & 0.0255 & 0.0392 & 0.0274 & 0.0260 & 0.0084 \\
64 & {\bf 0.0010} & {\bf 0.0253} & {\bf 0.0389} & {\bf 0.0272} & {\bf 0.0255} & 0.0082 \\
128 & 0.0014 & 0.0254 & 0.0390 & 0.0272 & 0.0256 & {\bf 0.0082} \\
64 \& 128 & 0.0013 & 0.0253 & 0.0389 & 0.0272 & 0.0255 & 0.0083 \\
\tableline\tableline
\end{tabular}
\end{table*}

Furthermore, we present the performance of the E3 ensemble model 
with 32 and 128 uniform redshift bins as well as 
the results obtained from the runs of the adopted 64 bins. Moreover, we consider merging 
the single models trained with 64 and 128 redshift bins rather than 
only the results with the same number of redshift bins. 
In particular, four single models sharing the same number of bins and trained 
with different anchor loss $\gamma$s --- 0, 0.2, 0.5, and 1 --- are combined 
for the ensemble models of 32, 64, and 128 bins, respectively. For the merged ensemble 
model of 64 and 128 bins, however, eight single models are combined; four 
each of the 64 and 128 bin models trained with different $\gamma$ values. Merging the models 
with the different number of redshift bins is conducted by interpolating their probability 
outputs to a higher number of redshift bins while maintaining the probability sum equal to 1.

Table \ref{tab:EnsembleDiffBins} compares the point estimation metrics of 
the ensemble models with 32, 64, and 128 bins, and the case of combining models 
generated with 64 and 128 redshift bins. 
We find no advantage or improvement 
in these cases over the adopted ensemble model 
of combing the runs with 64 redshift bins. 

\begin{table*}
\centering
\caption{Metrics for ensembling cases with the different anchor loss $\gamma$ values. The set ${\cal{G}}$ contains the models trained with anchor loss $\gamma$ of 0, 0.2, 0.5, and 1. \label{tab:EnsembleDiffGamma}}
\hspace*{-2.5cm}\begin{tabular}{c|c c c c c c}
\tableline\tableline
Set of $\gamma$ & Bias & MAD & $\sigma$ & $\sigma_{68}$ & {\it NMAD} & $R_{cat}$ \\
\tableline\tableline
${\cal{G}}$ & {\bf 0.0010} & {\bf 0.0253} & 0.0389 & {\bf 0.0272} & {\bf 0.0255} & {\bf 0.0082} \\
${\cal{G}} \cup \{2\}$ & 0.0014 & 0.0253 & 0.0389 & 0.0272 & 0.0256 & 0.0083 \\
${\cal{G}} \cup \{2, 5\}$ & 0.0024 & 0.0255 & {\bf 0.0389} & 0.0274 & 0.0260 & 0.0082 \\
${\cal{G}} \cup \{2, 5, 8\}$ & 0.0040 & 0.0258 & 0.0389 & 0.0279 & 0.0268 & 0.0083 \\
\tableline\tableline
\end{tabular}
\end{table*}

Moreover, we check how the high $\gamma$ values of the anchor loss 
affect the performance of the ensemble model. We attempt the E3 ensemble method 
with 64 uniform redshift bins while combining the additional results with $\gamma$ 
in the order of 2, 5, and 8 into the set of models 
used in the adopted E3 model (${\cal{G}}$), 
which is described in Section \ref{subsec:ensemble} as the model combining 
the results with $\gamma 
~ = ~ 0, 0.2, 0.5,$ and $1$. Table \ref{tab:EnsembleDiffGamma} illustrates 
the metrics of the models for the trial configurations. 
Compared with the other runs, the ensemble model with the set ${\cal{G}}$ 
achieves the highest accuracy in general.

\section{Examination of The Ensemble Model Calibration \label{AppendSec:ModelCalibration}}
\begin{figure*}
\centering

\includegraphics[]{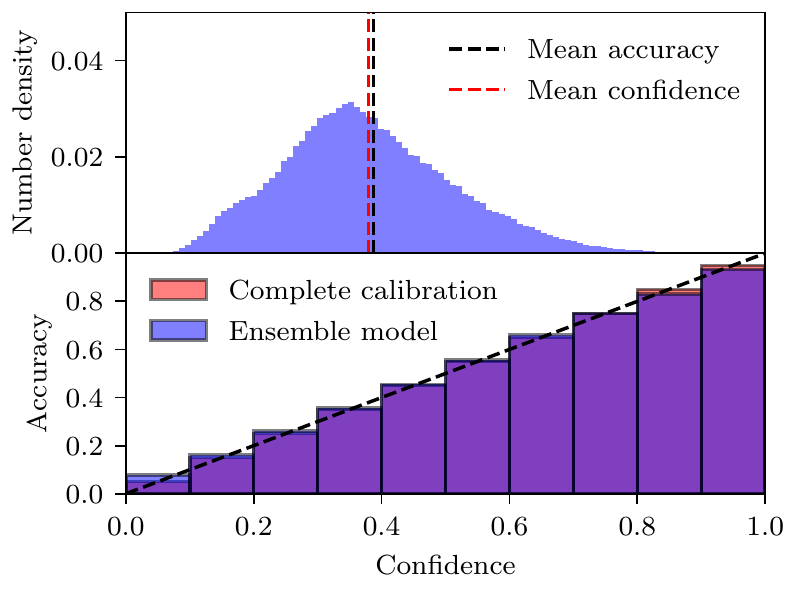}
\caption{
  {\it Top}: Confidence distribution in the E3 model. Black and red vertical lines indicate the 
  mean accuracy and confidence of the E3 model. {\it Bottom}: Reliability diagrams of 
  the E3 model. Red and blue bars represent the distribution of the complete calibration and 
  the E3 model result, respectively. The black dashed line marks a line with a slope of 1, meaning 
  the perfect calibration. 
\label{fig:EnsembleCalibration}}
\end{figure*}

Many typical modern NNs are not well-calibrated 
\citep{pmlr-v70-guo17a}. Hence, we examine the calibration of our E3 
ensemble model. A well-calibrated model should have similar mean accuracy and 
confidence values. 
Figure \ref{fig:EnsembleCalibration} shows that our ensemble model E3 is 
well-calibrated. 
The difference between the mean accuracy and confidence of our E3 ensemble model is 
$\sim$0.0078, which is close to 0. 
A reliability diagram depicting the accuracy variation as per 
the confidence confirms the adequate calibration of the ensemble model, 
almost identical to the completely calibrated case.

\section{Effects of The Galactic Extinction Correction \label{AppndSec:EBVcorr}}

We apply the galactic extinction correction to the input colors using $E(B-V)$
and a simple correction rule, and then we train our model with corrected colors. 
Therefore, the model learns without $E(B-V)$ as a training input feature in 
this experiment. Three different correction rules are tested with the given 
$E(B-V)$, and the three different correction results (I, II, and III cases in 
Table \ref{tab:extinction}) are compared in terms of model performances. 
Case I adopts the correction given as equations (7) to (13) 
in \citet{2012ApJ...750...99T} with the observed apparent $(g - i)$ color. 
We note that this correction is fundamentally incorrect because the correction 
is valid only with the intrinsic color $(g - i)$, which 
we do not know beforehand, rather than the observed color 
\citep[see][for discussions]{2017A&A...598A..20G}. The correction in 
cases II and III do not depend on $(g - i)$. Their correction rules adopt 
the representative values of the extinction $A$ given as $A$ at the pivot in 
\citet{2017A&A...598A..20G} and $A$ values used in 
\citet{2019ApJS..243....5S} for cases II and III, respectively. 
Although the metrics in case II are the lowest overall, 
there are no significant differences among these three cases.

\begin{figure*}
\centering
\plottwo{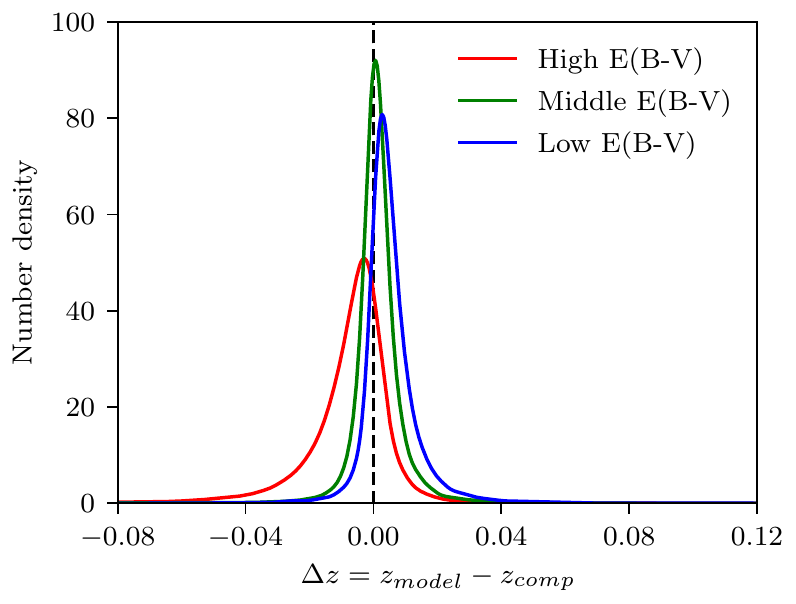}{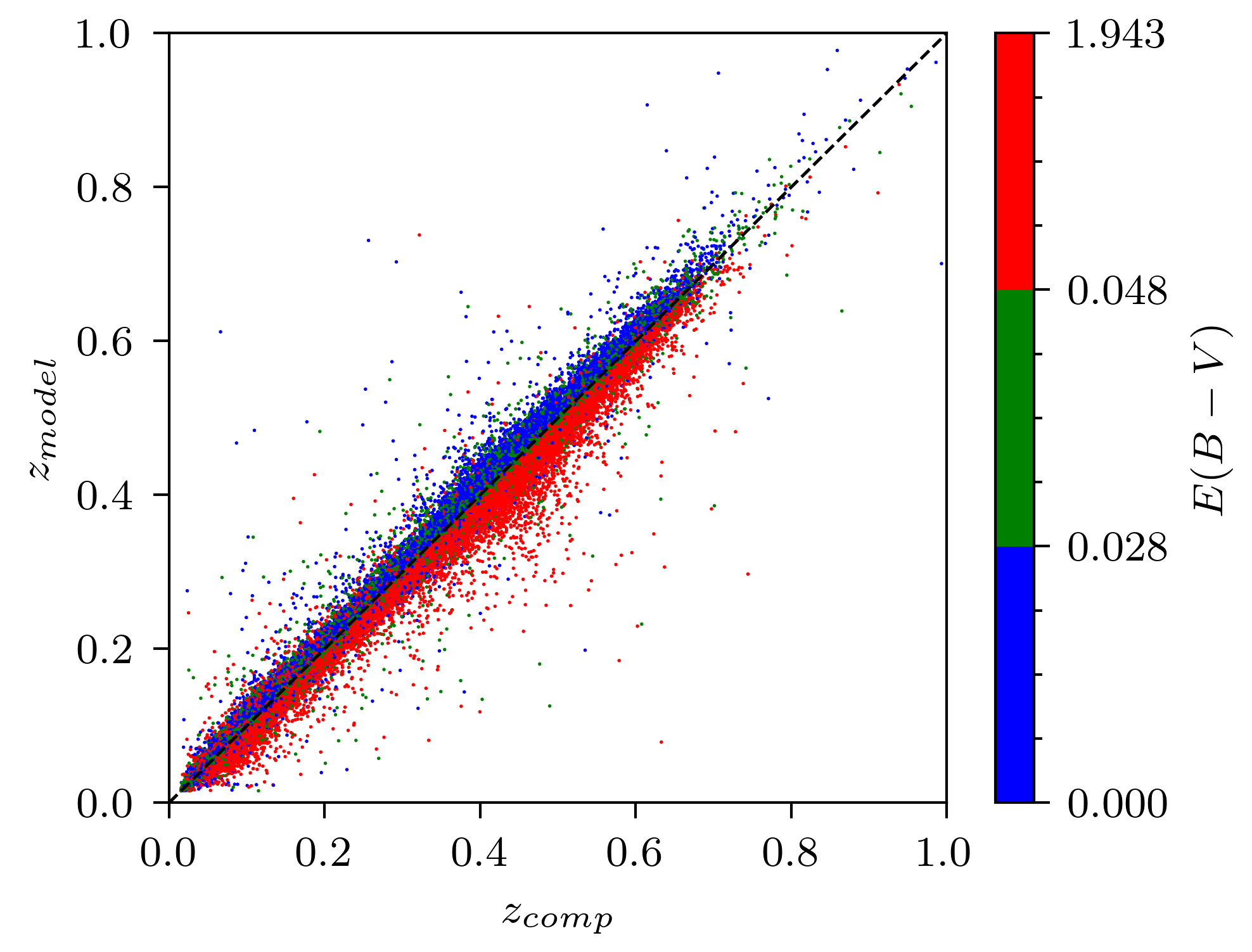}
  \caption{Distribution of the difference between photometric redshifts ($z_{model}$) 
  from the model trained with $E(B-V)$ and 
  those ($z_{comp}$) from the model trained without $E(B-V)$ but with 
  the Galactic extinction-corrected colors ({\it left}). 
  The samples are grouped into the low, medium, and high $E(B-V)$ value groups 
  in which each group has approximately one-third of the number of samples. 
  $\Delta z$ shows systematic differences depending on 
  the $E(B-V)$ values. This pattern is also found in the redshift 
  comparison plot ({\it right}). The redshifts of the low and high $E(B-V)$ 
  samples are mostly in the under- and over-estimation regions with respect to the 
  correspondence line, which is represented by the dashed line, respectively.
  \label{fig:model_best_comp}}
\end{figure*}

In case II, we analyze the effect of the correction on 
the photometric redshifts inferred by the trained model. 
We split $E(B-V)$ values into three different ranges, and each range 
has $\sim$33\% of samples as  low, middle, and high 
$E(B-V)$ values. Then, we compare photometric redshifts derived from 
our main model trained with $E(B - V)$ 
as an input feature ($z_{model}$) with case II results acquired with 
the extinction-corrected color data ($z_{comp}$). 
The left panel of Figure \ref{fig:model_best_comp} shows that the redshift difference 
($\Delta z ~=~ z_{model} ~-~ z_{comp}$) distribution. Interestingly, the distributions of redshifts for 
the low and high $E(B-V)$ values are positively and negatively shifted 
from a non-bias line ($\Delta z = 0$), respectively. This pattern is conspicuous 
in the right panel of the figure, which shows a comparison between $z_{comp}$ and $z_{model}$.
These results indicate that the model trained with the extinction-corrected colors 
tends to under- and over-estimate redshifts compared to $z_{model}$ 
for low and high $E(B-V)$ objects, respectively, indicating that the systematic bias is
induced by the Galactic extinction correction on colors.

\begin{figure*}
\centering
\includegraphics[width=1.0\textwidth]{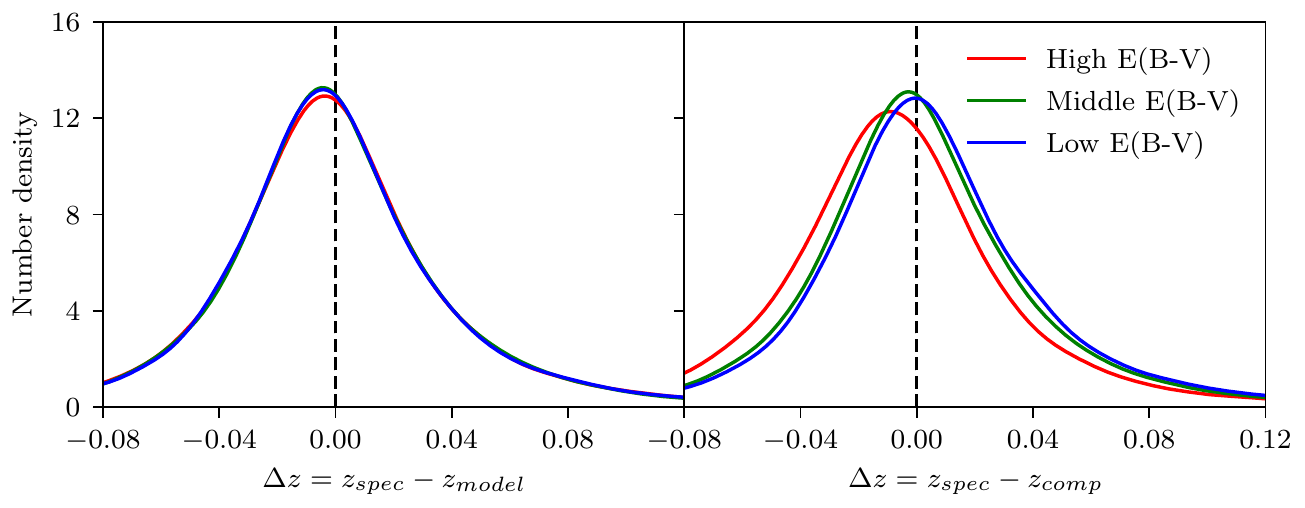}
  \caption{Distributions of the difference between spectroscopic redshift ($z_{spec}$) 
  and photometric redshifts ($z_{model}$ and $z_{comp}$). 
  The samples are binned with respect to $E(B-V)$ as done for Figure \ref{fig:model_best_comp}. 
  While the difference distributions for the $z_{model}$ ({\it left}) are unbiased regarding 
  $E(B-V)$, the distributions of the $z_{comp}$ with higher $E(B-V)$ are shifted towards the lower 
  $\Delta z$ direction ({\it right}).\label{fig:model_spec_model_best}}
\end{figure*}

Figure~\ref{fig:model_spec_model_best} 
shows the distribution of the $z_{spec} - z_{model}$ and $z_{spec} - z_{comp}$, 
where $z_{spec}$ corresponds to true spectroscopic redshifts. 
The distribution reported in the model with $E(B - V)$ as 
input features does not vary with respect to $E(B-V)$. 
However, the distribution of $z_{spec} - z_{comp}$ reveals that 
the trend found in Figure \ref{fig:model_best_comp} can be explained by the 
fact that the extinction correction introduces bias into the redshift inference. 


\begin{table*}
\begin{center}
\caption{Metrics for the experiments with three different Galactic extinction correction methods. \label{tab:extinction}}
\hspace*{-2.5cm}\begin{tabular}{c|c c c c c c}
\tableline\tableline
Case & Bias & MAD & $\sigma$ & $\sigma_{68}$ & {\it NMAD} & $R_{cat}$ \\
\tableline\tableline
I & {\bf 0.0019} & 0.0261 & {\bf 0.0403} & 0.0280 & 0.0265 & 0.0090 \\
II & 0.0019 & {\bf 0.0261} & 0.0404 & {\bf 0.0280} & {\bf 0.0264} & {\bf 0.0089} \\
III & 0.0019 & 0.0262 & 0.0405 & 0.0281 & 0.0265 & 0.0092 \\
\tableline\tableline
\end{tabular}
\end{center}
\end{table*}

\bibliography{astro,ml}{}

\begin{thebibliography}{}
\expandafter\ifx\csname natexlab\endcsname\relax\def\natexlab#1{#1}\fi
\providecommand{\url}[1]{\href{#1}{#1}}
\providecommand{\dodoi}[1]{doi:~\href{http://doi.org/#1}{\nolinkurl{#1}}}
\providecommand{\doeprint}[1]{\href{http://ascl.net/#1}{\nolinkurl{http://ascl.net/#1}}}
\providecommand{\doarXiv}[1]{\href{https://arxiv.org/abs/#1}{\nolinkurl{https://arxiv.org/abs/#1}}}

\bibitem[{{Abdalla} {et~al.}(2011){Abdalla}, {Banerji}, {Lahav}, \&
  {Rashkov}}]{2011MNRAS.417.1891A}
{Abdalla}, F.~B., {Banerji}, M., {Lahav}, O., \& {Rashkov}, V. 2011, \mnras,
  417, 1891, \dodoi{10.1111/j.1365-2966.2011.19375.x}

\bibitem[{{Aguado} {et~al.}(2019){Aguado}, {Ahumada}, {Almeida}, {Anderson},
  {Andrews}, {Anguiano}, {Aquino Ort{\'\i}z}, {Arag{\'o}n-Salamanca},
  {Argudo-Fern{\'a}ndez}, {Aubert}, {Avila-Reese}, {Badenes}, {Barboza
  Rembold}, {Barger}, {Barrera-Ballesteros}, {Bates}, {Bautista}, {Beaton},
  {Beers}, {Belfiore}, {Bernardi}, {Bershady}, {Beutler}, {Bird}, {Bizyaev},
  {Blanc}, {Blanton}, {Blomqvist}, {Bolton}, {Boquien}, {Borissova}, {Bovy},
  {Brandt}, {Brinkmann}, {Brownstein}, {Bundy}, {Burgasser}, {Byler}, {Cano
  Diaz}, {Cappellari}, {Carrera}, {Cervantes Sodi}, {Chen}, {Cherinka}, {Choi},
  {Chung}, {Coffey}, {Comerford}, {Comparat}, {Covey}, {da Silva Ilha}, {da
  Costa}, {Dai}, {Damke}, {Darling}, {Davies}, {Dawson}, {de Sainte Agathe},
  {Deconto Machado}, {Del Moro}, {De Lee}, {Diamond-Stanic}, {Dom{\'\i}nguez
  S{\'a}nchez}, {Donor}, {Drory}, {du Mas des Bourboux}, {Duckworth}, {Dwelly},
  {Ebelke}, {Emsellem}, {Escoffier}, {Fern{\'a}ndez-Trincado}, {Feuillet},
  {Fischer}, {Fleming}, {Fraser-McKelvie}, {Freischlad}, {Frinchaboy}, {Fu},
  {Galbany}, {Garcia-Dias}, {Garc{\'\i}a-Hern{\'a}ndez}, {Garma Oehmichen},
  {Geimba Maia}, {Gil-Mar{\'\i}n}, {Grabowski}, {Gu}, {Guo}, {Ha},
  {Harrington}, {Hasselquist}, {Hayes}, {Hearty}, {Hernandez Toledo}, {Hicks},
  {Hogg}, {Holley-Bockelmann}, {Holtzman}, {Hsieh}, {Hunt}, {Hwang},
  {Ibarra-Medel}, {Jimenez Angel}, {Johnson}, {Jones}, {J{\"o}nsson},
  {Kinemuchi}, {Kollmeier}, {Krawczyk}, {Kreckel}, {Kruk}, {Lacerna}, {Lan},
  {Lane}, {Law}, {Lee}, {Li}, {Lian}, {Lin}, {Lin}, {Lintott}, {Long},
  {Longa-Pe{\~n}a}, {Mackereth}, {de la Macorra}, {Majewski}, {Malanushenko},
  {Manchado}, {Maraston}, {Mariappan}, {Marinelli}, {Marques-Chaves},
  {Masseron}, {Masters}, {McDermid}, {Medina Pe{\~n}a}, {Meneses-Goytia},
  {Merloni}, {Merrifield}, {Meszaros}, {Minniti}, {Minsley}, {Muna}, {Myers},
  {Nair}, {Correa do Nascimento}, {Newman}, {Nitschelm}, {Olmstead}, {Oravetz},
  {Oravetz}, {Ortega Minakata}, {Pace}, {Padilla}, {Palicio}, {Pan}, {Pan},
  {Parikh}, {Parker}, {Peirani}, {Penny}, {Percival}, {Perez-Fournon},
  {Peterken}, {Pinsonneault}, {Prakash}, {Raddick}, {Raichoor}, {Riffel},
  {Riffel}, {Rix}, {Robin}, {Roman-Lopes}, {Rose}, {Ross}, {Rossi}, {Rowlands},
  {Rubin}, {S{\'a}nchez}, {S{\'a}nchez-Gallego}, {Sayres}, {Schaefer},
  {Schiavon}, {Schimoia}, {Schlafly}, {Schlegel}, {Schneider}, {Schultheis},
  {Seo}, {Shamsi}, {Shao}, {Shen}, {Shetty}, {Simonian}, {Smethurst}, {Sobeck},
  {Souter}, {Spindler}, {Stark}, {Stassun}, {Steinmetz}, {Storchi-Bergmann},
  {Stringfellow}, {Su{\'a}rez}, {Sun}, {Taghizadeh-Popp}, {Talbot}, {Tayar},
  {Thakar}, {Thomas}, {Tissera}, {Tojeiro}, {Troup}, {Unda-Sanzana},
  {Valenzuela}, {Vargas-Maga{\~n}a}, {V{\'a}zquez-Mata}, {Wake}, {Weaver},
  {Weijmans}, {Westfall}, {Wild}, {Wilson}, {Woods}, {Yan}, {Yang}, {Zamora},
  {Zasowski}, {Zhang}, {Zheng}, {Zheng}, {Zhu}, {Zinn}, \&
  {Zou}}]{2019ApJS..240...23A}
{Aguado}, D.~S., {Ahumada}, R., {Almeida}, A., {et~al.} 2019, \apjs, 240, 23,
  \dodoi{10.3847/1538-4365/aaf651}

\bibitem[{{Alam} {et~al.}(2015){Alam}, {Albareti}, {Allende Prieto}, {Anders},
  {Anderson}, {Anderton}, {Andrews}, {Armengaud}, {Aubourg}, {Bailey}, {Basu},
  {Bautista}, {Beaton}, {Beers}, {Bender}, {Berlind}, {Beutler}, {Bhardwaj},
  {Bird}, {Bizyaev}, {Blake}, {Blanton}, {Blomqvist}, {Bochanski}, {Bolton},
  {Bovy}, {Shelden Bradley}, {Brandt}, {Brauer}, {Brinkmann}, {Brown},
  {Brownstein}, {Burden}, {Burtin}, {Busca}, {Cai}, {Capozzi}, {Carnero
  Rosell}, {Carr}, {Carrera}, {Chambers}, {Chaplin}, {Chen}, {Chiappini},
  {Chojnowski}, {Chuang}, {Clerc}, {Comparat}, {Covey}, {Croft}, {Cuesta},
  {Cunha}, {da Costa}, {Da Rio}, {Davenport}, {Dawson}, {De Lee}, {Delubac},
  {Deshpande}, {Dhital}, {Dutra-Ferreira}, {Dwelly}, {Ealet}, {Ebelke},
  {Edmondson}, {Eisenstein}, {Ellsworth}, {Elsworth}, {Epstein}, {Eracleous},
  {Escoffier}, {Esposito}, {Evans}, {Fan}, {Fern{\'a}ndez-Alvar}, {Feuillet},
  {Filiz Ak}, {Finley}, {Finoguenov}, {Flaherty}, {Fleming}, {Font-Ribera},
  {Foster}, {Frinchaboy}, {Galbraith-Frew}, {Garc{\'\i}a},
  {Garc{\'\i}a-Hern{\'a}ndez}, {Garc{\'\i}a P{\'e}rez}, {Gaulme}, {Ge},
  {G{\'e}nova-Santos}, {Georgakakis}, {Ghezzi}, {Gillespie}, {Girardi},
  {Goddard}, {Gontcho}, {Gonz{\'a}lez Hern{\'a}ndez}, {Grebel}, {Green},
  {Grieb}, {Grieves}, {Gunn}, {Guo}, {Harding}, {Hasselquist}, {Hawley},
  {Hayden}, {Hearty}, {Hekker}, {Ho}, {Hogg}, {Holley-Bockelmann}, {Holtzman},
  {Honscheid}, {Huber}, {Huehnerhoff}, {Ivans}, {Jiang}, {Johnson},
  {Kinemuchi}, {Kirkby}, {Kitaura}, {Klaene}, {Knapp}, {Kneib}, {Koenig},
  {Lam}, {Lan}, {Lang}, {Laurent}, {Le Goff}, {Leauthaud}, {Lee}, {Lee},
  {Licquia}, {Liu}, {Long}, {L{\'o}pez-Corredoira}, {Lorenzo-Oliveira},
  {Lucatello}, {Lundgren}, {Lupton}, {Mack}, {Mahadevan}, {Maia}, {Majewski},
  {Malanushenko}, {Malanushenko}, {Manchado}, {Manera}, {Mao}, {Maraston},
  {Marchwinski}, {Margala}, {Martell}, {Martig}, {Masters}, {Mathur},
  {McBride}, {McGehee}, {McGreer}, {McMahon}, {M{\'e}nard}, {Menzel},
  {Merloni}, {M{\'e}sz{\'a}ros}, {Miller}, {Miralda-Escud{\'e}}, {Miyatake},
  {Montero-Dorta}, {More}, {Morganson}, {Morice-Atkinson}, {Morrison},
  {Mosser}, {Muna}, {Myers}, {Nand ra}, {Newman}, {Neyrinck}, {Nguyen},
  {Nichol}, {Nidever}, {Noterdaeme}, {Nuza}, {O'Connell}, {O'Connell},
  {O'Connell}, {Ogando}, {Olmstead}, {Oravetz}, {Oravetz}, {Osumi}, {Owen},
  {Padgett}, {Padmanabhan}, {Paegert}, {Palanque-Delabrouille}, {Pan},
  {Parejko}, {P{\^a}ris}, {Park}, {Pattarakijwanich}, {Pellejero-Ibanez},
  {Pepper}, {Percival}, {P{\'e}rez-Fournon}, {Ṕrez-Ra`fols}, {Petitjean},
  {Pieri}, {Pinsonneault}, {Porto de Mello}, {Prada}, {Prakash},
  {Price-Whelan}, {Protopapas}, {Raddick}, {Rahman}, {Reid}, {Rich}, {Rix},
  {Robin}, {Rockosi}, {Rodrigues}, {Rodr{\'\i}guez-Torres}, {Roe}, {Ross},
  {Ross}, {Rossi}, {Ruan}, {Rubi{\~n}o-Mart{\'\i}n}, {Rykoff},
  {Salazar-Albornoz}, {Salvato}, {Samushia}, {S{\'a}nchez}, {Santiago},
  {Sayres}, {Schiavon}, {Schlegel}, {Schmidt}, {Schneider}, {Schultheis},
  {Schwope}, {Sc{\'o}ccola}, {Scott}, {Sellgren}, {Seo}, {Serenelli}, {Shane},
  {Shen}, {Shetrone}, {Shu}, {Silva Aguirre}, {Sivarani}, {Skrutskie},
  {Slosar}, {Smith}, {Sobreira}, {Souto}, {Stassun}, {Steinmetz}, {Stello},
  {Strauss}, {Streblyanska}, {Suzuki}, {Swanson}, {Tan}, {Tayar}, {Terrien},
  {Thakar}, {Thomas}, {Thomas}, {Thompson}, {Tinker}, {Tojeiro}, {Troup},
  {Vargas-Maga{\~n}a}, {Vazquez}, {Verde}, {Viel}, {Vogt}, {Wake}, {Wang},
  {Weaver}, {Weinberg}, {Weiner}, {White}, {Wilson}, {Wisniewski},
  {Wood-Vasey}, {Ye`che}, {York}, {Zakamska}, {Zamora}, {Zasowski}, {Zehavi},
  {Zhao}, {Zheng}, {Zhou}, {Zhou}, {Zou}, \& {Zhu}}]{2015ApJS..219...12A}
{Alam}, S., {Albareti}, F.~D., {Allende Prieto}, C., {et~al.} 2015, \apjs, 219,
  12, \dodoi{10.1088/0067-0049/219/1/12}

\bibitem[{Altman(1992)}]{doi:10.108000031305.1992.10475879}
Altman, N.~S. 1992, The American Statistician, 46, 175,
  \dodoi{10.1080/00031305.1992.10475879}

\bibitem[{Amon {et~al.}(2018)Amon, Blake, Heymans, Leonard, Asgari, Bilicki,
  Choi, Erben, Glazebrook, Harnois-Déraps, Hildebrandt, Hoekstra, Joachimi,
  Joudaki, Kuijken, Lidman, Loveday, Parkinson, Valentijn, \&
  Wolf}]{10.1093/mnras/sty1624}
Amon, A., Blake, C., Heymans, C., {et~al.} 2018, Monthly Notices of the Royal
  Astronomical Society, 479, 3422, \dodoi{10.1093/mnras/sty1624}

\bibitem[{Ball {et~al.}(2008)Ball, Brunner, Myers, Strand, Alberts, \&
  Tcheng}]{ball2008robust}
Ball, N.~M., Brunner, R.~J., Myers, A.~D., {et~al.} 2008, The Astrophysical
  Journal, 683, 12

\bibitem[{Banerji {et~al.}(2008)Banerji, Abdalla, Lahav, \&
  Lin}]{10.1111/j.1365-2966.2008.13095.x}
Banerji, M., Abdalla, F.~B., Lahav, O., \& Lin, H. 2008, Monthly Notices of the
  Royal Astronomical Society, 386, 1219,
  \dodoi{10.1111/j.1365-2966.2008.13095.x}

\bibitem[{Barron(2019)}]{Barron_2019_CVPR}
Barron, J.~T. 2019, in Proceedings of the IEEE/CVF Conference on Computer
  Vision and Pattern Recognition (CVPR)

\bibitem[{{Beck} {et~al.}(2017){Beck}, {Lin}, {Ishida}, {Gieseke}, {de Souza},
  {Costa-Duarte}, {Hattab}, \& {Krone-Martins}}]{2017MNRAS.468.4323B}
{Beck}, R., {Lin}, C.~A., {Ishida}, E.~E.~O., {et~al.} 2017, \mnras, 468, 4323,
  \dodoi{10.1093/mnras/stx687}

\bibitem[{{Beck} {et~al.}(2021){Beck}, {Szapudi}, {Flewelling}, {Holmberg},
  {Magnier}, \& {Chambers}}]{2021MNRAS.500.1633B}
{Beck}, R., {Szapudi}, I., {Flewelling}, H., {et~al.} 2021, \mnras, 500, 1633,
  \dodoi{10.1093/mnras/staa2587}

\bibitem[{Bilicki {et~al.}(2018)Bilicki, Hoekstra, Brown, Amaro, Blake,
  Cavuoti, De~Jong, Georgiou, Hildebrandt, Wolf,
  {et~al.}}]{bilicki2018photometric}
Bilicki, M., Hoekstra, H., Brown, M., {et~al.} 2018, Astronomy \& Astrophysics,
  616, A69

\bibitem[{Blake \& Bridle(2005)}]{10.1111/j.1365-2966.2005.09526.x}
Blake, C., \& Bridle, S. 2005, Monthly Notices of the Royal Astronomical
  Society, 363, 1329, \dodoi{10.1111/j.1365-2966.2005.09526.x}

\bibitem[{{Bolzonella} {et~al.}(2000){Bolzonella}, {Miralles}, \&
  {Pell{\'o}}}]{2000AA...363..476B}
{Bolzonella}, M., {Miralles}, J.~M., \& {Pell{\'o}}, R. 2000, \aap, 363, 476.
\newblock \doarXiv{astro-ph/0003380}

\bibitem[{Bottou(2010)}]{pub.1017229575}
Bottou, L. 2010, Large-Scale Machine Learning with Stochastic Gradient Descent,
  177--186, \dodoi{10.1007/978-3-7908-2604-3_16}

\bibitem[{Breiman(2001)}]{breiman2001random}
Breiman, L. 2001, Machine Learning, 45, 5, \dodoi{10.1023/A:1010933404324}

\bibitem[{Brescia {et~al.}(2013)Brescia, Cavuoti, D'Abrusco, Longo, \&
  Mercurio}]{Brescia_2013}
Brescia, M., Cavuoti, S., D'Abrusco, R., Longo, G., \& Mercurio, A. 2013, The
  Astrophysical Journal, 772, 140, \dodoi{10.1088/0004-637x/772/2/140}

\bibitem[{Brown {et~al.}(2020)Brown, Mann, Ryder, Subbiah, Kaplan, Dhariwal,
  Neelakantan, Shyam, Sastry, Askell, Agarwal, Herbert-Voss, Krueger, Henighan,
  Child, Ramesh, Ziegler, Wu, Winter, \& Amodei}]{gpt-3}
Brown, T., Mann, B., Ryder, N., {et~al.} 2020, Language Models are Few-Shot
  Learners

\bibitem[{{Cavuoti} {et~al.}(2017){Cavuoti}, {Amaro}, {Brescia}, {Vellucci},
  {Tortora}, \& {Longo}}]{2017MNRAS.465.1959C}
{Cavuoti}, S., {Amaro}, V., {Brescia}, M., {et~al.} 2017, \mnras, 465, 1959,
  \dodoi{10.1093/mnras/stw2930}

\bibitem[{{Chambers} {et~al.}(2016){Chambers}, {Magnier}, {Metcalfe},
  {Flewelling}, {Huber}, {Waters}, {Denneau}, {Draper}, {Farrow}, {Finkbeiner},
  {Holmberg}, {Koppenhoefer}, {Price}, {Rest}, {Saglia}, {Schlafly}, {Smartt},
  {Sweeney}, {Wainscoat}, {Burgett}, {Chastel}, {Grav}, {Heasley}, {Hodapp},
  {Jedicke}, {Kaiser}, {Kudritzki}, {Luppino}, {Lupton}, {Monet}, {Morgan},
  {Onaka}, {Shiao}, {Stubbs}, {Tonry}, {White}, {Ba{\~n}ados}, {Bell},
  {Bender}, {Bernard}, {Boegner}, {Boffi}, {Botticella}, {Calamida},
  {Casertano}, {Chen}, {Chen}, {Cole}, {Deacon}, {Frenk}, {Fitzsimmons},
  {Gezari}, {Gibbs}, {Goessl}, {Goggia}, {Gourgue}, {Goldman}, {Grant},
  {Grebel}, {Hambly}, {Hasinger}, {Heavens}, {Heckman}, {Henderson}, {Henning},
  {Holman}, {Hopp}, {Ip}, {Isani}, {Jackson}, {Keyes}, {Koekemoer}, {Kotak},
  {Le}, {Liska}, {Long}, {Lucey}, {Liu}, {Martin}, {Masci}, {McLean}, {Mindel},
  {Misra}, {Morganson}, {Murphy}, {Obaika}, {Narayan}, {Nieto-Santisteban},
  {Norberg}, {Peacock}, {Pier}, {Postman}, {Primak}, {Rae}, {Rai}, {Riess},
  {Riffeser}, {Rix}, {R{\"o}ser}, {Russel}, {Rutz}, {Schilbach}, {Schultz},
  {Scolnic}, {Strolger}, {Szalay}, {Seitz}, {Small}, {Smith}, {Soderblom},
  {Taylor}, {Thomson}, {Taylor}, {Thakar}, {Thiel}, {Thilker}, {Unger},
  {Urata}, {Valenti}, {Wagner}, {Walder}, {Walter}, {Watters}, {Werner},
  {Wood-Vasey}, \& {Wyse}}]{2016arXiv161205560C}
{Chambers}, K.~C., {Magnier}, E.~A., {Metcalfe}, N., {et~al.} 2016, arXiv
  e-prints, arXiv:1612.05560.
\newblock \doarXiv{1612.05560}

\bibitem[{{Childress} {et~al.}(2017){Childress}, {Lidman}, {Davis}, {Tucker},
  {Asorey}, {Yuan}, {Abbott}, {Abdalla}, {Allam}, {Annis}, {Banerji},
  {Benoit-L{\'e}vy}, {Bernard}, {Bertin}, {Brooks}, {Buckley-Geer}, {Burke},
  {Carnero Rosell}, {Carollo}, {Carrasco Kind}, {Carretero}, {Castander},
  {Cunha}, {da Costa}, {D'Andrea}, {Doel}, {Eifler}, {Evrard}, {Flaugher},
  {Foley}, {Fosalba}, {Frieman}, {Garc{\'\i}a-Bellido}, {Glazebrook},
  {Goldstein}, {Gruen}, {Gruendl}, {Gschwend}, {Gupta}, {Gutierrez}, {Hinton},
  {Hoormann}, {James}, {Kessler}, {Kim}, {King}, {Kovacs}, {Kuehn}, {Kuhlmann},
  {Kuropatkin}, {Lagattuta}, {Lewis}, {Li}, {Lima}, {Lin}, {Macaulay}, {Maia},
  {Marriner}, {March}, {Marshall}, {Martini}, {McMahon}, {Menanteau}, {Miquel},
  {Moller}, {Morganson}, {Mould}, {Mudd}, {Muthukrishna}, {Nichol}, {Nord},
  {Ogando}, {Ostrovski}, {Parkinson}, {Plazas}, {Reed}, {Reil}, {Romer},
  {Rykoff}, {Sako}, {Sanchez}, {Scarpine}, {Schindler}, {Schubnell}, {Scolnic},
  {Sevilla-Noarbe}, {Seymour}, {Sharp}, {Smith}, {Soares-Santos}, {Sobreira},
  {Sommer}, {Spinka}, {Suchyta}, {Sullivan}, {Swanson}, {Tarle}, {Uddin},
  {Walker}, {Wester}, \& {Zhang}}]{2017MNRAS.472..273C}
{Childress}, M.~J., {Lidman}, C., {Davis}, T.~M., {et~al.} 2017, \mnras, 472,
  273, \dodoi{10.1093/mnras/stx1872}

\bibitem[{{Chong} \& {Yang}(2019)}]{2019EPJWC.20609006C}
{Chong}, De~Wei, K., \& {Yang}, A. 2019, in European Physical Journal Web of
  Conferences, Vol. 206, European Physical Journal Web of Conferences, 09006,
  \dodoi{10.1051/epjconf/201920609006}

\bibitem[{{Clarke} {et~al.}(2020){Clarke}, {Scaife}, {Greenhalgh}, \&
  {Griguta}}]{2020A&A...639A..84C}
{Clarke}, A.~O., {Scaife}, A.~M.~M., {Greenhalgh}, R., \& {Griguta}, V. 2020,
  \aap, 639, A84, \dodoi{10.1051/0004-6361/201936770}

\bibitem[{{Colless} {et~al.}(2001){Colless}, {Dalton}, {Maddox}, {Sutherland},
  {Norberg}, {Cole}, {Bland-Hawthorn}, {Bridges}, {Cannon}, {Collins}, {Couch},
  {Cross}, {Deeley}, {De Propris}, {Driver}, {Efstathiou}, {Ellis}, {Frenk},
  {Glazebrook}, {Jackson}, {Lahav}, {Lewis}, {Lumsden}, {Madgwick}, {Peacock},
  {Peterson}, {Price}, {Seaborne}, \& {Taylor}}]{2001MNRAS.328.1039C}
{Colless}, M., {Dalton}, G., {Maddox}, S., {et~al.} 2001, \mnras, 328, 1039,
  \dodoi{10.1046/j.1365-8711.2001.04902.x}

\bibitem[{{Cool} {et~al.}(2013){Cool}, {Moustakas}, {Blanton}, {Burles},
  {Coil}, {Eisenstein}, {Wong}, {Zhu}, {Aird}, {Bernstein}, {Bolton}, {Hogg},
  \& {Mendez}}]{2013ApJ...767..118C}
{Cool}, R.~J., {Moustakas}, J., {Blanton}, M.~R., {et~al.} 2013, \apj, 767,
  118, \dodoi{10.1088/0004-637X/767/2/118}

\bibitem[{{Csabai} {et~al.}(2007){Csabai}, {Dobos}, {Trencs{\'e}ni},
  {Herczegh}, {J{\'o}zsa}, {Purger}, {Budav{\'a}ri}, \&
  {Szalay}}]{2007AN....328..852C}
{Csabai}, I., {Dobos}, L., {Trencs{\'e}ni}, M., {et~al.} 2007, Astronomische
  Nachrichten, 328, 852, \dodoi{10.1002/asna.200710817}

\bibitem[{{Cui} {et~al.}(2012){Cui}, {Zhao}, {Chu}, {Li}, {Li}, {Zhang}, {Su},
  {Yao}, {Wang}, {Xing}, {Li}, {Zhu}, {Wang}, {Gu}, {Luo}, {Xu}, {Zhang},
  {Liu}, {Zhang}, {Yang}, {Cao}, {Chen}, {Chen}, {Chen}, {Chen}, {Chu}, {Feng},
  {Gong}, {Hou}, {Hu}, {Hu}, {Hu}, {Jia}, {Jiang}, {Jiang}, {Jiang}, {Jin},
  {Li}, {Li}, {Li}, {Liu}, {Liu}, {Lu}, {Mao}, {Men}, {Qi}, {Qi}, {Shi},
  {Tang}, {Tao}, {Wang}, {Wang}, {Wang}, {Wang}, {Wang}, {Wang}, {Wang},
  {Wang}, {Wang}, {Wang}, {Wang}, {Wang}, {Xu}, {Xu}, {Yang}, {Yu}, {Yuan},
  {Yuan}, {Zhai}, {Zhang}, {Zhang}, {Zhang}, {Zhao}, {Zhou}, {Zhou}, {Zhu}, \&
  {Zou}}]{2012RAA....12.1197C}
{Cui}, X.-Q., {Zhao}, Y.-H., {Chu}, Y.-Q., {et~al.} 2012, Research in Astronomy
  and Astrophysics, 12, 1197, \dodoi{10.1088/1674-4527/12/9/003}

\bibitem[{{D'Isanto} {et~al.}(2018){D'Isanto}, {Cavuoti}, {Gieseke}, \&
  {Polsterer}}]{2018A&A...616A..97D}
{D'Isanto}, A., {Cavuoti}, S., {Gieseke}, F., \& {Polsterer}, K.~L. 2018, \aap,
  616, A97, \dodoi{10.1051/0004-6361/201833103}

\bibitem[{Dosovitskiy {et~al.}(2017)Dosovitskiy, Ros, Codevilla, Lopez, \&
  Koltun}]{pmlr-v78-dosovitskiy17a}
Dosovitskiy, A., Ros, G., Codevilla, F., Lopez, A., \& Koltun, V. 2017, in
  Proceedings of Machine Learning Research, Vol.~78, Proceedings of the 1st
  Annual Conference on Robot Learning, ed. S.~Levine, V.~Vanhoucke, \&
  K.~Goldberg (PMLR), 1--16.
\newblock \url{http://proceedings.mlr.press/v78/dosovitskiy17a.html}

\bibitem[{{Euclid Collaboration} {et~al.}(2019){Euclid Collaboration}, {Adam},
  {Vannier}, {Maurogordato}, {Biviano}, {Adami}, {Ascaso}, {Bellagamba},
  {Benoist}, {Cappi}, {D{\'\i}az-S{\'a}nchez}, {Durret}, {Farrens}, {Gonzalez},
  {Iovino}, {Licitra}, {Maturi}, {Mei}, {Merson}, {Munari}, {Pell{\'o}},
  {Ricci}, {Rocci}, {Roncarelli}, {Sarron}, {Amoura}, {Andreon}, {Apostolakos},
  {Arnaud}, {Bardelli}, {Bartlett}, {Baugh}, {Borgani}, {Brodwin}, {Castander},
  {Castignani}, {Cucciati}, {De Lucia}, {Dubath}, {Fosalba}, {Giocoli},
  {Hoekstra}, {Mamon}, {Melin}, {Moscardini}, {Paltani}, {Radovich},
  {Sartoris}, {Schultheis}, {Sereno}, {Weller}, {Burigana}, {Carvalho},
  {Corcione}, {Kurki-Suonio}, {Lilje}, {Sirri}, {Toledo-Moreo}, \&
  {Zamorani}}]{2019A&A...627A..23E}
{Euclid Collaboration}, {Adam}, R., {Vannier}, M., {et~al.} 2019, \aap, 627,
  A23, \dodoi{10.1051/0004-6361/201935088}

\bibitem[{Firth {et~al.}(2003)Firth, Lahav, \&
  Somerville}]{10.1046/j.1365-8711.2003.06271.x}
Firth, A.~E., Lahav, O., \& Somerville, R.~S. 2003, Monthly Notices of the
  Royal Astronomical Society, 339, 1195,
  \dodoi{10.1046/j.1365-8711.2003.06271.x}

\bibitem[{{Flewelling} {et~al.}(2020){Flewelling}, {Magnier}, {Chambers},
  {Heasley}, {Holmberg}, {Huber}, {Sweeney}, {Waters}, {Calamida}, {Casertano},
  {Chen}, {Farrow}, {Hasinger}, {Henderson}, {Long}, {Metcalfe}, {Narayan},
  {Nieto-Santisteban}, {Norberg}, {Rest}, {Saglia}, {Szalay}, {Thakar},
  {Tonry}, {Valenti}, {Werner}, {White}, {Denneau}, {Draper}, {Hodapp},
  {Jedicke}, {Kaiser}, {Kudritzki}, {Price}, {Wainscoat}, {Chastel}, {McLean},
  {Postman}, \& {Shiao}}]{2020ApJS..251....7F}
{Flewelling}, H.~A., {Magnier}, E.~A., {Chambers}, K.~C., {et~al.} 2020, \apjs,
  251, 7, \dodoi{10.3847/1538-4365/abb82d}

\bibitem[{{Fotopoulou} \& {Paltani}(2018)}]{2018A&A...619A..14F}
{Fotopoulou}, S., \& {Paltani}, S. 2018, \aap, 619, A14,
  \dodoi{10.1051/0004-6361/201730763}

\bibitem[{Francisco~Massa \& Aubry(2016)}]{BMVC2016_91}
Francisco~Massa, R.~M., \& Aubry, M. 2016, in Proceedings of the British
  Machine Vision Conference (BMVC), ed. E.~R.~H. Richard C.~Wilson \& W.~A.~P.
  Smith (BMVA Press), 91.1--91.12, \dodoi{10.5244/C.30.91}

\bibitem[{{Galametz} {et~al.}(2017){Galametz}, {Saglia}, {Paltani},
  {Apostolakos}, \& {Dubath}}]{2017A&A...598A..20G}
{Galametz}, A., {Saglia}, R., {Paltani}, S., {Apostolakos}, N., \& {Dubath}, P.
  2017, \aap, 598, A20, \dodoi{10.1051/0004-6361/201629333}

\bibitem[{{Golob} {et~al.}(2021){Golob}, {Sawicki}, {Goulding}, \&
  {Coupon}}]{2021MNRAS.503.4136G}
{Golob}, A., {Sawicki}, M., {Goulding}, A.~D., \& {Coupon}, J. 2021, \mnras,
  503, 4136, \dodoi{10.1093/mnras/stab719}

\bibitem[{Guo {et~al.}(2017)Guo, Pleiss, Sun, \& Weinberger}]{pmlr-v70-guo17a}
Guo, C., Pleiss, G., Sun, Y., \& Weinberger, K.~Q. 2017, in Proceedings of
  Machine Learning Research, Vol.~70, Proceedings of the 34th International
  Conference on Machine Learning, ed. D.~Precup \& Y.~W. Teh (PMLR),
  1321--1330.
\newblock \url{http://proceedings.mlr.press/v70/guo17a.html}

\bibitem[{{Hasinger} {et~al.}(2018){Hasinger}, {Capak}, {Salvato}, {Barger},
  {Cowie}, {Faisst}, {Hemmati}, {Kakazu}, {Kartaltepe}, {Masters}, {Mobasher},
  {Nayyeri}, {Sanders}, {Scoville}, {Suh}, {Steinhardt}, \&
  {Yang}}]{2018ApJ...858...77H}
{Hasinger}, G., {Capak}, P., {Salvato}, M., {et~al.} 2018, \apj, 858, 77,
  \dodoi{10.3847/1538-4357/aabacf}

\bibitem[{Hendrycks \& Gimpel(2017)}]{DBLP:conf/iclr/HendrycksG17}
Hendrycks, D., \& Gimpel, K. 2017, in 5th International Conference on Learning
  Representations, {ICLR} 2017, Toulon, France, April 24-26, 2017, Conference
  Track Proceedings (OpenReview.net).
\newblock \url{https://openreview.net/forum?id=Hkg4TI9xl}

\bibitem[{Hendrycks {et~al.}(2019)Hendrycks, Mazeika, \&
  Dietterich}]{DBLP:conf/iclr/HendrycksMD19}
Hendrycks, D., Mazeika, M., \& Dietterich, T.~G. 2019, in 7th International
  Conference on Learning Representations, {ICLR} 2019, New Orleans, LA, USA,
  May 6-9, 2019 (OpenReview.net).
\newblock \url{https://openreview.net/forum?id=HyxCxhRcY7}

\bibitem[{Hopfield(1982)}]{Hopfield2554}
Hopfield, J.~J. 1982, Proceedings of the National Academy of Sciences, 79,
  2554, \dodoi{10.1073/pnas.79.8.2554}

\bibitem[{Ioffe \& Szegedy(2015)}]{pmlr-v37-ioffe15}
Ioffe, S., \& Szegedy, C. 2015, in Proceedings of Machine Learning Research,
  Vol.~37, Proceedings of the 32nd International Conference on Machine
  Learning, ed. F.~Bach \& D.~Blei (Lille, France: PMLR), 448--456.
\newblock \url{http://proceedings.mlr.press/v37/ioffe15.html}

\bibitem[{{Ivezi{\'c}} {et~al.}(2019){Ivezi{\'c}}, {Kahn}, {Tyson}, {Abel},
  {Acosta}, {Allsman}, {Alonso}, {AlSayyad}, {Anderson}, {Andrew}, {Angel},
  {Angeli}, {Ansari}, {Antilogus}, {Araujo}, {Armstrong}, {Arndt}, {Astier},
  {Aubourg}, {Auza}, {Axelrod}, {Bard}, {Barr}, {Barrau}, {Bartlett}, {Bauer},
  {Bauman}, {Baumont}, {Bechtol}, {Bechtol}, {Becker}, {Becla}, {Beldica},
  {Bellavia}, {Bianco}, {Biswas}, {Blanc}, {Blazek}, {Blandford}, {Bloom},
  {Bogart}, {Bond}, {Booth}, {Borgland}, {Borne}, {Bosch}, {Boutigny},
  {Brackett}, {Bradshaw}, {Brandt}, {Brown}, {Bullock}, {Burchat}, {Burke},
  {Cagnoli}, {Calabrese}, {Callahan}, {Callen}, {Carlin}, {Carlson},
  {Chandrasekharan}, {Charles-Emerson}, {Chesley}, {Cheu}, {Chiang}, {Chiang},
  {Chirino}, {Chow}, {Ciardi}, {Claver}, {Cohen-Tanugi}, {Cockrum}, {Coles},
  {Connolly}, {Cook}, {Cooray}, {Covey}, {Cribbs}, {Cui}, {Cutri}, {Daly},
  {Daniel}, {Daruich}, {Daubard}, {Daues}, {Dawson}, {Delgado}, {Dellapenna},
  {de Peyster}, {de Val-Borro}, {Digel}, {Doherty}, {Dubois},
  {Dubois-Felsmann}, {Durech}, {Economou}, {Eifler}, {Eracleous}, {Emmons},
  {Fausti Neto}, {Ferguson}, {Figueroa}, {Fisher-Levine}, {Focke}, {Foss},
  {Frank}, {Freemon}, {Gangler}, {Gawiser}, {Geary}, {Gee}, {Geha}, {Gessner},
  {Gibson}, {Gilmore}, {Glanzman}, {Glick}, {Goldina}, {Goldstein}, {Goodenow},
  {Graham}, {Gressler}, {Gris}, {Guy}, {Guyonnet}, {Haller}, {Harris},
  {Hascall}, {Haupt}, {Hernandez}, {Herrmann}, {Hileman}, {Hoblitt}, {Hodgson},
  {Hogan}, {Howard}, {Huang}, {Huffer}, {Ingraham}, {Innes}, {Jacoby}, {Jain},
  {Jammes}, {Jee}, {Jenness}, {Jernigan}, {Jevremovi{\'c}}, {Johns}, {Johnson},
  {Johnson}, {Jones}, {Juramy-Gilles}, {Juri{\'c}}, {Kalirai}, {Kallivayalil},
  {Kalmbach}, {Kantor}, {Karst}, {Kasliwal}, {Kelly}, {Kessler}, {Kinnison},
  {Kirkby}, {Knox}, {Kotov}, {Krabbendam}, {Krughoff}, {Kub{\'a}nek},
  {Kuczewski}, {Kulkarni}, {Ku}, {Kurita}, {Lage}, {Lambert}, {Lange},
  {Langton}, {Le Guillou}, {Levine}, {Liang}, {Lim}, {Lintott}, {Long},
  {Lopez}, {Lotz}, {Lupton}, {Lust}, {MacArthur}, {Mahabal}, {Mandelbaum},
  {Markiewicz}, {Marsh}, {Marshall}, {Marshall}, {May}, {McKercher}, {McQueen},
  {Meyers}, {Migliore}, {Miller}, {Mills}, {Miraval}, {Moeyens}, {Moolekamp},
  {Monet}, {Moniez}, {Monkewitz}, {Montgomery}, {Morrison}, {Mueller},
  {Muller}, {Mu{\~n}oz Arancibia}, {Neill}, {Newbry}, {Nief}, {Nomerotski},
  {Nordby}, {O'Connor}, {Oliver}, {Olivier}, {Olsen}, {O'Mullane}, {Ortiz},
  {Osier}, {Owen}, {Pain}, {Palecek}, {Parejko}, {Parsons}, {Pease},
  {Peterson}, {Peterson}, {Petravick}, {Libby Petrick}, {Petry},
  {Pierfederici}, {Pietrowicz}, {Pike}, {Pinto}, {Plante}, {Plate}, {Plutchak},
  {Price}, {Prouza}, {Radeka}, {Rajagopal}, {Rasmussen}, {Regnault}, {Reil},
  {Reiss}, {Reuter}, {Ridgway}, {Riot}, {Ritz}, {Robinson}, {Roby}, {Roodman},
  {Rosing}, {Roucelle}, {Rumore}, {Russo}, {Saha}, {Sassolas}, {Schalk},
  {Schellart}, {Schindler}, {Schmidt}, {Schneider}, {Schneider}, {Schoening},
  {Schumacher}, {Schwamb}, {Sebag}, {Selvy}, {Sembroski}, {Seppala}, {Serio},
  {Serrano}, {Shaw}, {Shipsey}, {Sick}, {Silvestri}, {Slater}, {Smith},
  {Smith}, {Sobhani}, {Soldahl}, {Storrie-Lombardi}, {Stover}, {Strauss},
  {Street}, {Stubbs}, {Sullivan}, {Sweeney}, {Swinbank}, {Szalay}, {Takacs},
  {Tether}, {Thaler}, {Thayer}, {Thomas}, {Thornton}, {Thukral}, {Tice},
  {Trilling}, {Turri}, {Van Berg}, {Vanden Berk}, {Vetter}, {Virieux},
  {Vucina}, {Wahl}, {Walkowicz}, {Walsh}, {Walter}, {Wang}, {Wang}, {Warner},
  {Wiecha}, {Willman}, {Winters}, {Wittman}, {Wolff}, {Wood-Vasey}, {Wu},
  {Xin}, {Yoachim}, \& {Zhan}}]{2019ApJ...873..111I}
{Ivezi{\'c}}, {\v{Z}}., {Kahn}, S.~M., {Tyson}, J.~A., {et~al.} 2019, \apj,
  873, 111, \dodoi{10.3847/1538-4357/ab042c}

\bibitem[{{Jones} {et~al.}(2009){Jones}, {Read}, {Saunders}, {Colless},
  {Jarrett}, {Parker}, {Fairall}, {Mauch}, {Sadler}, {Watson}, {Burton},
  {Campbell}, {Cass}, {Croom}, {Dawe}, {Fiegert}, {Frankcombe}, {Hartley},
  {Huchra}, {James}, {Kirby}, {Lahav}, {Lucey}, {Mamon}, {Moore}, {Peterson},
  {Prior}, {Proust}, {Russell}, {Safouris}, {Wakamatsu}, {Westra}, \&
  {Williams}}]{2009MNRAS.399..683J}
{Jones}, D.~H., {Read}, M.~A., {Saunders}, W., {et~al.} 2009, \mnras, 399, 683,
  \dodoi{10.1111/j.1365-2966.2009.15338.x}

\bibitem[{{Jones} \& {Heavens}(2019)}]{2019MNRAS.483.2487J}
{Jones}, D.~M., \& {Heavens}, A.~F. 2019, \mnras, 483, 2487,
  \dodoi{10.1093/mnras/sty3279}

\bibitem[{{Kaiser} {et~al.}(2010){Kaiser}, {Burgett}, {Chambers}, {Denneau},
  {Heasley}, {Jedicke}, {Magnier}, {Morgan}, {Onaka}, \&
  {Tonry}}]{2010SPIE.7733E..0EK}
{Kaiser}, N., {Burgett}, W., {Chambers}, K., {et~al.} 2010, Society of
  Photo-Optical Instrumentation Engineers (SPIE) Conference Series, Vol. 7733,
  {The Pan-STARRS wide-field optical/NIR imaging survey}, 77330E,
  \dodoi{10.1117/12.859188}

\bibitem[{{Kalmbach} \& {Connolly}(2017)}]{2017AJ....154..277K}
{Kalmbach}, J.~B., \& {Connolly}, A.~J. 2017, \aj, 154, 277,
  \dodoi{10.3847/1538-3881/aa9933}

\bibitem[{{Keller} {et~al.}(2007){Keller}, {Schmidt}, {Bessell}, {Conroy},
  {Francis}, {Granlund}, {Kowald}, {Oates}, {Martin-Jones}, {Preston},
  {Tisserand}, {Vaccarella}, \& {Waterson}}]{2007PASA...24....1K}
{Keller}, S.~C., {Schmidt}, B.~P., {Bessell}, M.~S., {et~al.} 2007, \pasa, 24,
  1, \dodoi{10.1071/AS07001}

\bibitem[{{Kim} {et~al.}(2015){Kim}, {Brunner}, \& {Carrasco
  Kind}}]{2015MNRAS.453..507K}
{Kim}, E.~J., {Brunner}, R.~J., \& {Carrasco Kind}, M. 2015, \mnras, 453, 507,
  \dodoi{10.1093/mnras/stv1608}

\bibitem[{{Korytov} {et~al.}(2019){Korytov}, {Hearin}, {Kovacs}, {Larsen},
  {Rangel}, {Hollowed}, {Benson}, {Heitmann}, {Mao}, {Bahmanyar}, {Chang},
  {Campbell}, {DeRose}, {Finkel}, {Frontiere}, {Gawiser}, {Habib}, {Joachimi},
  {Lanusse}, {Li}, {Mandelbaum}, {Morrison}, {Newman}, {Pope}, {Rykoff},
  {Simet}, {To}, {Vikraman}, {Wechsler}, {White}, \& {(The LSST Dark Energy
  Science Collaboration}}]{2019ApJS..245...26K}
{Korytov}, D., {Hearin}, A., {Kovacs}, E., {et~al.} 2019, \apjs, 245, 26,
  \dodoi{10.3847/1538-4365/ab510c}

\bibitem[{Krizhevsky(2012)}]{cifar}
Krizhevsky, A. 2012, University of Toronto

\bibitem[{Krizhevsky {et~al.}(2012)Krizhevsky, Sutskever, \&
  Hinton}]{NIPS2012_c399862d}
Krizhevsky, A., Sutskever, I., \& Hinton, G.~E. 2012, in Advances in Neural
  Information Processing Systems, ed. F.~Pereira, C.~J.~C. Burges, L.~Bottou,
  \& K.~Q. Weinberger, Vol.~25 (Curran Associates, Inc.), 1097--1105.
\newblock
  \url{https://proceedings.neurips.cc/paper/2012/file/c399862d3b9d6b76c8436e924a68c45b-Paper.pdf}

\bibitem[{Kundu {et~al.}(2018)Kundu, Li, \& Rehg}]{3DRCNN_CVPR18}
Kundu, A., Li, Y., \& Rehg, J.~M. 2018, in CVPR

\bibitem[{Laigle {et~al.}(2017)Laigle, Pichon, Arnouts, McCracken, Dubois,
  Devriendt, Slyz, Le~Borgne, Benoit-Lévy, Hwang, Ilbert, Kraljic, Malavasi,
  Park, \& Vibert}]{10.1093/mnras/stx3055}
Laigle, C., Pichon, C., Arnouts, S., {et~al.} 2017, Monthly Notices of the
  Royal Astronomical Society, 474, 5437, \dodoi{10.1093/mnras/stx3055}

\bibitem[{{Le F{\`e}vre} {et~al.}(2013){Le F{\`e}vre}, {Cassata}, {Cucciati},
  {Garilli}, {Ilbert}, {Le Brun}, {Maccagni}, {Moreau}, {Scodeggio}, {Tresse},
  {Zamorani}, {Adami}, {Arnouts}, {Bardelli}, {Bolzonella}, {Bondi},
  {Bongiorno}, {Bottini}, {Cappi}, {Charlot}, {Ciliegi}, {Contini}, {de la
  Torre}, {Foucaud}, {Franzetti}, {Gavignaud}, {Guzzo}, {Iovino}, {Lemaux},
  {L{\'o}pez-Sanjuan}, {McCracken}, {Marano}, {Marinoni}, {Mazure}, {Mellier},
  {Merighi}, {Merluzzi}, {Paltani}, {Pell{\`o}}, {Pollo}, {Pozzetti},
  {Scaramella}, {Tasca}, {Vergani}, {Vettolani}, {Zanichelli}, \&
  {Zucca}}]{2013AA...559A..14L}
{Le F{\`e}vre}, O., {Cassata}, P., {Cucciati}, O., {et~al.} 2013, \aap, 559,
  A14, \dodoi{10.1051/0004-6361/201322179}

\bibitem[{LeCun \& Cortes(2010)}]{lecun-mnisthandwrittendigit-2010}
LeCun, Y., \& Cortes, C. 2010.
\newblock \url{http://yann.lecun.com/exdb/mnist/}

\bibitem[{Lee \& Shin(2021)}]{lee_and_shin_2021_5529452}
Lee, J., \& Shin, M.-S. 2021, MBRNN: Multiple bin regression with neural
  network for photometric redshift estimation, 1.0,  Zenodo,
  \dodoi{10.5281/zenodo.5529452}

\bibitem[{Lee {et~al.}(2018)Lee, Lee, Lee, \& Shin}]{10.5555/3327757.3327819}
Lee, K., Lee, K., Lee, H., \& Shin, J. 2018, in Proceedings of the 32nd
  International Conference on Neural Information Processing Systems, NIPS'18
  (Red Hook, NY, USA: Curran Associates Inc.), 7167–7177

\bibitem[{Levinson {et~al.}(2011)Levinson, Askeland, Becker, Dolson, Held,
  Kammel, Kolter, Langer, Pink, Pratt, {et~al.}}]{levinson2011towards}
Levinson, J., Askeland, J., Becker, J., {et~al.} 2011, in 2011 IEEE Intelligent
  Vehicles Symposium (IV), IEEE, 163--168

\bibitem[{Liang {et~al.}(2018)Liang, Li, \& Srikant}]{liang2017enhancing}
Liang, S., Li, Y., \& Srikant, R. 2018, International Conference on Learning
  Representations (ICLR)

\bibitem[{{Lilly} {et~al.}(2007){Lilly}, {Le F{\`e}vre}, {Renzini}, {Zamorani},
  {Scodeggio}, {Contini}, {Carollo}, {Hasinger}, {Kneib}, {Iovino}, {Le Brun},
  {Maier}, {Mainieri}, {Mignoli}, {Silverman}, {Tasca}, {Bolzonella},
  {Bongiorno}, {Bottini}, {Capak}, {Caputi}, {Cimatti}, {Cucciati}, {Daddi},
  {Feldmann}, {Franzetti}, {Garilli}, {Guzzo}, {Ilbert}, {Kampczyk}, {Kovac},
  {Lamareille}, {Leauthaud}, {Le Borgne}, {McCracken}, {Marinoni}, {Pello},
  {Ricciardelli}, {Scarlata}, {Vergani}, {Sanders}, {Schinnerer}, {Scoville},
  {Taniguchi}, {Arnouts}, {Aussel}, {Bardelli}, {Brusa}, {Cappi}, {Ciliegi},
  {Finoguenov}, {Foucaud}, {Franceschini}, {Halliday}, {Impey}, {Knobel},
  {Koekemoer}, {Kurk}, {Maccagni}, {Maddox}, {Marano}, {Marconi}, {Meneux},
  {Mobasher}, {Moreau}, {Peacock}, {Porciani}, {Pozzetti}, {Scaramella},
  {Schiminovich}, {Shopbell}, {Smail}, {Thompson}, {Tresse}, {Vettolani},
  {Zanichelli}, \& {Zucca}}]{2007ApJS..172...70L}
{Lilly}, S.~J., {Le F{\`e}vre}, O., {Renzini}, A., {et~al.} 2007, \apjs, 172,
  70, \dodoi{10.1086/516589}

\bibitem[{{Lilly} {et~al.}(2009){Lilly}, {Le Brun}, {Maier}, {Mainieri},
  {Mignoli}, {Scodeggio}, {Zamorani}, {Carollo}, {Contini}, {Kneib}, {Le
  F{\`e}vre}, {Renzini}, {Bardelli}, {Bolzonella}, {Bongiorno}, {Caputi},
  {Coppa}, {Cucciati}, {de la Torre}, {de Ravel}, {Franzetti}, {Garilli},
  {Iovino}, {Kampczyk}, {Kovac}, {Knobel}, {Lamareille}, {Le Borgne}, {Pello},
  {Peng}, {P{\'e}rez-Montero}, {Ricciardelli}, {Silverman}, {Tanaka}, {Tasca},
  {Tresse}, {Vergani}, {Zucca}, {Ilbert}, {Salvato}, {Oesch}, {Abbas},
  {Bottini}, {Capak}, {Cappi}, {Cassata}, {Cimatti}, {Elvis}, {Fumana},
  {Guzzo}, {Hasinger}, {Koekemoer}, {Leauthaud}, {Maccagni}, {Marinoni},
  {McCracken}, {Memeo}, {Meneux}, {Porciani}, {Pozzetti}, {Sanders},
  {Scaramella}, {Scarlata}, {Scoville}, {Shopbell}, \&
  {Taniguchi}}]{2009ApJS..184..218L}
{Lilly}, S.~J., {Le Brun}, V., {Maier}, C., {et~al.} 2009, \apjs, 184, 218,
  \dodoi{10.1088/0067-0049/184/2/218}

\bibitem[{Liu(2019)}]{L2Reg}
Liu, P.~L. 2019, Multimodal Regression — Beyond L1 and L2 Loss.
\newblock
  \url{https://towardsdatascience.com/anchors-and-multi-bin-loss-for-multi-modal-target-regression-647ea1974617}

\bibitem[{{LSST Dark Energy Science Collaboration (LSST DESC)}
  {et~al.}(2021){LSST Dark Energy Science Collaboration (LSST DESC)},
  {Abolfathi}, {Alonso}, {Armstrong}, {Aubourg}, {Awan}, {Babuji}, {Bauer},
  {Bean}, {Beckett}, {Biswas}, {Bogart}, {Boutigny}, {Chard}, {Chiang},
  {Claver}, {Cohen-Tanugi}, {Combet}, {Connolly}, {Daniel}, {Digel},
  {Drlica-Wagner}, {Dubois}, {Gangler}, {Gawiser}, {Glanzman}, {Gris}, {Habib},
  {Hearin}, {Heitmann}, {Hernandez}, {Hlo{\v{z}}ek}, {Hollowed}, {Ishak},
  {Ivezi{\'c}}, {Jarvis}, {Jha}, {Kahn}, {Kalmbach}, {Kelly}, {Kovacs},
  {Korytov}, {Krughoff}, {Lage}, {Lanusse}, {Larsen}, {Le Guillou}, {Li},
  {Longley}, {Lupton}, {Mandelbaum}, {Mao}, {Marshall}, {Meyers}, {Moniez},
  {Morrison}, {Nomerotski}, {O'Connor}, {Park}, {Park}, {Peloton}, {Perrefort},
  {Perry}, {Plaszczynski}, {Pope}, {Rasmussen}, {Reil}, {Roodman}, {Rykoff},
  {S{\'a}nchez}, {Schmidt}, {Scolnic}, {Stubbs}, {Tyson}, {Uram}, {Villarreal},
  {Walter}, {Wiesner}, {Wood-Vasey}, \& {Zuntz}}]{2021ApJS..253...31L}
{LSST Dark Energy Science Collaboration (LSST DESC)}, {Abolfathi}, B.,
  {Alonso}, D., {et~al.} 2021, \apjs, 253, 31, \dodoi{10.3847/1538-4365/abd62c}

\bibitem[{{Masters} {et~al.}(2017){Masters}, {Stern}, {Cohen}, {Capak},
  {Rhodes}, {Castander}, \& {Paltani}}]{2017ApJ...841..111M}
{Masters}, D.~C., {Stern}, D.~K., {Cohen}, J.~G., {et~al.} 2017, \apj, 841,
  111, \dodoi{10.3847/1538-4357/aa6f08}

\bibitem[{{Masters} {et~al.}(2019){Masters}, {Stern}, {Cohen}, {Capak},
  {Stanford}, {Hernitschek}, {Galametz}, {Davidzon}, {Rhodes}, {Sanders},
  {Mobasher}, {Castander}, {Pruett}, \& {Fotopoulou}}]{2019ApJ...877...81M}
---. 2019, \apj, 877, 81, \dodoi{10.3847/1538-4357/ab184d}

\bibitem[{Mousavian {et~al.}(2017)Mousavian, Anguelov, Flynn, \&
  Kosecka}]{Mousavian_2017_CVPR}
Mousavian, A., Anguelov, D., Flynn, J., \& Kosecka, J. 2017, in Proceedings of
  the IEEE Conference on Computer Vision and Pattern Recognition (CVPR)

\bibitem[{{Newman} {et~al.}(2013){Newman}, {Cooper}, {Davis}, {Faber}, {Coil},
  {Guhathakurta}, {Koo}, {Phillips}, {Conroy}, {Dutton}, {Finkbeiner}, {Gerke},
  {Rosario}, {Weiner}, {Willmer}, {Yan}, {Harker}, {Kassin}, {Konidaris},
  {Lai}, {Madgwick}, {Noeske}, {Wirth}, {Connolly}, {Kaiser}, {Kirby},
  {Lemaux}, {Lin}, {Lotz}, {Luppino}, {Marinoni}, {Matthews}, {Metevier}, \&
  {Schiavon}}]{2013ApJS..208....5N}
{Newman}, J.~A., {Cooper}, M.~C., {Davis}, M., {et~al.} 2013, \apjs, 208, 5,
  \dodoi{10.1088/0067-0049/208/1/5}

\bibitem[{{Nishizawa} {et~al.}(2020){Nishizawa}, {Hsieh}, {Tanaka}, \&
  {Takata}}]{2020arXiv200301511N}
{Nishizawa}, A.~J., {Hsieh}, B.-C., {Tanaka}, M., \& {Takata}, T. 2020, arXiv
  e-prints, arXiv:2003.01511.
\newblock \doarXiv{2003.01511}

\bibitem[{{Pasquet} {et~al.}(2019){Pasquet}, {Bertin}, {Treyer}, {Arnouts}, \&
  {Fouchez}}]{2019AA...621A..26P}
{Pasquet}, J., {Bertin}, E., {Treyer}, M., {Arnouts}, S., \& {Fouchez}, D.
  2019, \aap, 621, A26, \dodoi{10.1051/0004-6361/201833617}

\bibitem[{{Pasquet-Itam} \& {Pasquet}(2018)}]{2018AA...611A..97P}
{Pasquet-Itam}, J., \& {Pasquet}, J. 2018, \aap, 611, A97,
  \dodoi{10.1051/0004-6361/201731106}

\bibitem[{Paszke {et~al.}(2019)Paszke, Gross, Massa, Lerer, Bradbury, Chanan,
  Killeen, Lin, Gimelshein, Antiga, Desmaison, Kopf, Yang, DeVito, Raison,
  Tejani, Chilamkurthy, Steiner, Fang, Bai, \& Chintala}]{NEURIPS2019_9015}
Paszke, A., Gross, S., Massa, F., {et~al.} 2019, in Advances in Neural
  Information Processing Systems 32, ed. H.~Wallach, H.~Larochelle,
  A.~Beygelzimer, F.~d\textquotesingle Alch\'{e}-Buc, E.~Fox, \& R.~Garnett
  (Curran Associates, Inc.), 8024--8035.
\newblock
  \url{http://papers.neurips.cc/paper/9015-pytorch-an-imperative-style-high-performance-deep-learning-library.pdf}

\bibitem[{{Planck Collaboration} {et~al.}(2014){Planck Collaboration},
  {Abergel}, {Ade}, {Aghanim}, {Alves}, {Aniano}, {Armitage-Caplan}, {Arnaud},
  {Ashdown}, {Atrio-Barandela}, {Aumont}, {Baccigalupi}, {Banday}, {Barreiro},
  {Bartlett}, {Battaner}, {Benabed}, {Beno{\^\i}t}, {Benoit-L{\'e}vy},
  {Bernard}, {Bersanelli}, {Bielewicz}, {Bobin}, {Bock}, {Bonaldi}, {Bond},
  {Borrill}, {Bouchet}, {Boulanger}, {Bridges}, {Bucher}, {Burigana}, {Butler},
  {Cardoso}, {Catalano}, {Chamballu}, {Chary}, {Chiang}, {Chiang},
  {Christensen}, {Church}, {Clemens}, {Clements}, {Colombi}, {Colombo},
  {Combet}, {Couchot}, {Coulais}, {Crill}, {Curto}, {Cuttaia}, {Danese},
  {Davies}, {Davis}, {de Bernardis}, {de Rosa}, {de Zotti}, {Delabrouille},
  {Delouis}, {D{\'e}sert}, {Dickinson}, {Diego}, {Dole}, {Donzelli},
  {Dor{\'e}}, {Douspis}, {Draine}, {Dupac}, {Efstathiou}, {En{\ss}lin},
  {Eriksen}, {Falgarone}, {Finelli}, {Forni}, {Frailis}, {Fraisse},
  {Franceschi}, {Galeotta}, {Ganga}, {Ghosh}, {Giard}, {Giardino},
  {Giraud-H{\'e}raud}, {Gonz{\'a}lez-Nuevo}, {G{\'o}rski}, {Gratton},
  {Gregorio}, {Grenier}, {Gruppuso}, {Guillet}, {Hansen}, {Hanson}, {Harrison},
  {Helou}, {Henrot-Versill{\'e}}, {Hern{\'a}ndez-Monteagudo}, {Herranz},
  {Hildebrandt}, {Hivon}, {Hobson}, {Holmes}, {Hornstrup}, {Hovest},
  {Huffenberger}, {Jaffe}, {Jaffe}, {Jewell}, {Joncas}, {Jones}, {Juvela},
  {Keih{\"a}nen}, {Keskitalo}, {Kisner}, {Knoche}, {Knox}, {Kunz},
  {Kurki-Suonio}, {Lagache}, {L{\"a}hteenm{\"a}ki}, {Lamarre}, {Lasenby},
  {Laureijs}, {Lawrence}, {Leonardi}, {Le{\'o}n-Tavares}, {Lesgourgues},
  {Levrier}, {Liguori}, {Lilje}, {Linden-V{\o}rnle}, {L{\'o}pez-Caniego},
  {Lubin}, {Mac{\'\i}as-P{\'e}rez}, {Maffei}, {Maino}, {Mandolesi}, {Maris},
  {Marshall}, {Martin}, {Mart{\'\i}nez-Gonz{\'a}lez}, {Masi}, {Massardi},
  {Matarrese}, {Matthai}, {Mazzotta}, {McGehee}, {Melchiorri}, {Mendes},
  {Mennella}, {Migliaccio}, {Mitra}, {Miville-Desch{\^e}nes}, {Moneti},
  {Montier}, {Morgante}, {Mortlock}, {Munshi}, {Murphy}, {Naselsky}, {Nati},
  {Natoli}, {Netterfield}, {N{\o}rgaard-Nielsen}, {Noviello}, {Novikov},
  {Novikov}, {Osborne}, {Oxborrow}, {Paci}, {Pagano}, {Pajot}, {Paladini},
  {Paoletti}, {Pasian}, {Patanchon}, {Perdereau}, {Perotto}, {Perrotta},
  {Piacentini}, {Piat}, {Pierpaoli}, {Pietrobon}, {Plaszczynski},
  {Pointecouteau}, {Polenta}, {Ponthieu}, {Popa}, {Poutanen}, {Pratt},
  {Pr{\'e}zeau}, {Prunet}, {Puget}, {Rachen}, {Reach}, {Rebolo}, {Reinecke},
  {Remazeilles}, {Renault}, {Ricciardi}, {Riller}, {Ristorcelli}, {Rocha},
  {Rosset}, {Roudier}, {Rowan-Robinson}, {Rubi{\~n}o-Mart{\'\i}n}, {Rusholme},
  {Sandri}, {Santos}, {Savini}, {Scott}, {Seiffert}, {Shellard}, {Spencer},
  {Starck}, {Stolyarov}, {Stompor}, {Sudiwala}, {Sunyaev}, {Sureau}, {Sutton},
  {Suur-Uski}, {Sygnet}, {Tauber}, {Tavagnacco}, {Terenzi}, {Toffolatti},
  {Tomasi}, {Tristram}, {Tucci}, {Tuovinen}, {T{\"u}rler}, {Umana},
  {Valenziano}, {Valiviita}, {Van Tent}, {Verstraete}, {Vielva}, {Villa},
  {Vittorio}, {Wade}, {Wandelt}, {Welikala}, {Ysard}, {Yvon}, {Zacchei}, \&
  {Zonca}}]{2014A&A...571A..11P}
{Planck Collaboration}, {Abergel}, A., {Ade}, P.~A.~R., {et~al.} 2014, \aap,
  571, A11, \dodoi{10.1051/0004-6361/201323195}

\bibitem[{{Rafelski} {et~al.}(2015){Rafelski}, {Teplitz}, {Gardner}, {Coe},
  {Bond}, {Koekemoer}, {Grogin}, {Kurczynski}, {McGrath}, {Bourque}, {Atek},
  {Brown}, {Colbert}, {Codoreanu}, {Ferguson}, {Finkelstein}, {Gawiser},
  {Giavalisco}, {Gronwall}, {Hanish}, {Lee}, {Mehta}, {de Mello},
  {Ravindranath}, {Ryan}, {Scarlata}, {Siana}, {Soto}, \&
  {Voyer}}]{2015AJ....150...31R}
{Rafelski}, M., {Teplitz}, H.~I., {Gardner}, J.~P., {et~al.} 2015, \aj, 150,
  31, \dodoi{10.1088/0004-6256/150/1/31}

\bibitem[{Ren {et~al.}(2019)Ren, Liu, Fertig, Snoek, Poplin, Depristo, Dillon,
  \& Lakshminarayanan}]{NIPS2019_9611}
Ren, J., Liu, P.~J., Fertig, E., {et~al.} 2019, in Advances in Neural
  Information Processing Systems 32, ed. H.~Wallach, H.~Larochelle,
  A.~Beygelzimer, F.~d'Alch\'{e} Buc, E.~Fox, \& R.~Garnett (Curran Associates,
  Inc.), 14707--14718.
\newblock
  \url{http://papers.nips.cc/paper/9611-likelihood-ratios-for-out-of-distribution-detection.pdf}

\bibitem[{{Rines} {et~al.}(2013){Rines}, {Geller}, {Diaferio}, \&
  {Kurtz}}]{2013ApJ...767...15R}
{Rines}, K., {Geller}, M.~J., {Diaferio}, A., \& {Kurtz}, M.~J. 2013, \apj,
  767, 15, \dodoi{10.1088/0004-637X/767/1/15}

\bibitem[{{Rivera} {et~al.}(2018){Rivera}, {Moraes}, {Merson}, {Jouvel},
  {Abdalla}, \& {Abdalla}}]{2018MNRAS.477.4330R}
{Rivera}, J.~D., {Moraes}, B., {Merson}, A.~I., {et~al.} 2018, \mnras, 477,
  4330, \dodoi{10.1093/mnras/sty880}

\bibitem[{Rosenblatt(1958)}]{Rosenblatt58theperceptron:}
Rosenblatt, F. 1958, Psychological Review, 65

\bibitem[{Ryou {et~al.}(2019)Ryou, Jeong, \& Perona}]{9009013}
Ryou, S., Jeong, S., \& Perona, P. 2019, in 2019 IEEE/CVF International
  Conference on Computer Vision (ICCV) (Los Alamitos, CA, USA: IEEE Computer
  Society), 5991--6000, \dodoi{10.1109/ICCV.2019.00609}

\bibitem[{{Salvato} {et~al.}(2019){Salvato}, {Ilbert}, \&
  {Hoyle}}]{2019NatAs...3..212S}
{Salvato}, M., {Ilbert}, O., \& {Hoyle}, B. 2019, Nature Astronomy, 3, 212,
  \dodoi{10.1038/s41550-018-0478-0}

\bibitem[{{Sawicki} {et~al.}(1997){Sawicki}, {Lin}, \&
  {Yee}}]{1997AJ....113....1S}
{Sawicki}, M.~J., {Lin}, H., \& {Yee}, H.~K.~C. 1997, \aj, 113, 1,
  \dodoi{10.1086/118231}

\bibitem[{{Schindler} {et~al.}(2019){Schindler}, {Fan}, {Huang}, {Yue}, {Yang},
  {Hall}, {Wenzl}, {Hughes}, {Litke}, \& {Rees}}]{2019ApJS..243....5S}
{Schindler}, J.-T., {Fan}, X., {Huang}, Y.-H., {et~al.} 2019, \apjs, 243, 5,
  \dodoi{10.3847/1538-4365/ab20d0}

\bibitem[{{Scodeggio} {et~al.}(2018){Scodeggio}, {Guzzo}, {Garilli}, {Granett},
  {Bolzonella}, {de la Torre}, {Abbas}, {Adami}, {Arnouts}, {Bottini}, {Cappi},
  {Coupon}, {Cucciati}, {Davidzon}, {Franzetti}, {Fritz}, {Iovino}, {Krywult},
  {Le Brun}, {Le F{\`e}vre}, {Maccagni}, {Ma{\l}ek}, {Marchetti}, {Marulli},
  {Polletta}, {Pollo}, {Tasca}, {Tojeiro}, {Vergani}, {Zanichelli}, {Bel},
  {Branchini}, {De Lucia}, {Ilbert}, {McCracken}, {Moutard}, {Peacock},
  {Zamorani}, {Burden}, {Fumana}, {Jullo}, {Marinoni}, {Mellier}, {Moscardini},
  \& {Percival}}]{2018AA...609A..84S}
{Scodeggio}, M., {Guzzo}, L., {Garilli}, B., {et~al.} 2018, \aap, 609, A84,
  \dodoi{10.1051/0004-6361/201630114}

\bibitem[{Senior {et~al.}(2020)Senior, Evans, Jumper, Kirkpatrick, Sifre,
  Green, Qin, Žídek, Nelson, Bridgland, Penedones, Petersen, Simonyan,
  Crossan, Kohli, Jones, Silver, Kavukcuoglu, \& Hassabis}]{alphafold}
Senior, A., Evans, R., Jumper, J., {et~al.} 2020, Nature, 577, 1,
  \dodoi{10.1038/s41586-019-1923-7}

\bibitem[{Serr{\`a} {et~al.}(2020)Serr{\`a}, {\'A}lvarez, G\'{o}mez,
  Slizovskaia, N{\'u}{\~n}ez, \& Luque}]{Serra2020Input}
Serr{\`a}, J., {\'A}lvarez, D., G\'{o}mez, V., {et~al.} 2020, in International
  Conference on Learning Representations.
\newblock \url{https://openreview.net/forum?id=SyxIWpVYvr}

\bibitem[{{Shin} {et~al.}(2018){Shin}, {Chang}, {Yi}, {Kim}, {Kim}, \&
  {Byun}}]{2018AJ....156..201S}
{Shin}, M.-S., {Chang}, S.-W., {Yi}, H., {et~al.} 2018, \aj, 156, 201,
  \dodoi{10.3847/1538-3881/aae263}

\bibitem[{{Singal} {et~al.}(2011){Singal}, {Shmakova}, {Gerke}, {Griffith}, \&
  {Lotz}}]{2011PASP..123..615S}
{Singal}, J., {Shmakova}, M., {Gerke}, B., {Griffith}, R.~L., \& {Lotz}, J.
  2011, \pasp, 123, 615, \dodoi{10.1086/660155}

\bibitem[{Snoek {et~al.}(2012)Snoek, Larochelle, \&
  Adams}]{10.5555/2999325.2999464}
Snoek, J., Larochelle, H., \& Adams, R.~P. 2012, in Proceedings of the 25th
  International Conference on Neural Information Processing Systems - Volume 2,
  NIPS'12 (Red Hook, NY, USA: Curran Associates Inc.), 2951–2959

\bibitem[{{Sola} \& {Sevilla}(1997)}]{589532}
{Sola}, J., \& {Sevilla}, J. 1997, IEEE Transactions on Nuclear Science, 44,
  1464

\bibitem[{Su {et~al.}(2015)Su, Qi, Li, \& Guibas}]{7410665}
Su, H., Qi, C.~R., Li, Y., \& Guibas, L.~J. 2015, in 2015 IEEE International
  Conference on Computer Vision (ICCV), 2686--2694,
  \dodoi{10.1109/ICCV.2015.308}

\bibitem[{{Szegedy} {et~al.}(2015){Szegedy}, {Wei Liu}, {Yangqing Jia},
  {Sermanet}, {Reed}, {Anguelov}, {Erhan}, {Vanhoucke}, \&
  {Rabinovich}}]{7298594}
{Szegedy}, C., {Wei Liu}, {Yangqing Jia}, {et~al.} 2015, in 2015 IEEE
  Conference on Computer Vision and Pattern Recognition (CVPR), 1--9,
  \dodoi{10.1109/CVPR.2015.7298594}

\bibitem[{Sánchez {et~al.}(2014)Sánchez, Carrasco~Kind, Lin, Miquel, Abdalla,
  Amara, Banerji, Bonnett, Brunner, Capozzi, Carnero, Castander, da~Costa,
  Cunha, Fausti, Gerdes, Greisel, Gschwend, Hartley, Jouvel, Lahav, Lima, Maia,
  Martí, Ogando, Ostrovski, Pellegrini, Rau, Sadeh, Seitz, Sevilla-Noarbe,
  Sypniewski, de~Vicente, Abbot, Allam, Atlee, Bernstein, Bernstein,
  Buckley-Geer, Burke, Childress, Davis, DePoy, Dey, Desai, Diehl, Doel,
  Estrada, Evrard, Fernández, Finley, Flaugher, Frieman, Gaztanaga,
  Glazebrook, Honscheid, Kim, Kuehn, Kuropatkin, Lidman, Makler, Marshall,
  Nichol, Roodman, Sánchez, Santiago, Sako, Scalzo, Smith, Swanson, Tarle,
  Thomas, Tucker, Uddin, Valdés, Walker, Yuan, \&
  Zuntz}]{10.1093/mnras/stu1836}
Sánchez, C., Carrasco~Kind, M., Lin, H., {et~al.} 2014, Monthly Notices of the
  Royal Astronomical Society, 445, 1482, \dodoi{10.1093/mnras/stu1836}

\bibitem[{{Tachibana} \& {Miller}(2018)}]{2018PASP..130l8001T}
{Tachibana}, Y., \& {Miller}, A.~A. 2018, \pasp, 130, 128001,
  \dodoi{10.1088/1538-3873/aae3d9}

\bibitem[{{Tanaka}(2015)}]{2015ApJ...801...20T}
{Tanaka}, M. 2015, \apj, 801, 20, \dodoi{10.1088/0004-637X/801/1/20}

\bibitem[{{Tanaka} {et~al.}(2018){Tanaka}, {Coupon}, {Hsieh}, {Mineo},
  {Nishizawa}, {Speagle}, {Furusawa}, {Miyazaki}, \&
  {Murayama}}]{2018PASJ...70S...9T}
{Tanaka}, M., {Coupon}, J., {Hsieh}, B.-C., {et~al.} 2018, \pasj, 70, S9,
  \dodoi{10.1093/pasj/psx077}

\bibitem[{{Tonry} {et~al.}(2012){Tonry}, {Stubbs}, {Lykke}, {Doherty},
  {Shivvers}, {Burgett}, {Chambers}, {Hodapp}, {Kaiser}, {Kudritzki},
  {Magnier}, {Morgan}, {Price}, \& {Wainscoat}}]{2012ApJ...750...99T}
{Tonry}, J.~L., {Stubbs}, C.~W., {Lykke}, K.~R., {et~al.} 2012, \apj, 750, 99,
  \dodoi{10.1088/0004-637X/750/2/99}

\bibitem[{Torrisi {et~al.}(2020)Torrisi, Pollastri, \& Le}]{TORRISI20201301}
Torrisi, M., Pollastri, G., \& Le, Q. 2020, Computational and Structural
  Biotechnology Journal, 18, 1301 ,
  \dodoi{https://doi.org/10.1016/j.csbj.2019.12.011}

\bibitem[{{Trump} {et~al.}(2009){Trump}, {Impey}, {Elvis}, {McCarthy},
  {Huchra}, {Brusa}, {Salvato}, {Capak}, {Cappelluti}, {Civano}, {Comastri},
  {Gabor}, {Hao}, {Hasinger}, {Jahnke}, {Kelly}, {Lilly}, {Schinnerer},
  {Scoville}, \& {Smol{\v{c}}i{\'c}}}]{2009ApJ...696.1195T}
{Trump}, J.~R., {Impey}, C.~D., {Elvis}, M., {et~al.} 2009, \apj, 696, 1195,
  \dodoi{10.1088/0004-637X/696/2/1195}

\bibitem[{{Tyson}(2002)}]{2002SPIE.4836...10T}
{Tyson}, J.~A. 2002, in Society of Photo-Optical Instrumentation Engineers
  (SPIE) Conference Series, Vol. 4836, Survey and Other Telescope Technologies
  and Discoveries, ed. J.~A. {Tyson} \& S.~{Wolff}, 10--20,
  \dodoi{10.1117/12.456772}

\bibitem[{{Urrutia} {et~al.}(2019){Urrutia}, {Wisotzki}, {Kerutt}, {Schmidt},
  {Herenz}, {Klar}, {Saust}, {Werhahn}, {Diener}, {Caruana}, {Krajnovi{\'c}},
  {Bacon}, {Boogaard}, {Brinchmann}, {Enke}, {Maseda}, {Nanayakkara},
  {Richard}, {Steinmetz}, \& {Weilbacher}}]{2019AA...624A.141U}
{Urrutia}, T., {Wisotzki}, L., {Kerutt}, J., {et~al.} 2019, \aap, 624, A141,
  \dodoi{10.1051/0004-6361/201834656}

\bibitem[{Vaswani {et~al.}(2017)Vaswani, Shazeer, Parmar, Uszkoreit, Jones,
  Gomez, Kaiser, \& Polosukhin}]{NIPS2017_3f5ee243}
Vaswani, A., Shazeer, N., Parmar, N., {et~al.} 2017, in Advances in Neural
  Information Processing Systems, ed. I.~Guyon, U.~V. Luxburg, S.~Bengio,
  H.~Wallach, R.~Fergus, S.~Vishwanathan, \& R.~Garnett, Vol.~30 (Curran
  Associates, Inc.), 5998--6008.
\newblock
  \url{https://proceedings.neurips.cc/paper/2017/file/3f5ee243547dee91fbd053c1c4a845aa-Paper.pdf}

\bibitem[{{Walcher} {et~al.}(2011){Walcher}, {Groves}, {Budav{\'a}ri}, \&
  {Dale}}]{2011ApSS.331....1W}
{Walcher}, J., {Groves}, B., {Budav{\'a}ri}, T., \& {Dale}, D. 2011, \apss,
  331, 1, \dodoi{10.1007/s10509-010-0458-z}

\bibitem[{Yu \& Aizawa(2019)}]{Yu_2019_ICCV}
Yu, Q., \& Aizawa, K. 2019, in Proceedings of the IEEE/CVF International
  Conference on Computer Vision (ICCV)

\bibitem[{Yu-yan(2010)}]{5631385}
Yu-yan, J. 2010, in 2010 International Conference on Electrical and Control
  Engineering, 1859--1862, \dodoi{10.1109/iCECE.2010.457}

\bibitem[{{Zhang} {et~al.}(2013){Zhang}, {Ma}, {Peng}, {Zhao}, \&
  {Wu}}]{2013AJ....146...22Z}
{Zhang}, Y., {Ma}, H., {Peng}, N., {Zhao}, Y., \& {Wu}, X.-b. 2013, \aj, 146,
  22, \dodoi{10.1088/0004-6256/146/2/22}

\bibitem[{{Zhang} \& {Zhao}(2015)}]{2015DatSJ..14...11Z}
{Zhang}, Y., \& {Zhao}, Y. 2015, Data Science Journal, 14, 11,
  \dodoi{10.5334/dsj-2015-011}

\bibitem[{Zheng {et~al.}(2015)Zheng, Yang, Liu, Liang, \& Li}]{7280459}
Zheng, H., Yang, Z., Liu, W., Liang, J., \& Li, Y. 2015, in 2015 International
  Joint Conference on Neural Networks (IJCNN), 1--4,
  \dodoi{10.1109/IJCNN.2015.7280459}

\bibitem[{Zhou(2009)}]{Zhou2009}
Zhou, Z.-H. 2009, Ensemble Learning, ed. S.~Z. Li \& A.~Jain (Boston, MA:
  Springer US), 270--273, \dodoi{10.1007/978-0-387-73003-5_293}

\end{thebibliography}
\bibliographystyle{aasjournal}

\end{document}